\documentclass[a4paper,12pt]{article}

\usepackage{amsmath}
\usepackage{amssymb}
\usepackage{authblk}
\usepackage{graphicx}
\usepackage{geometry}
\usepackage[square]{natbib}

\begin{document}

\title{Variation-preserving normalization unveils \\
blind spots in gene expression profiling}

\author[1,2,*]{Carlos P.\ Roca}
\author[3]{Susana I.\ L.\ Gomes}
\author[3]{M\'{o}nica J.\ B.\ Amorim}
\author[2,*]{Janeck J.\ Scott-Fordsmand}

\affil[1]{Department of Chemical Engineering, Universitat Rovira i Virgili, 
43007 Tarragona, Spain}
\affil[2]{Department of Bioscience, Aarhus University, 8600 Silkeborg, Denmark}
\affil[3]{Department of Biology \& CESAM, University of Aveiro, 3810-193 
Aveiro, Portugal}

\affil[*]{Correspondence should be addressed to 
C.P.R.\ \texttt{(carlosproca@gmail.com)} and 
J.J.S.-F.\ \texttt{(jsf@bios.au.dk)}}

\maketitle

\newpage

\begin{abstract}

RNA-Seq and gene expression microarrays provide comprehensive profiles 
of gene activity, 
but lack of reproducibility has hindered their application. 
A key challenge in the data analysis is the normalization of 
gene expression levels, 
which is currently performed following the implicit assumption that most 
genes are not differentially expressed. 
Here, we present a mathematical approach to normalization that makes no 
assumption of this sort. 
We have found that variation in gene expression is much larger than 
currently believed, 
and that it can be measured with available assays. 
Our results also explain, at least partially, the reproducibility problems 
encountered in transcriptomics studies. 
We expect that this improvement in detection will help efforts to realize the 
full potential of gene expression profiling, 
especially in analyses of cellular processes involving complex 
modulations of gene expression. 

\end{abstract}

\vspace{3ex}

\textbf{Keywords} \\
differential gene expression \\
gene expression microarrays \\
RNA-Seq \\
normalization of high-throughput data

\vspace{3ex}

\textbf{Abbreviations} \\
DEG: Differentially Expressed Gene \\
FDR: False Discovery Rate \\
MedianCD normalization: Median Condition-Decomposition normalization \\
SVCD normalization: Standard-Vector Condition-Decomposition normalization

\newpage

\section*{Introduction}

Since the discovery of DNA structure by Watson and Crick, 
molecular biology has progressed increasingly fast, 
with rapid advances in sequencing and related genomic technologies. 
Among these, DNA microarrays and RNA-Seq have been widely adopted to obtain 
gene expression profiles, 
by measuring the concentration of tens of thousands of mRNA molecules in 
single assays 
\citep{schena:1995,lockhart:1996,duggan:1999,mortazavi:2008,wang:2009}. 
Despite their enormous potential 
\citep{golub:1999,veer:2002,ivanova:2002,chi:2003}, 
problems of reproducibility and reliability 
\citep{tan:2003,frantz:2005,couzin:2006} 
have discouraged their use in some areas, 
e.g.\ biomedicine 
\citep{michiels:2005,weigelt:2010,brettingham-moore:2011,boutros:2015}. 

The normalization of gene expression, 
which is required to set a common reference level among samples 
\citep{irizarry:2003,tarca:2006,garber:2011,conesa:2016}, 
has been reported to be problematic, 
affecting the reproducibility of results with both microarray 
\citep{shi:2006,shippy:2006,draghici:2006} and RNA-Seq 
\citep{bullard:2010,dillies:2013,su:2014,lin:2016}. 
Batch effects and their influence on normalization have 
recently received a great deal of attention 
\citep{leek:2010,reese:2013,li:2014}, 
resulting in approaches aiming to remove unwanted technical variation 
caused by differences between batches of samples or 
by other sources of expression heterogeneity 
\citep{listgarten:2010,gagnon-bartsch:2012,risso:2014}. 
A different issue, however, 
is the underlying assumption made by the most widely 
used normalization methods to date, 
such as Median and Quantile normalization \citep{bolstad:2003} 
for microarrays, 
or RPKM (Reads Per Kilobase per Million mapped reads) \citep{mortazavi:2008}, 
TMM (Trimmed Mean of M-values) \citep{robinson:2010}, 
and DESeq \citep{anders:2010} normalization for RNA-Seq, 
which posit that all or most genes are not differentially expressed 
\citep{peppel:2003,hannah:2008,loven:2012,dillies:2013,hicks:2015}. 
Although it may seem reasonable for many applications, 
this \emph{lack-of-variation assumption} has not been confirmed. 
Moreover, results obtained with external controls 
\citep{peppel:2003,hannah:2005,hannah:2008,loven:2012} 
or with RT-qPCR \citep{shi:2006,bullard:2010} 
suggest that it may not be valid. 

Some methods have been proposed to address this issue, 
based on the use of spike-ins \citep{peppel:2003,hannah:2008,loven:2012}, 
negative control probes (SQN, Subset Quantile normalization) \citep{wu:2010}, 
or negative control genes (RUV-2, Remove Unwanted Variation, 2-step) 
\citep{gagnon-bartsch:2012}. 
These methods use external or internal controls that are \emph{known a priori} 
not to be differentially expressed \citep{lippa:2010}. 
Their applicability, however, 
has been limited by this requirement of a priori knowledge, 
which is rarely available for a sufficiently large number of 
controls. 
In addition, other methods have been proposed to address the lack-of-variation 
assumption by identifying a subset of non-differentially expressed genes from 
the assay data, 
such as Cross-Correlation normalization \citep{chua:2006}, 
LVS (Least-Variant Set) normalization \citep{calza:2008}, 
and NVAS (Nonparametric Variable Selection and Approximation) normalization 
\citep{ni:2008}. 
While LVS normalization requires setting in advance a number for the fraction 
of genes to be considered as non-differentially expressed, 
with values in the range 40--60\% \citep{calza:2008}, 
Cross-Correlation and NVAS normalization are expected to degrade in performance 
when more than 50\% of genes are differentially expressed 
\citep{chua:2006,ni:2008}. 
More recently, CrossNorm has been introduced \citep{cheng:2016}, 
based on the mixture of gene expression distributions from the experimental 
conditions. 
This method, however, has been proposed for two experimental conditions, 
and specially for paired samples. 
The extension of this approach to experimental designs with unpaired samples 
and more than a few experimental conditions would lead, 
as far as we can hypothesize, 
to an unmanageable size of the data matrix to process. 

Thus, to clarify and overcome the limitations imposed by the lack-of-variation 
assumption, 
we have developed an approach to normalization that does not assume 
lack-of-variation and that is suitable to most real-world applications. 
Hence, we aimed to avoid the need of spike-ins, a priori knowledge of control 
genes, or assumptions on the number of differentially expressed genes. 
The analysis of several gene expression datasets using this approach 
confirmed that our methods reached these goals. 
Furthermore, our results show that assuming lack-of-variation can severely 
undermine the detection of gene expression variation in real assays. 
We have found that large numbers of differentially expressed genes, 
with substantial expression changes, are missed or misidentified when data are 
normalized with methods that assume lack-of-variation.

\section*{Results}

\subsection*{\emph{E. crypticus} and Synthetic Datasets}

A large gene expression dataset was obtained from biological triplicates of 
\emph{Enchytraeus crypticus} (a globally distributed soil organism used 
in standard ecotoxicity tests), sampled under 51 experimental conditions 
(42 treatments and 9 controls), 
involving exposure to several substances, 
at several concentrations and durations according to a factorial design 
(Supplementary Table~S1). 
Gene expression was measured using a customized high-density 
oligonucleotide microarray \citep{castro-ferreira:2014}, 
resulting in a dataset with 18,339 gene probes featuring good hybridization 
signal in all 153 samples. 
Taking into account the design of the microarray \citep{castro-ferreira:2014}, 
we refer to these gene probes as \emph{genes} in what follows. 

To further explore and compare outcomes between normalization methods, 
two synthetic random datasets were built and analyzed. 
One of them was generated with identical means and variances 
gene-by-gene to the real \emph{E. crypticus} dataset, 
and under the assumption that no gene was differentially expressed. 
In addition, normalization factors were applied, equal to those 
obtained from the real dataset. 
Thus, this synthetic dataset was similar to the real one, 
while complying by construction with the lack-of-variation assumption. 
The other synthetic dataset was also generated with comparable means and 
variances to the real dataset and with normalization factors, 
but in this case differential expression was added. 
Depending on the experimental condition, 
several numbers of differentially expressed genes and 
ratios between over- and under-expressed genes were introduced (see Methods). 
Together, these synthetic datasets with and without differential gene 
expression represent, respectively, the alternative and null hypotheses for a 
statistical test of differential gene expression. 

\subsection*{Normalization Methods}

The gene expression datasets were normalized with four methods. 
Two of these methods are the most widely used procedures for microarrays, 
namely Median (or Scale) normalization and Quantile normalization 
\citep{bolstad:2003}. 
(Note that current methods of normalization for RNA-Seq, 
such as RPKM \citep{mortazavi:2008}, TMM \citep{robinson:2010}, 
and DESeq \citep{anders:2010}, 
perform between-sample normalization by introducing a scaling per sample 
obtained with some form of mean or median, using all or a large set of genes. 
Thus their performance, in what concerns the issues addressed here, 
is expected to be similar to that of Median normalization for microarrays.) 

The other two normalization methods were developed for this study, 
they being called \emph{Median Condition-Decomposition normalization} and 
\emph{Standard-Vector Condition-Decomposition normalization}, 
respectively MedianCD and SVCD normalization in what follows. 

With the exception of Quantile normalization, all used methods apply 
a multiplicative factor to the expression levels of each sample, 
equivalent to the addition of a number in the usual 
$\log_2$-scale for gene expression levels. 
Solving the \emph{normalization problem} consists of finding these 
correction factors. 
This problem can be exactly and linearly decomposed into several 
sub-problems: 
one within-condition normalization for each experimental condition and 
one final between-condition normalization for the condition averages 
(see Methods). 
In the within-condition normalizations, the samples (replicates) 
subjected to each experimental condition are normalized separately, 
whereas in the final between-condition normalization 
average levels for all conditions are normalized together. 
Because there are no genes with differential expression in any of the 
within-condition normalizations, 
the lack-of-variation assumption only affects the final between-condition 
normalization. 
The assumption is avoided by using, in this normalization, 
expression levels only from \emph{no-variation genes}, 
defined as genes that show no evidence of differential expression 
under a statistical test. 
An important detail is that the within-condition normalizations ensure good 
estimates of the within-condition variances, 
which are required by the statistical test for identifying no-variation genes. 
This requisite also implies that a minimum of two samples is required per 
experimental condition. 
Both methods of normalization proposed here, 
MedianCD and SVCD normalization, 
follow this condition-decomposition approach. 

With MedianCD normalization, 
all normalizations are performed with median values, 
as in conventional Median normalization, 
but only no-variation genes are employed in the between-condition step. 
Otherwise, if all genes were used in this final step, 
the resulting total normalization factors would be exactly the same as 
those obtained with conventional Median normalization. 

For SVCD normalization, 
a vectorial procedure was developed to carry out each normalization step, 
called Standard-Vector normalization. 
The samples of any experimental condition, 
in a properly normalized dataset, 
must be \emph{exchangeable}. 
In mathematical terms, the expression levels of each gene can be 
considered as an $s$-dimensional vector, 
where $s$ is the number of samples for the experimental condition. 
After standardization (mean subtraction and variance scaling), 
these standard vectors are located in a $(s-2)$-dimensional 
hypersphere. 
The exchangeability mentioned above implies that, when properly normalized, 
the distribution of standard vectors must be invariant with respect to 
permutations of the samples and must have zero expected value. 
These properties allow to obtain a robust estimator of the normalization 
factors, 
under fairly general assumptions that do not imply any particular 
distribution of gene expression (see Methods). 

It is worth mentioning that the limit case when the number of samples is two 
($s=2$) represents a degenerate case for Standard-Vector normalization, 
in which the space of standard vectors reduces to a 0-dimensional space with 
only two points. 
In this degenerate case, Standard-Vector normalization is equivalent to 
global Loess normalization \citep{yang:2002,smyth:2003}, 
i.e.\ Loess normalization without correction for non-linearities with respect 
to the level of gene expression or microarray print-tips. 
In this sense, Standard-Vector normalization is a generalization to any 
number of samples of the approach underlying the different types of Loess 
normalization.

\subsection*{Normalization Results}

Figure~\ref{fig:01} displays the results of applying the four 
normalization methods to the real and two synthetic datasets. 
Each panel shows the interquartile range of expression levels for the 
153 samples, 
grouped in triplicates exposed to each experimental condition. 
Both Median and Quantile normalization (second and third rows) yielded 
similar outputs for the three datasets. 
In contrast, MedianCD and SVCD normalization (fourth and fifth rows) 
detected much greater variation between conditions in the real dataset and 
the synthetic dataset with differential gene expression. 
Conventional Median normalization makes, by design, the median of all 
samples to be the same, 
while Quantile normalization makes the full distribution of gene expression 
of all samples to be the same. 
Hence, if there were differences in medians or distributions between 
experimental conditions, 
both methods would have removed them. 
Such variation was indeed present in the synthetic dataset with differential 
gene expression (Fig.~\ref{fig:01}k,n), 
and hence we can hypothesize the same for the real dataset 
(Fig.~\ref{fig:01}j,m). 

\subsection*{Influence of No-Variation Genes on Normalization}

To clarify how MedianCD and SVCD normalization preserved the variation between 
conditions, 
we studied the influence of the choice of no-variation genes in the 
final between-condition normalization. 
To this end, we obtained the between-condition variation as a function of the 
number of no-variation genes, 
in two families of cases. 
In one family, no-variation genes were chosen in decreasing order of $p$-values 
from the statistical test used to analyze variation between conditions. 
In the other family, genes were chosen at random. 
The first option was similar to the approach implemented to obtain 
the results presented in Fig.~\ref{fig:01}j--o, 
with the difference that, there, 
the no-variation genes were chosen automatically, 
by a subsequent statistical test performed on the distribution of $p$-values 
(see Methods). 

For the real dataset (Fig.~\ref{fig:02}a), 
the random choice of genes resulted in $n^{-1/2}$ decays 
($n$ being the number of chosen genes), followed by a plateau. 
The $n^{-1/2}$ decays reflect the error in the estimation of normalization 
factors. 
Selecting the genes by decreasing $p$-values, however, 
yielded a completely different result. 
Up to a certain number of genes, the variance remained similar, 
but for larger numbers of genes the variance dropped rapidly. 
Figure~\ref{fig:02}a shows, therefore, 
that between-condition variation was removed as soon as the 
between-condition normalizations used genes that changed in expression 
level across experimental conditions. 
The big circles in Fig.~\ref{fig:02}a indicate the working points of the 
normalizations used for the results displayed in Fig.~\ref{fig:01}j,m. 
In fact, these points slightly underestimated the variation between 
conditions. 
Although the statistical test for identifying no-variation genes ensured that 
no evidence of variation was found, 
the expression of some selected genes varied across conditions. 

The results obtained for the synthetic dataset with differential gene 
expression (Fig.~\ref{fig:02}b) were qualitatively similar to those of the real 
dataset, 
but with two important differences. 
The amount of between-condition variation detected 
(by selecting no-variation genes by decreasing $p$-values) 
was smaller than with the real dataset, 
implying that the real dataset had larger differential gene expression. 
Additionally, the variation detected in the synthetic dataset had a simpler 
dependency on the number of genes, 
an indication that the differential gene expression introduced in the synthetic 
dataset had a simpler structure than that of the real dataset. 

Figure~\ref{fig:02}c shows the results for the synthetic dataset without 
differential gene expression. 
There were no plateaus when no-variation genes were chosen randomly, 
only $n^{-1/2}$ decays, 
and differences were small when no-variation genes were selected by 
decreasing $p$-values. 
Big circles show that the working points of Fig.~\ref{fig:01}l,o were 
selected with all available genes as no-variation genes, 
which is the optimum choice when there is no differential gene expression. 

Overall, Fig.~\ref{fig:02} shows that the between-condition variation displayed 
in Fig.~\ref{fig:01}j,k,m,n is not an artifact caused by using an exceedingly 
small or extremely particular set of genes in the final between-condition 
normalization, 
but that this variation originated from the datasets. 
The positions of the big circles in Fig.~\ref{fig:02} highlight the 
good performance of the statistical approach for choosing no-variation 
genes in the normalizations carried out for Fig.~\ref{fig:01}j--o. 
Besides, the residual variation displayed by the $n^{-1/2}$ decays implies that, 
as estimators of the normalization factors, 
SVCD normalization features smaller error than MedianCD normalization. 

\subsection*{Differential Gene Expression}

In what follows, 
we call \emph{detected positives} the differentially expressed genes (DEGs) 
resulting from the statistical analyses, 
\emph{treatment positives} the DEGs introduced in the synthetic dataset with 
differential gene expression, 
\emph{true positives} the detected positives which were also treatment 
positives, 
and \emph{false positives} the detected positives which were not treatment 
positives. 
Corresponding terms for \emph{negatives} refer to genes which were not DEGs. 

Figure~\ref{fig:03} shows the numbers of DEGs detected in the real and 
synthetic datasets, 
for each of the 42 experimental treatments compared to the corresponding 
control (Supplementary Table~S2), 
after normalizing with the four methods. 
For the real dataset (Fig.~\ref{fig:03}a), 
the number of DEGs identified after MedianCD and SVCD normalization were much 
larger for most treatments, 
in some cases by more than one order of magnitude. 
For the synthetic dataset with differential gene expression 
(Fig.~\ref{fig:03}b), results were qualitatively similar, 
but with less differential gene expression detected, 
consistently with Fig.~\ref{fig:02}a,b. 
The number of treatment positives can be displayed in this case 
(empty black down triangles, Fig.~\ref{fig:03}b), 
showing a better correlation, with MedianCD and SVCD normalization, 
between the number of treatment positives and detected positives. 
For the synthetic dataset without differential gene expression 
(Fig.~\ref{fig:03}c), 
no DEG was found but for one or two DEGs in two conditions. 
Given that the false discovery rate was controlled to be less than 5\% 
per treatment, 
this is expected to happen when evaluating 42 treatments. 

Figure~\ref{fig:03}a reports, 
for the real dataset, 
statistically significant changes of gene expression, 
that is, changes that cannot be explained by chance. 
Equally important is the effect size, 
i.e.\ the scale of detected variation in DEGs, 
which is displayed by Fig.~\ref{fig:04}. 
The boxplots show absolute fold changes of expression level for all DEGs 
detected after applying each normalization method. 
MedianCD and SVCD normalization allowed to detect smaller changes of gene 
expression, 
which were otherwise missed when using Median and Quantile normalization. 
This differential gene expression detected with MedianCD and SVCD 
normalization can hardly be considered negligible, 
given that, for all treatments, 
the interquartile range of absolute fold changes was above 1.5-fold, 
and, for more than 28 (67\%) treatments, 
the median absolute fold change was greater than 2-fold. 
Interestingly, 
the scale of differential gene expression detected with MedianCD and SVCD 
normalization in this assay is of similar magnitude to those reported by 
studies of global mRNA changes using external controls with microarrays 
and/or RNA-Seq \citep{peppel:2003,loven:2012}. 

Figure~\ref{fig:05} displays the balance of differential gene expression, 
i.e.\ the comparison between the number of over- and under-expressed genes, 
for the real dataset. 
The quantity in the $y$-axes is the mean of an indicator variable $B$, 
which assigns $+1$ to each over-expressed DEG and $-1$ to each under-expressed 
DEG. 
Hence, 
balance of differential gene expression corresponds to $\overline{B}=0$, 
all DEGs over-expressed to $\overline{B}=+1$, 
and, for example, 60\% DEGs under-expressed to $\overline{B}=-0.2$. 
As discussed below, 
and as it has been reported before 
\citep{irizarry:2006,calza:2008,ni:2008,zhu:2010}, 
the balance of differential gene expression has a strong impact on the 
performance of normalization methods. 
Figure~\ref{fig:05} shows that, 
regardless of the normalization method used, 
the unbalance of differential gene expression detected in the real dataset was 
substantial for most conditions. 
Detected unbalances were (in absolute value) larger with MedianCD and SVCD 
normalization, 
in both cases with more than 30 (71\%) treatments having $|\overline{B}|>0.5$, 
that is, more than 75\% of over- or under-expressed genes. 
Moreover, the differences between the unbalances detected with Median and 
Quantile normalization, on one hand, 
and MedianCD and SVCD normalization, on the other, 
were specially notorious for the treatments with more DEGs 
(treatments 26--42, Fig.~\ref{fig:03}a). 
In those cases, Median and Quantile normalization resulted in the smallest 
detected unbalances, 
whereas MedianCD and SVCD normalization yielded the largest ones, 
with values near $\overline{B}=\pm1$ for all but two treatments.

True differential expression was known, by construction, 
for the synthetic dataset with differential gene expression. 
Thus, Fig.~\ref{fig:06} shows for this dataset the true positive rate 
(ratio between true positives and treatment positives, 
also known as statistical power or sensitivity) 
and the false discovery rate 
(FDR, ratio between false positives and detected positives). 
With conditions 1 to 20, 
which correspond to those conditions with less than approximately 10\% of 
treatment positives 
(Fig.~\ref{fig:03}b, empty black down triangles), 
the true positive rate was similarly low for all normalizations. 
Regarding the FDR, 
when the (total) number of detected positives was up to a few tens, 
variability of the FDR around the target bound at 0.05 is to be expected, 
given that the bound is defined over an average of repetitions of the 
multiple-hypothesis test. 
Yet, the FDR obtained after Median and Quantile normalization 
was higher than the 0.05 bound for most conditions. 
More striking, however, was the behavior for conditions 21 to 42 
(more than 10\% of treatment positives). 
The true positive rates obtained after Median and Quantile normalization 
were much lower than those obtained with MedianCD and SVCD normalization, 
while the FDR after Median and Quantile normalization was 
clearly over the bound at 0.05. 
In comparison, MedianCD and SVCD normalization, 
besides offering better sensitivity of differential gene expression, 
maintained the FDR consistently below the desired bound. 

Figure~\ref{fig:07} further explores these results, 
by representing the true positive rate and false discovery rate (FDR) as a 
function of the unbalance between over- and under-expressed genes. 
Figure~\ref{fig:07} shows that the unbalance of differential gene expression 
was a key factor in the results obtained with Median and Quantile 
normalization. 
When most DEGs were over- or under-expressed, 
both the true positive rate and FDR degraded markedly after using Median or 
Quantile normalization. 
In contrast, the true positive rate and FDR were not affected by the unbalance 
of differential gene expression when using MedianCD or SVCD normalization. 

Concerning the identification of no-variation genes, 
both MedianCD and SVCD normalization performed well. 
In the synthetic dataset without differential gene expression, 
both methods identified all genes as no-variation genes, 
which is the best possible result. 
In the synthetic dataset with differential gene expression, 
1,834 genes (10\% of a total of 18,339 genes) were, by construction, 
negatives across all treatments. 
MedianCD and SVCD normalization detected, respectively, 1,723 and 1,827 
no-variation genes, among which 96.9\% and 95.2\% were true negatives. 

\subsection*{Analysis of the Golden Spike and Platinum Spike Datasets}

To provide additional evidence of the performance of MedianCD and SVCD 
normalization, 
we analyzed the Golden Spike \citep{choe:2005} and Platinum Spike 
\citep{zhu:2010} datasets. 
Both of them are artificial real datasets, 
the largest ones for which true DEGs are known. 
Hence, they have been widely used to benchmark normalization methods 
\citep{choe:2005,schuster:2007,pearson:2008,calza:2008,ni:2008,zhu:2010,cheng:2016}. 

The design of the Golden Spike dataset was questioned for reasons concerning, 
among others, 
the anomalous null distribution of $p$-values, \
the lack of biological replicates, 
and the high concentration of spike-ins 
\citep{dabney:2006,irizarry:2006,gaile:2007}. 
Nevertheless, this dataset is worth considering here because it challenges what 
we claim are key capabilities of our approach, that is, 
to correctly normalize gene expression data when many genes are differentially 
expressed, 
even with large unbalance between over- and under-expression. 
This dataset consists of microarray data obtained with the 
Affymetrix GeneChip DrosGenome1, 
with two experimental conditions and three technical replicates per condition. 
Excluding Affymetrix internal control probes, 
the dataset contains a total of 13,966 gene probe sets, 
of which 3,876 were spiked-in, 
which we call \emph{known} in what follows. 
Among these, 1,328 (34.3\%) were over-expressed (known positives) to 
varying degrees between 1.1- and 4-fold, 
while the remaining 2,535 (65.4\%) were spiked-in at the same concentration 
in both conditions (known negatives). 
(Percentages do not add up to 100\% because of a very small number 
of probe sets with weak matching to multiple clones  \citep{choe:2005}.) 

In addition to the normalization methods used above, 
we included Cyclic Loess normalization \citep{yang:2002,ballman:2004} in this 
case, 
because it facilitates a better comparison of results with previous studies 
\citep{choe:2005,schuster:2007,pearson:2008,calza:2008,ni:2008,zhu:2010,cheng:2016}. 
Figure~\ref{fig:08} summarizes the results obtained for the Golden Spike 
dataset, 
by displaying Receiver Operating Characteristic (ROC) curves for the detection 
of differential gene expression. 
The upper panel shows the true positive rate 
(as before, ratio between true positives and treatment positives) 
versus the false positive rate 
(ratio between false positives and treatment negatives), 
while the lower panel shows the number of true positives versus the number of 
false positives. 
In both cases, 
detected and treatment positives/negatives were restricted to known genes, 
following previous studies \citep{gaile:2007,schuster:2007,pearson:2008}. 
Doing otherwise would have given an excessively dominant role to the issue of 
cross-hybridization in the analysis of differential gene expression 
\citep{schuster:2007}. 
Additionally, the analysis was performed using only probe sets with 
hybridization signal in all samples, 
with the aim of factoring out differences between normalization methods caused 
by the response to missing data. 
Results obtained without this restriction (Supplementary Fig.~S3) 
or with $t$-tests instead of limma \citep{ritchie:2015} analysis 
(Supplementary Fig.~S4) were very similar to those of Fig.~\ref{fig:08}. 

The comparison of ROC curves shown in Fig.~\ref{fig:08} highlight the 
superior performance of MedianCD and, in particular, SVCD normalization. 
Dashed lines show results when the list of known negatives was given as an 
input to some of the normalization methods 
(something than cannot be done in real assays). 
It is remarkable that SVCD normalization featured equally well 
with or without this information. 

Points in Fig.~\ref{fig:08} indicate the results when controlling the 
false discovery rate (FDR) to be below 0.01 (left point on each curve) or 0.05 
(right point). 
Figure~\ref{fig:08}b shows reference lines for actual FDR equal to 0.01, 0.05, 
0.1, 0.2 and 0.5 (from left to right). 
In all cases, 
the FDR was not adequately controlled, 
although the difference between intended and actual FDR was notably smaller 
with MedianCD and SVCD normalization. 
Lack of control of the FDR in the analysis of this dataset has been previously 
reported \citep{choe:2005,pearson:2008}. 
It is caused by the non-uniform (hence anomalous) distribution of $p$-values 
for negative genes, 
which results from the analysis of differential gene expression 
\citep{dabney:2006,gaile:2007,fodor:2007,pearson:2008}. 
It has been argued that this anomalous distribution of $p$-values is, in turn, 
a consequence of the own experimental design of the dataset, 
in particular the lack of biological replication and the way clone aliquots 
were mixed to produce each gene group with a given fold change 
\citep{dabney:2006}. 
Later studies have attributed this issue mostly to non-linear or 
intensity-dependent effects, 
not properly corrected in the within-sample normalization step 
(e.g.\ background correction) of the analysis pipeline 
\citep{gaile:2007,fodor:2007,pearson:2008,zhu:2010}. 

Concerning the identification of no-variation genes, 
both MedianCD and SVCD normalization worked correctly. 
MedianCD normalization identified 561 no-variation genes, 
of which 93.9\% were known, 
and among which 84.1\% were known negatives. 
SVCD normalization, in comparison, featured better detection, 
with 1,224 no-variation genes identified, of which 94.4\% were known, 
and among which 90.0\% were known negatives.

The design of the Platinum Spike dataset \citep{zhu:2010} took into account the 
concerns raised by the Golden Spike dataset, 
offering a dataset with two experimental conditions and nine 
(three biological $\times$ three technical) replicates per condition, 
and including near 50\% more spike-ins. 
Besides, differential gene expression was balanced, with respect to both total 
mRNA amount and extent of over- and under-expression. 
Gene expression data was obtained with Affymetrix Drosophila Genome 2.0 
microarrays. 
Excluding Affymetrix internal control probes, 
the dataset contained a total of 18,769 probe sets, 
of which 5,587 were spiked-in, 
called \emph{known} as above. 
Among these, 1,940 (34.7\%) were differentially expressed (known positives) to 
varying degrees between 1.2- and 4-fold 
(1,057 over-expressed, 883 under-expressed), 
while the remaining 3,406 (61.0\%) were spiked-in at the same concentration 
in both conditions (known negatives). 

Figure~\ref{fig:09} shows ROC curves for the Platinum Spike dataset. 
As above, only known genes were considered for detected and treatment 
positives/negatives. 
Additionally, gene probes were restricted to those with signal in all samples. 
Results obtained without this restriction (Supplementary Fig.~S5) 
or with $t$-tests instead of limma analysis (Supplementary Fig.~S6) 
were again very similar. 
In contrast to the Golden Spike dataset (Fig.~\ref{fig:08}), 
the performance concerning true and false positives resulting from the 
different normalization methods was much more comparable. 
In this case, MedianCD and SVCD normalization were only marginally better. 
Note, however, that the FDR was again not properly controlled 
(Fig.~\ref{fig:09}b). 
Similarly to the Golden Spike dataset, 
and despite biological replication and a different experimental setup, 
obtained distributions of $p$-values for negative genes have been reported to 
be non-uniform \citep{zhu:2010}. 
This fact is consistent with previous arguments relating the lack of control of 
the FDR to a general problem concerning the correction of non-linearities 
in the preprocessing of microarray data 
\citep{gaile:2007,fodor:2007,pearson:2008,zhu:2010}. 

Regarding the identification of no-variation genes, 
MedianCD and SVCD normalization also worked correctly with this dataset. 
MedianCD normalization identified 2,090 no-variation genes, 
of which 95.4\% were known, 
and among which 98.7\% were known negatives. 
SVCD normalization featured slightly better, 
with 2,232 no-variation genes identified, of which 95.3\% were known, 
and among which 98.3\% were known negatives.

\section*{Discussion}

The lack-of-variation assumption underlying current methods of normalization 
was self-fulfilling, 
removing variation in gene expression that was actually present. 
Moreover, it had negative consequences for downstream analyses, 
as it removed potentially important biological information 
and introduced errors in the detection of gene expression. 
The resulting decrease in statistical power or sensitivity is a handicap, 
which can be addressed by increasing the number of samples per experimental 
condition. 
However, degradation of the (already weak) control of the false discovery rate 
when using Median or Quantile normalization is a major issue for real-world 
applications. 

The removal of variation can be understood as additive errors in the estimation 
of normalization factors. 
Considering data and errors vectorially (see Methods), 
the length of each vector equals, after centering and up to a constant 
factor, the standard deviation of the data or error. 
Errors of small magnitude, compared to the data variance, 
would only have minor effects. 
However, 
errors of similar or greater magnitude than the data variance may, 
depending on the vector lengths and the angle between the vectors, 
severely distort the observed data variance. 
This will, in turn, cause spurious results in the statistical analyses. 
Furthermore, the angles between the data and the correct normalization 
factors (considered as vectors) are random, 
given that expression data reflect biological variation while normalization 
factors respond to technical variation. 
If the assay is repeated, 
even with exactly the same experimental setup, 
the errors in the normalization factors will vary randomly, 
causing random spurious results in the downstream analyses. 
This explains, at least partially, the lack of reproducibility 
found in transcriptomics studies, 
especially for the detection of changes in gene expression of small-to-medium 
magnitude (up to 2-fold), 
because variation of this size is more likely to be distorted by errors 
in the estimation of normalization factors. 
Accordingly, the largest differences in numbers of differentially expressed 
genes detected by Median and Quantile normalization, 
compared to MedianCD and SVCD normalization, 
occurred in the treatments with the smallest magnitudes of 
gene expression changes (Figs.~\ref{fig:03}a, \ref{fig:04}). 

The variation between medians displayed in Fig.~\ref{fig:01}j,k,m,n 
may seem surprising, given routine expectations based on current methods 
(Fig.~\ref{fig:01}d,e,g,h). 
Nevertheless, this variation inevitably results from the unbalance 
between over- and under-expressed genes. 
As an illustration of this issue, let us consider a case with two experimental 
conditions, in which the average expression of a given gene is less than 
the distribution median under one condition, 
but greater than the median under the other. 
The variation of this gene alone will change the value of the median to 
the expression level of the next ranked gene. 
Therefore, if the number of over-expressed genes is different 
from the number of under-expressed genes, 
and enough changes cross the median boundary, 
then the median will substantially differ between conditions. 
Only when differential expression is negligible or is balanced with the respect 
to the median, 
will the median stay the same. 
Note that this is a related but different requirement from the number of 
over- and under-expressed genes being the same. 
This argument applies equally to any other quantile in the distribution 
of gene expression. 
The case of Quantile normalization is the least favorable, 
because it requires that changes of gene expression are balanced with respect 
to all distribution quantiles. 

Compared with other normalization approaches that try to identify 
no-variation genes from expression data, 
such as Cross-Correlation \citep{chua:2006}, LVS \citep{calza:2008}, or NVAS 
\citep{ni:2008} normalization, 
our proposal is able to work correctly with higher degrees of variation in gene 
expression, 
given that those methods are not expected to work correctly when more than 
50--60\% of genes vary. 
The reason for this difference in performance lies in that those methods use a 
binning strategy over the average expression between conditions 
(Cross-Correlation, NVAS), 
or need to assume an a priori fraction (usually 40-60\%) of non-differentially 
expressed genes (LVS). 
When the majority of genes are differentially expressed, 
very few of those bins may be suitable for normalization, 
or the assumed fraction of non-differentially expressed genes may not hold. 
In contrast, our approach makes one single search in a space of $p$-values, 
and without assuming any fraction of non-differentially expressed genes. 
As long as there are a sufficient number of non-differentially expressed genes, 
of the order of several hundreds, 
normalization is possible, 
including cases with global mRNA changes or transcriptional amplification 
\citep{peppel:2003,hannah:2005,hannah:2008,loven:2012}. 
In general, it is a matter of comparison between 
the magnitude of the error in the estimation of normalization factors and 
the amount of biological variation. 
The estimation error decreases with the number of no-variation genes 
detected (Fig.~\ref{fig:02}), 
and whenever normalization error is well below biological variation, 
normalization between samples will be correct and beneficial for downstream 
analyses. 

Our approach to normalization is based in four key ideas: 
first, decomposing the normalization by experimental conditions and normalizing 
separately each condition before normalizing the condition means; 
second, using the novel Standard-Vector normalization 
(or alternatively median scaling) to perform each normalization; 
third, identifying no-variation genes from the distribution of $p$-values 
resulting from a statistical test of variation between conditions; 
and fourth, employing only no-variation genes for the final between-condition 
normalization. 
These four ideas are grounded on rigorous mathematical statistics (see Methods 
and Supplementary Information). 
It is also worth noting that both Median and Standard-Vector normalization, 
as methods for each normalization step, 
are distribution-free methods; 
they do not assume Gaussianity or any other kind of probability distribution 
for the expression levels of genes. 
MedianCD and SVCD normalization are freely available in the R package 
\emph{cdnormbio}, installable from GitHub (see Code Availability). 

Previous assumptions that gene variation is rather limited could suggest that 
there is no need for more comprehensive normalization methods such as our 
proposal. 
In line with this, 
it could be argued that the amount of variation in our real 
(\emph{E.\ crypticus}) dataset is exceptional and much larger than the 
variation likely to be occur in most experiments. 
We think that this an invalid belief. 
Most of the available evidence concerning widespread variation in gene 
expression is inadequate, 
because it involves circular reasoning. 
We have shown here that current normalization methods, 
used by almost all studies to date, 
assume no variation in gene expression between experimental conditions, 
and they remove it if it exists, unless it is balanced. 
Therefore, these methods \emph{cannot} be used to discern the extent and 
balance of global variation in gene expression. 
Only methods that are able to normalize correctly, 
whatever these extent and balance are, 
can be trusted for this task. 
The fact that our methods perform well with large and unbalanced 
differential gene expression does not imply that they perform poorly when 
differential gene expression is more moderate or balanced. 
Our results show that this is not case. 
In the design of our methods, 
no compromise  was made to achieve good performance with high variation in 
exchange for not so good performance with low variation. 
The downside of our approach lies elsewhere, 
in a greater algorithmic complexity and a greater demand of computing 
resources. 
Yet, we consider this a minor demand, 
given the capabilities of today's computers and 
the resources required by current high-throughput assays. 

Our results have being obtained from microarray data, 
but similar effects are expected to be found in RNA-Seq assays. 
Current normalization procedures for RNA-Seq, 
such as RPKM \citep{mortazavi:2008}, TMM \citep{robinson:2010}, or DESeq 
\citep{anders:2010}, 
perform between-sample normalization based on some form of global scaling and 
under the assumption that most genes are not differentially expressed. 
This makes RPKM, TMM, and DESeq normalization, 
in what concerns between-sample normalization and the removal or distortion of 
variation discussed here, 
similar to conventional Median normalization. 
An example of this issue, 
including results from microarray and RNA-Seq assays, 
has been reported in a study of the transcriptional amplification mediated by 
the oncogene c-Myc \citep{loven:2012}.

Importantly, MedianCD and SVCD normalization were designed with no dependencies 
on any particular aspect of the technology used to globally measure gene 
expression, 
i.e.\ microarrays or RNA-Seq. 
The numbers in the input data are interpreted as steady state concentrations 
of mRNA molecules, 
in order to identify the normalization factors, 
and irrespectively of whether the concentrations were obtained from 
fluorescence intensities of hybridized cDNA (microarrays) or 
from counts of fragments of cDNA (RNA-Seq). 
Both technologies require between-sample normalization, 
because in some step of the assay the total mRNA or cDNA mass in each 
sample must be equalized within a given range required by the experimental 
platform, 
This equalization of total mass, 
together with other sources of variation in the total efficiency of the assay, 
amounts to a factor multiplying the concentration of each mRNA species. 
This factor is different for each sample, and it is what between-sample 
normalization aims to detect and correct for. 
Moreover, the total mRNA mass in each sample is, in many cases, mostly 
determined by a few highly expressed genes, 
rather than an unbiased average over the total mRNA population. 
This makes between-sample normalization critical regarding comparisons of gene 
expression between different experimental conditions, 
as our results illustrate. 
It is also important to highlight that this between-sample uncertainty in the 
measurement of mRNA concentrations is different from other issues, 
such as for example non-linearities. 
These other problems are usually more specific to each technology, 
and they are the scope of within-sample normalization 
(e.g.\ background correction for microarrays and gene-length normalization 
for RNA-Seq), 
which are obviously also necessary and should be applied \emph{before} 
between-sample normalization. 
Similarly, methods that address the influence of 
biological or technical confounding factors on downstream 
analyses, such as SVA \citep{leek:2007} or PEER \citep{stegle:2010}, 
should be applied when necessary, \emph{after} normalizing. 

Finally, the significance of widespread variation in gene expression merits 
consideration from the viewpoint of molecular and cell biology. 
Established understanding about the regulation of gene expression considers it 
as a set of processes that generally switch on or off the expression of genes, 
performed mostly at transcription initiation, 
by the combinatorial regulation of a large number of transcription factors, 
and with an emphasis on gene expression programs associated with cell 
differentiation and development. 
Recent studies, however, have expanded this understanding, 
offering a more complex perspective on the regulation of gene expression, 
by identifying other rate-limiting regulation points between transcription 
initiation and protein translation, 
such as transcription elongation and termination, as well as mRNA processing, 
transport and degradation. 
Promoter-proximal pausing of RNA polymerase II (in eukaryotes) 
\citep{core:2008, adelman:2012} and transcript elongation \citep{jonkers:2015}, 
in particular, have received a great deal of attention recently, 
in connection with gene products involved in signal transduction pathways. 
These mechanisms, which seem to be highly conserved among metazoans, 
would allow cells to tune the expression of activated genes 
in response to signals concerning, for example, homeostasis, environmental 
stress or immune response. 
As an illustration, 
studies about the transcription amplification mediated by the oncogene c-Myc 
have uncovered that it regulates the promoter-proximal pausing of RNA 
polymerase II, 
affecting a large number of genes already activated by other regulatory 
mechanisms \citep{lin:2012,nie:2012,littlewood:2012}. 
Our results for the toxicity experiment with \emph{E. crypticus} are consistent 
with regulatory capabilities for broad fine-tuning of gene expression levels, 
far beyond what conventional methods of normalization would allow to detect. 
This contrast underlines that normalization methods that truly preserve 
variation between experimental conditions are necessary for high-throughput 
assays exploring genome-wide regulation of gene expression, 
as required by current research in molecular and cell biology. 

In summary, this study proves that large numbers of genes can change in 
expression level across experimental conditions, 
and too extensively to ignore in the normalization of gene expression 
data. 
Current normalization methods for gene expression microarrays and RNA-Seq, 
because of a lack-of-variation assumption, 
likely remove and distort variation in gene expression. 
The normalization methods proposed here solve this problem, 
offering a means to investigate broad changes in gene expression that have 
remained hidden to date. 
We expect this to provide revealing insights about diverse biomolecular 
processes, 
particularly those involving substantial numbers of genes, 
and to assist efforts to realize the full potential of gene expression 
profiling.

\section*{Methods}

\subsection*{Test Organism and Exposure Media}

The test species was \emph{Enchytraeus crypticus}. 
Individuals were cultured in Petri dishes containing agar medium, 
in controlled conditions \citep{gomes:2015a}. 

For copper ($\mathrm{Cu}$) exposure, 
a natural soil collected at Hygum, Jutland, Denmark was 
used \citep{gomes:2015a,scott-fordsmand:2000}. 
For silver ($\mathrm{Ag}$) and nickel ($\mathrm{Ni}$) exposure, 
the natural standard soil LUFA 2.2 (LUFA Speyer, Germany) was 
used \citep{gomes:2015a}. 
The exposure to ultra-violet (UV) radiation was done in ISO reconstituted 
water \citep{oecd:2004a}.

\subsection*{Test Chemicals}
 
The tested $\mathrm{Cu}$ forms \citep{gomes:2015a} included copper 
nitrate ($\mathrm{Cu(NO_3)_2 \!\cdot\! 3H_2O} > 99\%$, Sigma Aldrich), 
$\mathrm{Cu}$ nanoparticles ($\mathrm{Cu}$-NPs, 20--30 nm, 
American Elements) and $\mathrm{Cu}$ nanowires ($\mathrm{Cu}$-Nwires, 
as synthesized \citep{chang:2005}). 

The tested Ag forms \citep{gomes:2015a} included silver nitratre 
($\mathrm{AgNO_3} > 99\%$, Sigma Aldrich), 
non-coated $\mathrm{Ag}$ nanoparticles ($\mathrm{Ag}$-NPs Non-Coated, 
20--30 nm, American Elements), 
Polyvinylpyrrolidone (PVP)-coated $\mathrm{Ag}$ nanoparticles 
($\mathrm{Ag}$-NPs PVP-Coated, 20--30 nm, American Elements), 
and $\mathrm{Ag}$ NM300K nanoparticles ($\mathrm{Ag}$ NM300K, 15 nm, 
JRC Repository). 
The $\mathrm{Ag}$ NM300K was dispersed in 4\% Polyoxyethylene Glycerol 
Triolaete and Polyoxyethylene (20) orbitan mono-Laurat (Tween 20), 
thus the dispersant was tested alone as control (CTdisp). 

The tested $\mathrm{Ni}$ forms included nickel nitrate 
($\mathrm{Ni(NO_3)_2 \!\cdot\! 6H_2O} \ge 98.5\%$, Fluka) and 
Ni nanoparticles ($\mathrm{Ni}$-NPs, 20 nm, American Elements).

\subsection*{Spiking Procedure}

Spiking for the $\mathrm{Cu}$ and  $\mathrm{Ag}$ materials was done as 
previously described \citep{gomes:2015a}. 
For the $\mathrm{Ni}$ materials, 
the $\mathrm{Ni}$-NPs were added to the soil as powder, 
following the same procedure as for the $\mathrm{Cu}$ materials. 
$\mathrm{NiNO_3}$, being soluble, 
was added to the pre-moistened soil as aqueous dispersions.

The concentrations tested were selected based on the reproduction effect 
concentrations $\mathrm{EC_{20}}$ and $\mathrm{EC_{50}}$, 
for \emph{E.\ crypticus}, 
within 95\% of confidence intervals, being: 
$\mathrm{CuNO_3}$ $\mathrm{EC_{20/50}}$ = 290/360 mg$\mathrm{Cu}$/kg, 
$\mathrm{Cu}$-NPs $\mathrm{EC_{20/50}}$ = 980/1760 mg$\mathrm{Cu}$/kg, 
$\mathrm{Cu}$-Nwires $\mathrm{EC_{20/50}}$ = 850/1610 mg$\mathrm{Cu}$/kg, 
$\mathrm{Cu}$-Field $\mathrm{EC_{20/50}}$ = 500/1400 mg$\mathrm{Cu}$/kg, 
$\mathrm{AgNO_3}$ $\mathrm{EC_{20/50}}$ = 45/60 mg$\mathrm{Ag}$/kg, 
$\mathrm{Ag}$-NP PVP-coated $\mathrm{EC_{20/50}}$ = 380/550 mg$\mathrm{Ag}$/kg, 
$\mathrm{Ag}$-NP Non-coated $\mathrm{EC_{20/50}}$ = 380/430 mg$\mathrm{Ag}$/kg, 
$\mathrm{Ag}$ NM300K $\mathrm{EC_{20/50}}$ = 60/170 mg$\mathrm{Ag}$/kg, 
CTdisp = 4\% w/w Tween 20,
$\mathrm{NiNO_3}$ $\mathrm{EC_{20/50}}$ = 40/60 mg$\mathrm{Ni}$/kg, 
$\mathrm{Ni}$-NPs $ \mathrm{EC_{20/50}}$ = 980/1760 mg$\mathrm{Ni}$/kg. 

Four biological replicates were performed per test condition, 
including controls. 
For $\mathrm{Cu}$ exposure, 
the control condition   for all the  treatments consisted of soil from a 
control area at Hygum site, 
which has a $\mathrm{Cu}$ background concentration of 
15 mg/kg \citep{scott-fordsmand:2000}. 
For $\mathrm{Ag}$ exposure, 
two control sets were performed: 
CT (un-spiked LUFA soil, to be the control condition for $\mathrm{AgNO_3}$, 
$\mathrm{Ag}$-NPs PVP-Coated and $\mathrm{Ag}$-NPs Non-Coated treatments) and 
CTdisp (LUFA soil spiked with the dispersant Tween 20, 
to be the control condition for the $\mathrm{Ag}$ NM300K treatments). 
For $\mathrm{Ni}$ exposure, the control consisted of un-spiked LUFA soil.

\subsection*{Exposure Details}

In soil (i.e.\ for $\mathrm{Cu}$, $\mathrm{Ag}$ and $\mathrm{Ni}$) 
exposure followed the standard ERT \citep{oecd:2004b} with adaptations 
as follows: 
twenty adults with well-developed clitellum were introduced in each 
test vessel, containing 20 g of moist soil (control or spiked). 
The organisms were exposed for three and seven days under controlled conditions 
of photoperiod (16:8 h light:dark) and temperature 
$20 \pm 1 \,^\circ\mathrm{C}$ without food. 
After the exposure period, the organisms were carefully removed from the 
soil, 
rinsed in deionized water and frozen in liquid nitrogen. 
The samples were stored at $-80\,^\circ\mathrm{C}$, until analysis.

For UV exposure, 
the test conditions \citep{oecd:2004a} were adapted for 
\emph{E.\ crypticus} \citep{gomes:2015b}. 
The exposure was performed in 24-well plates, 
where each well corresponded to a replicate and contained 1 ml of ISO water 
and five adult organisms with clitellum. 
The test duration was five days, at $20 \pm 1 \,^\circ\mathrm{C}$. 
The organisms were exposed to UV on a daily basis, 
during 15 minutes per day to two UV intensities (280--400nm) of 
$1669.25 \pm 50.83$ and $1804.08 \pm 43.10$ $\mathrm{mW/m^2}$, 
corresponding to total UV doses of 7511.6 and 8118.35 $\mathrm{J/m^2}$, 
respectively. 
The remaining time was spent under standard laboratory illumination 
(16:8 h photoperiod). 
UV radiation was provided by an UV lamp 
(Spectroline XX15F/B, Spectronics Corporation, NY, USA, peak emission at 
312 nm) and a cellulose acetate sheet was coupled to the lamp to cut-off 
UVC-range wavelengths \citep{gomes:2015b}. 
Thirty two replicates per test condition 
(including control without UV radiation) were performed to obtain 
4 biological replicates for RNA extraction, 
each one with 40 organisms. 
After the exposure period, 
the organisms were carefully removed from the water and 
frozen in liquid nitrogen. 
The samples were stored at $-80\,^\circ\mathrm{C}$, until analysis.

\subsection*{RNA Extraction, Labeling and Hybridization}

RNA was extracted from each replicate, which contained a pool of 20 and 40 
organisms, for soil and water exposure, respectively. 
Three biological replicates per test treatment (including controls) were 
used. 
Total RNA was extracted using SV Total RNA Isolation System (Promega). 
The quantity and purity were measured spectrophotometrically with a 
nanodrop (NanoDrop ND-1000 Spectrophotometer) 
and its quality checked by denaturing formaldehyde agarose gel 
electrophoresis.

500 ng of total RNA were amplified and labeled with Agilent Low Input Quick 
Amp Labeling Kit (Agilent Technologies, Palo Alto, CA, USA). 
Positive controls were added with the Agilent one-color RNA Spike-In Kit. 
Purification of the amplified and labeled cRNA was performed with RNeasy 
columns (Qiagen, Valencia, CA, USA).

The cRNA samples were hybridized on custom Gene Expression Agilent 
Microarrays (4 x 44k format), 
with a single-color design \citep{castro-ferreira:2014}. 
Hybridizations were performed using the Agilent Gene Expression 
Hybridization Kit and each biological replicate was individually 
hybridized on one array. 
The arrays were hybridized at $65\,^\circ\mathrm{C}$ with a rotation of 
10 rpm, during 17 h. 
Afterwards, microarrays were washed using Agilent Gene Expression Wash 
Buffer Kit and scanned with the Agilent DNA microarray scanner G2505B. 

\subsection*{Data Acquisition}

Fluorescence intensity data was obtained with Agilent Feature Extraction 
Software v.~10.7.3.1, 
using recommended protocol GE1\_107\_Sep09. 
Quality control was done by inspecting the reports on the Agilent Spike-in 
control probes. 

\subsection*{Data Analysis}

Analyses were performed with R \citep{r:2016} v.~3.3.1, 
using R packages plotrix \citep{lemon:2006} v.~3.6.3 and RColorBrewer 
\citep{neuwirth:2014} v.~1.1.2, 
and with Bioconductor \citep{huber:2015} v.~3.3 packages 
affy \citep{gautier:2004} v.~1.50.0, 
drosgenome1.db v~.3.2.3, 
drosophila2.db v.~3.2.3, 
genefilter v.~1.54.2, and 
limma \citep{ritchie:2015} v.~3.28.20. 
Background correction was carried out by Agilent Feature Extraction software 
for the real (\emph{E.\ crypticus}) dataset, 
while the Affymetrix MAS5 algorithm, as implemented in the limma package, 
was used for the Golden and Platinum Spike datasets. 

To ensure an optimal comparison between the different normalization 
methods, 
only gene probes with good signal quality (flag IsPosAndSignif = True)
in all samples were employed for the analysis of the \emph{E.\ crypticus} 
dataset. 
This implied the selection of 18,339 gene probes from a total of 43,750. 
For the Golden and Platinum Spike datasets, 
data were considered as missing when probe sets were not called present by the 
MAS5 algorithm.

The synthetic dataset without differential gene expression was generated gene 
by gene as normal variates with mean and variance equal, respectively, 
to the sample mean and sample variance of the expression levels for each gene, 
as detected from the real \emph{E.\ crypticus} dataset after SVCD 
normalization. 
The synthetic dataset with differential gene expression was generated 
equally, 
except for the introduction of differences in expression averages between 
treatments and controls. 
The magnitude of the difference in averages was equal, 
for each differentially expressed gene (DEG), 
to twice the sample variance. 
The percentage of DEGs for each treatment was chosen randomly, 
in logarithmic scale, 
from a range between 0.9\% and 90\%, 
while ensuring that 10\% of genes were not differentially expressed across the 
entire dataset. 
One third of the treatments were mostly over-expressed 
(for each treatment independently, the probability of a DEG being 
over-expressed was $O \sim 1-|\mathcal{N}(0,0.1^2)|$), 
one third of the treatments were mostly under-expressed 
($O \sim |\mathcal{N}(0,0.1^2)|$), 
and the remaining third had mostly balanced differential gene expression 
($O \sim \mathcal{N}(0.5,0.1^2)$). 
For both synthetic datasets, 
the applied normalization factors were those detected by SVCD normalization 
from the real \emph{E.\ crypticus} dataset. 

Median normalization was performed, for each sample, by subtracting the median 
of the distribution of expression levels, 
and then adding the overall median to preserve the global expression level. 
Quantile normalization was performed as implemented in the limma package. 

The two condition-decomposition normalizations, MedianCD and SVCD, proceeded in 
the same way: 
first, independent within-condition normalization for each experimental 
condition, using all genes. 
Then, one between-condition normalization, 
iteratively identifying no-variation genes and normalizing until convergence of 
the set of no-variation genes. 
And finally, another between-condition normalization using only the 
no-variation genes detected, 
to calculate the between-condition normalization factors.

The criterion for convergence of MedianCD normalization was to require that 
the relative changes in the 
standard deviation of the normalization factors were less than 0.1\%, 
or less than 10\% for 10 steps in a row. 
In the case of SVCD normalization, 
convergence required that numerical errors were, 
compared to estimated statistical errors (see below), 
less than 1\%, or less than 10\% for 10 steps in a row. 
Convergence of the set of no-variation genes was achieved by intersection of 
the sets found during 10 additional steps under convergence conditions. 
These default convergence parameters were used for all the MedianCD and SVCD 
normalizations reported, 
with the exception of MedianCD with the Golden Spike dataset, 
which used 30\% (instead of 10\%) of relative change for 10 steps in a row, 
to reach convergence. 

In SVCD normalization, 
the distribution of standard vectors was trimmed in each step 
to remove the 1\% more extreme values of variance. 

Differentially expressed genes were identified with limma analysis or t-tests, 
controlling the false discovery rate to be below 5\%, 
independently for each comparison of treatment versus control. 

The reference distributions with permutation symmetry shown in the polar plots 
of Supplementary Movies~S1--S3 were calculated through the six possible 
permutations of the empirical standard vectors. 
The Watson $U^2$ statistic was calculated with the two-sample test 
\citep{durbin:1973}, 
comparing with an equal number of samples obtained by sampling with replacement 
the permuted standard vectors. 

\subsection*{Condition Decomposition of the Normalization Problem}

In a gene expression dataset with $g$ genes, $c$ experimental 
conditions and $n$ samples per condition, 
the \emph{observed} expression levels of gene $j$ in condition $k$, 
$\mathbf{y}_j^{(k)} = ( y_{1j}^{(k)}, \dots , y_{nj}^{(k)} )'$, 
can be expressed in $\log_2$-scale as 
\begin{equation}
\label{eq:mm:norm-eq}
\mathbf{y}_j^{(k)} \;=\; \mathbf{x}_j^{(k)} + \mathbf{a}^{(k)} ,
\end{equation}
where $\mathbf{x}_j^{(k)}$ is the vector of \emph{true} gene expression 
levels and $\mathbf{a}^{(k)}$ is the vector of normalization factors. 

Given a sample vector $\mathbf{x}$, 
the mean vector is $ \overline{\mathbf{x}} = \bar{x} \mathbf{1} $, 
and the residual vector is $\widetilde{\mathbf{x}} = 
\mathbf{x} - \overline{\mathbf{x}}$. 
Then, \eqref{eq:mm:norm-eq} can be linearly decomposed into 
\begin{align}
\label{eq:mm:norm-vec-mean}
\overline{\mathbf{y}}_j^{(k)} 
    & \;=\; \overline{\mathbf{x}}_j^{(k)} + \overline{\mathbf{a}}^{(k)} , \\
\label{eq:mm:norm-vec-resid}
\widetilde{\mathbf{y}}_j^{(k)} 
    & \;=\; \widetilde{\mathbf{x}}_j^{(k)} + \widetilde{\mathbf{a}}^{(k)} .
\end{align}
Equations~\eqref{eq:mm:norm-vec-resid} define the within-condition 
normalizations for each condition $k$. 
The scalar values in \eqref{eq:mm:norm-vec-mean} are used to obtain 
the equations on condition means, 
\begin{align}
\label{eq:mm:norm-vec-cond-mean}
\overline{\mathbf{y}}_j^* 
    & \;=\; \overline{\mathbf{x}}_j^* + \overline{\mathbf{a}}^* , \\
\label{eq:mm:norm-vec-cond-resid}
\widetilde{\mathbf{y}}_j^* 
    & \;=\; \widetilde{\mathbf{x}}_j^* + \widetilde{\mathbf{a}}^* .
\end{align}
The between-condition normalization is defined by 
\eqref{eq:mm:norm-vec-cond-resid}. 
Equations~\eqref{eq:mm:norm-vec-cond-mean} reduce to a single number, 
which is irrelevant to the normalization. 
The complete solution for each condition is obtained with 
$\mathbf{a}^{(k)} \;=\; \overline{\mathbf{a}}^{(k)} + 
\widetilde{\mathbf{a}}^{(k)}$.

For full details about this condition-decomposition approach, 
see Supplementary Mathematical Methods in the Supplementary Information. 

\subsection*{Standard-Vector Normalization}

The $n$ samples of gene $j$ in a given condition can be modeled with the 
random vectors $\mathbf{X}_j, \mathbf{Y}_j \in \mathbb{R}^n$. 
Again, $\mathbf{Y}_j = \mathbf{X}_j + \mathbf{a}$, 
where $\mathbf{a}$ is a fixed vector of normalization factors. 
It can be proved under fairly general assumptions 
(see Supplementary Information), 
that the true standard vectors have zero expected value 
\begin{equation}
\mathrm{E} \left( \sqrt{n-1}\, \frac{ \widetilde{\mathbf{X}}_j }
    { \| \widetilde{\mathbf{X}}_j \| } \right) \;=\; \mathbf{0} ,
\end{equation}
whereas the observed standard vectors verify, 
as long as $\mathbf{a} \ne \mathbf{0}$, 
\begin{equation}
0 \;<\; \mathrm{E} \left( 
    \sqrt{n-1}\, \frac{ \widetilde{\mathbf{Y}}_j }
    { \| \widetilde{\mathbf{Y}}_j \| } \right)' 
    \frac{ \widetilde{\mathbf{a}} }{ \| \widetilde{\mathbf{a}} \| } 
\;<\; \mathrm{E} \left( 
    \sqrt{n-1}\, \frac{ 1 }{ \| \widetilde{\mathbf{Y}}_j \| } \right) 
    \| \widetilde{\mathbf{a}} \| . 
\end{equation}

This motivates the following iterative procedure to 
solve \eqref{eq:mm:norm-vec-resid} and \eqref{eq:mm:norm-vec-cond-resid} 
(\emph{standard-vector normalization}): 
\begin{align}
\widehat{\mathbf{y}}_j^{(0)} & \;=\; \widetilde{\mathbf{y}}_j , \\
\widehat{\mathbf{y}}_j^{(t)} & \;=\; \widehat{\mathbf{y}}_j^{(t-1)} - 
    \, \widehat{\mathbf{b}}^{(t-1)} ,
    \quad \text{for } t \ge 1 , \\
\widehat{\mathbf{b}}^{(t)} & \;=\; \frac{\displaystyle 
     \sum_{j=1}^g  \frac{ \widehat{\mathbf{y}}_j^{(t)} }
    { \| \widehat{\mathbf{y}}_j^{(t)} \| } }
    {\displaystyle \sum_{j=1}^g \frac{1}
    { \| \widehat{\mathbf{y}}_j^{(t)} \| } } ,
    \quad \text{for } t \ge 0 .
\end{align}

At convergence, 
$\lim_{t \to \infty} \widehat{\mathbf{b}}^{(t)} = \mathbf{0}$, 
which implies $\lim_{t \to \infty} \widehat{\mathbf{y}}_j^{(t)} = 
\widetilde{\mathbf{x}}_j$ and 
$\sum_{t=0}^\infty \widehat{\mathbf{b}}^{(t)} = 
    \widetilde{\mathbf{a}}$. 
Convergence is faster the more symmetric the empirical distribution of 
$ \widetilde{\mathbf{x}}_j / \| \widetilde{\mathbf{x}}_j \| $ 
is on the unit $(n-2)$-sphere. 
Convergence is optimal with spherically symmetric distributions, 
such as the Gaussian distribution, 
because in that case 
\begin{equation}
\mathrm{E} \left( 
    \frac{ \widetilde{\mathbf{Y}}_j }{ \| \widetilde{\mathbf{Y}}_j \| } 
    \right) \;=\; \lambda \widetilde{\mathbf{a}} , 
    \quad \text{with} \quad
    0 < \lambda < \mathrm{E} \left( 
    \frac{ 1 }{ \| \widetilde{\mathbf{Y}}_j \| } \right) . 
\end{equation} 

Assuming no dependencies between genes, 
an approximation of the statistical error at step $t$ can be obtained 
with 
\begin{equation}
\mathrm{E} \left(\, \| \widehat{\mathbf{b}}^{(t)} \| \,\right)
    \;\approx\; \frac{ \sqrt{g} }
    {\displaystyle \sum_{j=1}^g \frac{ 1 }
    { \| \widehat{\mathbf{y}}_j^{(t)} \| } } .
\end{equation}
This statistical error was compared with the numerical error to assess 
convergence. 

See Supplementary Mathematical Methods in the Supplementary Information 
for full details about this algorithm. 
See also Supplementary Movies S1--S3 for normalization examples. 

\subsection*{Identification of No-Variation Genes}

No-variation genes were identified with one-sided Kolmogorov-Smirnov tests, 
as goodness-of-fit tests against the uniform distribution, 
carried out on a distribution of $p$-values. 
These $p$-values were obtained from ANOVA tests on the expression levels of 
genes, grouped by experimental condition. 
The KS test was rejected at $\alpha = 0.001$. 

See Supplementary Mathematical Methods in the Supplementary Information for 
more details about this approach to identify no-variation genes. 
See also Supplementary Movies S4--S6 for examples of use.

\section*{Acknowledgements}

This work was supported by the European Union FP7 projects 
MODERN (Ref.\ 309314-2, to C.P.R., J.J.S.-F.), 
MARINA (Ref.\ 263215, to J.J.S.-F.), 
and SUN (Ref.\ 604305, to C.P.R., S.I.L.G., M.J.B.A., J.J.S.-F.), 
by FEDER funding through COMPETE (Programa Operacional Factores de 
Competitividade) and Portugal funding through FCT (Funda\c{c}\~{a}o para a 
Ci\^{e}ncia e Tecnologia) within the project 
bio-CHIP (Refs.\ FCOMP-01-0124-FEDER-041177, FCT EXPL/AAG-MAA/0180/2013, 
to S.I.L.G., M.J.B.A.), 
and by a post-doctoral grant (Ref.\ SFRH/BPD/95775/2013, to S.I.L.G.)

\section*{Author Contributions}

S.I.L.G., M.J.B.A.\ and J.J.S.-F. designed the toxicity experiment. 
S.I.L.G.\ carried out the experimental work and collected the microarray data. 
C.P.R.\ designed and implemented the novel normalization methods. 
C.P.R.\ performed the statistical analyses. 
All the authors jointly discussed the results. 
C.P.R.\ drafted the paper, with input from all the authors. 
All the authors edited the final version of the paper.

\section*{Additional Information}

\subsection*{Data Deposition and Code Availability}

MIAME-compliant microarray data were submitted to the Gene Expression Omnibus 
(GEO) repository at the NCBI website 
(platform: GPL20310; series: GSE69746, GSE69792, GSE69793 and GSE69794). 
The novel normalization methods were implemented as R functions. 
This code, together with R scripts that generate the synthetic datasets 
and reproduce all reported results starting from the raw microarray data, 
are available at the GitHub repository 
\texttt{https://github.com/carlosproca/gene-expr-norm-paper} and 
in the Supplementary Data File \texttt{gene-expr-norm.zip}. 
Additionally, MedianCD and SVCD normalization are available via the R package 
\emph{cdnormbio}, installable from the GitHub repository 
\texttt{https://github.com/carlosproca/cdnormbio}. 

\subsection*{Competing Financial Interests}

The authors declare that they have no competing financial interests.

\newpage

\section*{Figures}

\begin{figure}[ht!]
\centering
\includegraphics[width=160mm]{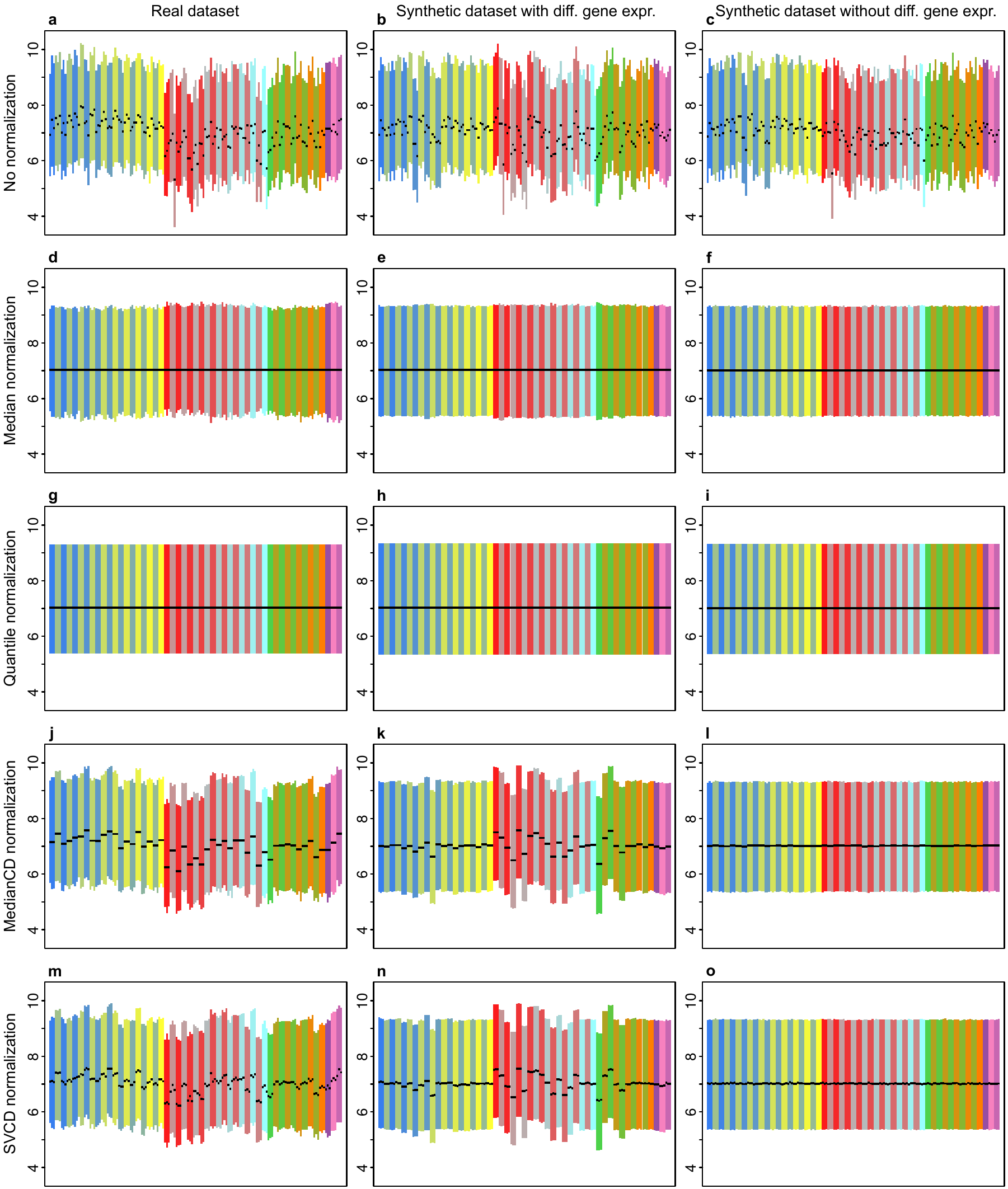}
\end{figure}

\begin{center}
Figure 1
\end{center}

\clearpage

\begin{figure}[ht]
\caption{
MedianCD and SVCD normalization resulted in the detection of much larger 
between-condition variation in the datasets with differential gene expression, 
compared to Median and Quantile normalization. 
Panels show interquartile ranges of expression levels for the 153 samples, 
grouped by the 51 experimental conditions 
(Ag, blue-yellow; Cu, red-cyan; Ni, green-orange; UV, purple; 
see Supplementary Table~S1). 
Black lines indicate medians. 
Rows and columns correspond to normalization methods and datasets, 
respectively, as labeled.
}
\label{fig:01}
\end{figure}

\clearpage

\begin{figure}[ht!]
\centering
\includegraphics[width=80mm]{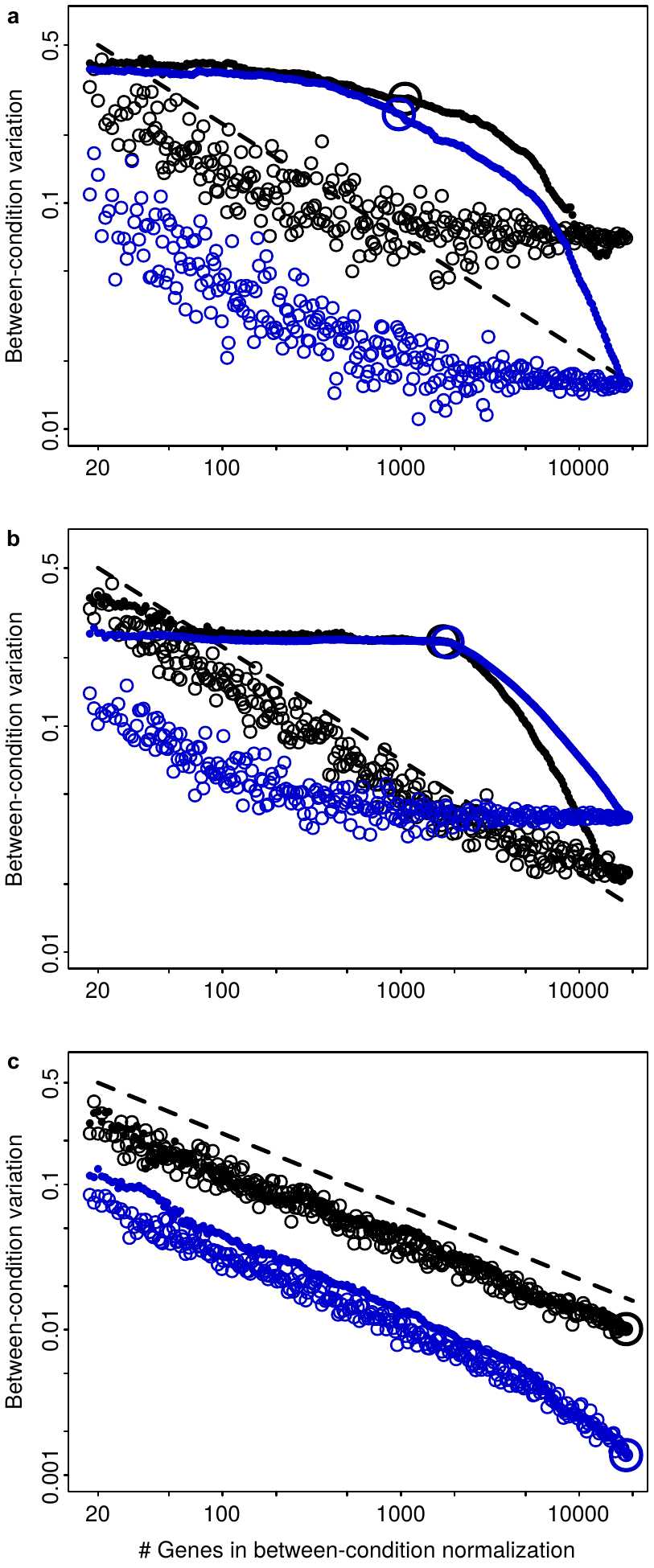}
\end{figure}

\begin{center}
Figure 2
\end{center}

\clearpage

\begin{figure}[ht]
\caption{
The selection of genes for the final between-condition normalization 
in MedianCD and SVCD normalization was crucial to preserve the variation 
between conditions. 
Panels show the detected variation as a function of the number of 
genes used in the between-condition normalization, 
for the real dataset (a), 
synthetic dataset with differential gene expression (b), 
and synthetic dataset without differential gene expression (c). 
Between-condition variation is represented as the standard deviation of 
the within-condition mean averages 
(averages of sample means, for all samples of the condition). 
See Supplementary Fig.~S1 for results using within-condition median 
averages, with similar behavior. 
Each point in each panel indicates the variation obtained with one complete 
normalization 
(black circles, MedianCD normalization; blue circles, SVCD normalization). 
Genes were selected in two ways: 
randomly (empty circles) or in decreasing order of $p$-values from a test for 
detecting no-variation genes (filled circles). 
Big circles show the working points corresponding to the results depicted 
in Fig.~1j--o, which were chosen automatically. 
Black dashed lines show references for $n^{-1/2}$ decays, 
with the same values in all panels.
}
\label{fig:02}
\end{figure}

\clearpage

\begin{figure}[ht!]
\centering
\includegraphics[width=80mm]{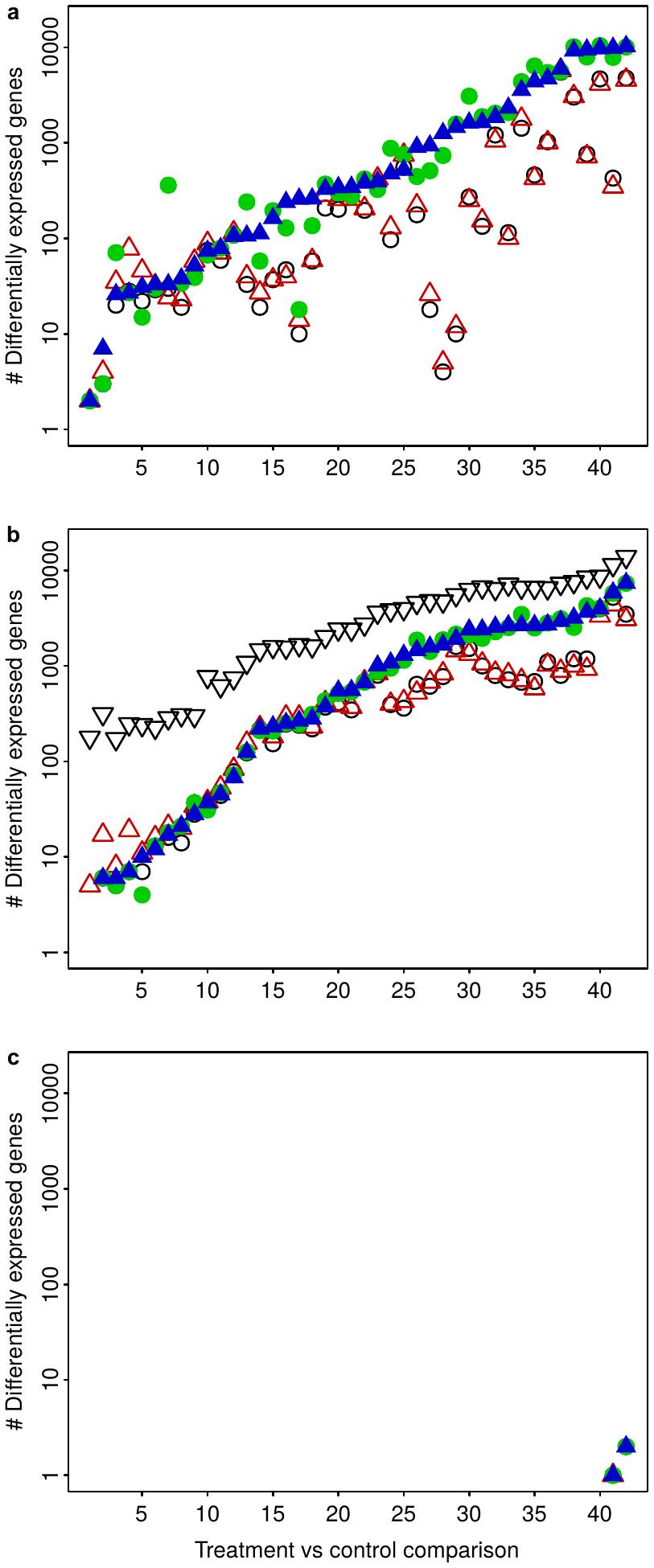}
\end{figure}

\begin{center}
Figure 3
\end{center}

\clearpage

\begin{figure}[ht]
\caption{
MedianCD and SVCD normalization allowed to detect much larger numbers 
of differentially expressed genes (DEGs) in the datasets with 
differential gene expression. 
Panels show results for the real dataset (a), 
synthetic dataset with differential gene expression (b), 
and synthetic dataset without differential gene expression (c). 
They display the number of DEGs for each treatment compared to the 
corresponding control, 
obtained after applying the four normalization methods 
(empty black circles, Median normalization; 
empty red up triangles, Quantile normalization; 
filled green circles, MedianCD normalization; 
filled blue up triangles, SVCD normalization). 
For the synthetic dataset with differential gene expression (b), 
the numbers of treatment positives are also shown, 
as empty black down triangles. 
In each panel, treatments are ordered according to the number of 
DEGs identified with SVCD normalization, 
increasing from left to right 
(see Supplementary Table~2, for real dataset). 
Differential gene expression was analyzed with R/Bioconductor package limma. 
Supplementary Fig.~S2 shows results obtained with $t$-tests, 
qualitatively similar but with much lower detection of differential gene 
expression. 
}
\label{fig:03}
\end{figure}

\clearpage

\begin{figure}[ht!]
\centering
\includegraphics[width=160mm]{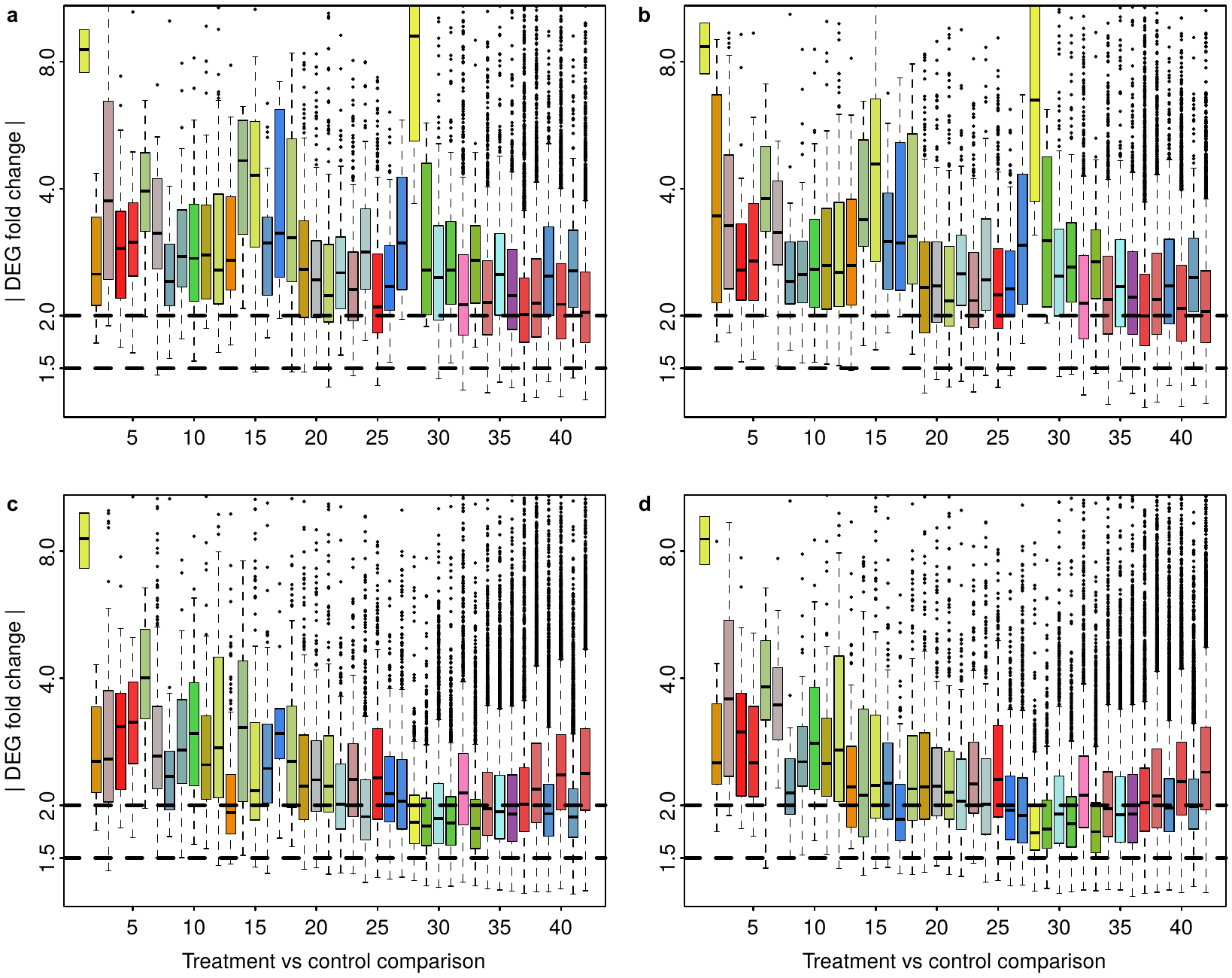}
\end{figure}

\begin{center}
Figure 4
\end{center}

\clearpage

\begin{figure}[ht]
\caption{
For the real dataset, 
MedianCD and SVCD normalization allowed to detect variation in gene expression 
of smaller magnitude than with Median and Quantile normalization. 
Boxplots display absolute values of DEG fold changes, 
for each treatment compared to the corresponding control, 
obtained after Median normalization (a), 
Quantile normalization (b), 
MedianCD normalization (c), 
and SVCD normalization (d). 
Boxplots are colored by treatment, with the same color code as in 
Figs.~\ref{fig:01}. 
All panels have the same order of treatments as in Fig.~\ref{fig:03}a, 
i.e.\ in increasing number of DEGs identified with SVCD normalization 
(Supplementary Table~2). 
Dashed horizontal lines indicate references of 1.5-fold and 2-fold changes. 
}
\label{fig:04}
\end{figure}

\clearpage

\begin{figure}[ht!]
\centering
\includegraphics[width=160mm]{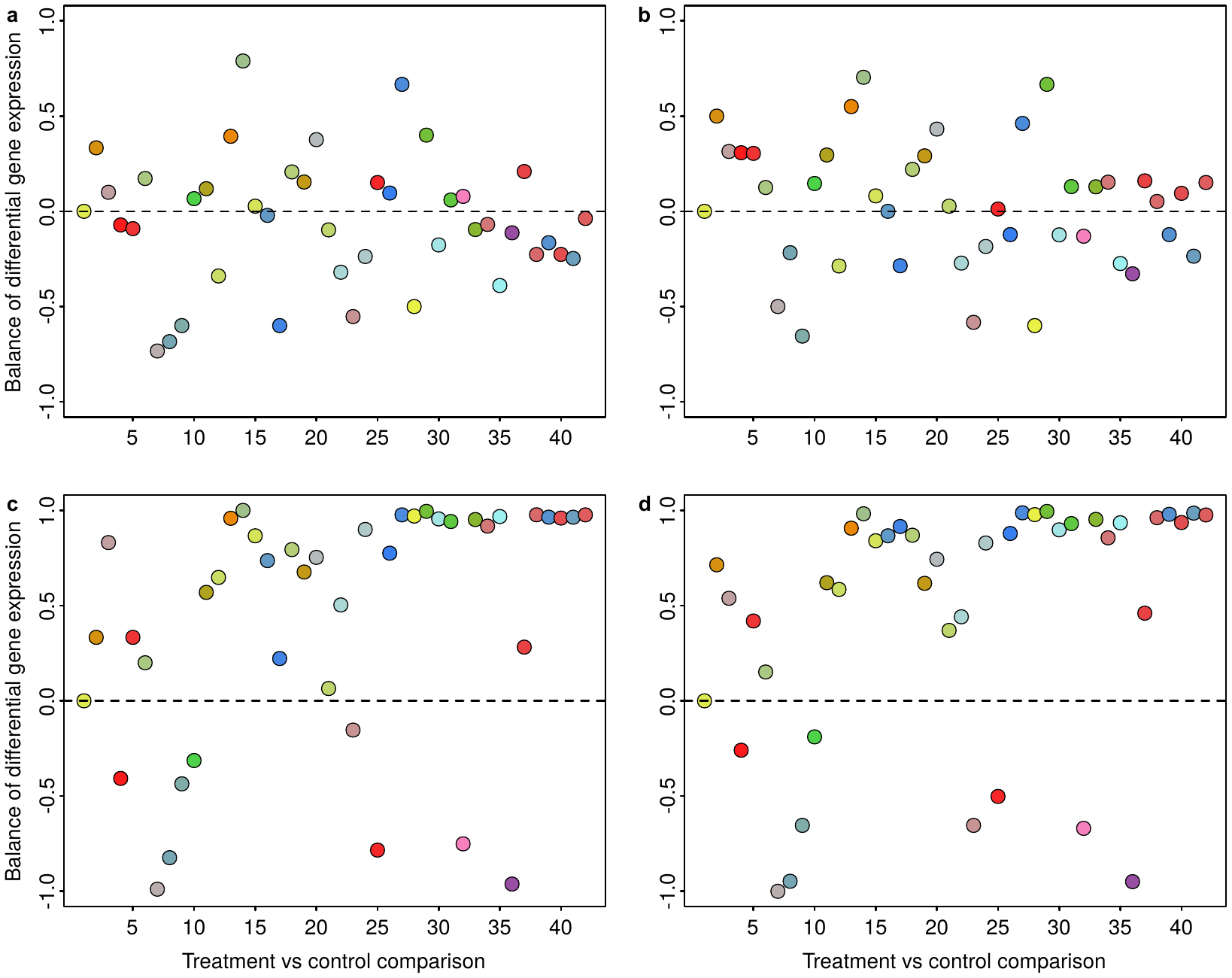}
\end{figure}

\begin{center}
Figure 5
\end{center}

\clearpage

\begin{figure}[ht]
\caption{
For the real dataset, detected differential gene expression was unbalanced, 
specially after using MedianCD and SVCD normalization and for the treatments 
with more DEGs (Fig.~\ref{fig:03}a). 
Panels show the balance of differential gene expression, 
for each treatment compared to the corresponding control, 
obtained after Median normalization (a), 
Quantile normalization (b), 
MedianCD normalization (c), 
and SVCD normalization (d). 
Each point represents the balance of differential gene expression, 
$\overline{B}$ 
($\overline{B}=0$, same number of over- and under-expressed genes; 
$\overline{B}=+1$, all DEGs over-expressed; 
$\overline{B}=-0.5$, 75\% DEGs under-expressed). 
Points are colored by treatment, with the same color code as in 
Figs.~\ref{fig:01}, \ref{fig:04}. 
All panels have the same order of treatments as in Figs.~\ref{fig:03}a, 
\ref{fig:04}, 
i.e.\ in increasing number of DEGs identified with SVCD normalization 
(Supplementary Table~2). 
Dashed horizontal lines indicate references for balanced differential gene 
expression ($\overline{B}=0$). 
}
\label{fig:05}
\end{figure}

\clearpage

\begin{figure}[ht!]
\centering
\includegraphics[width=80mm]{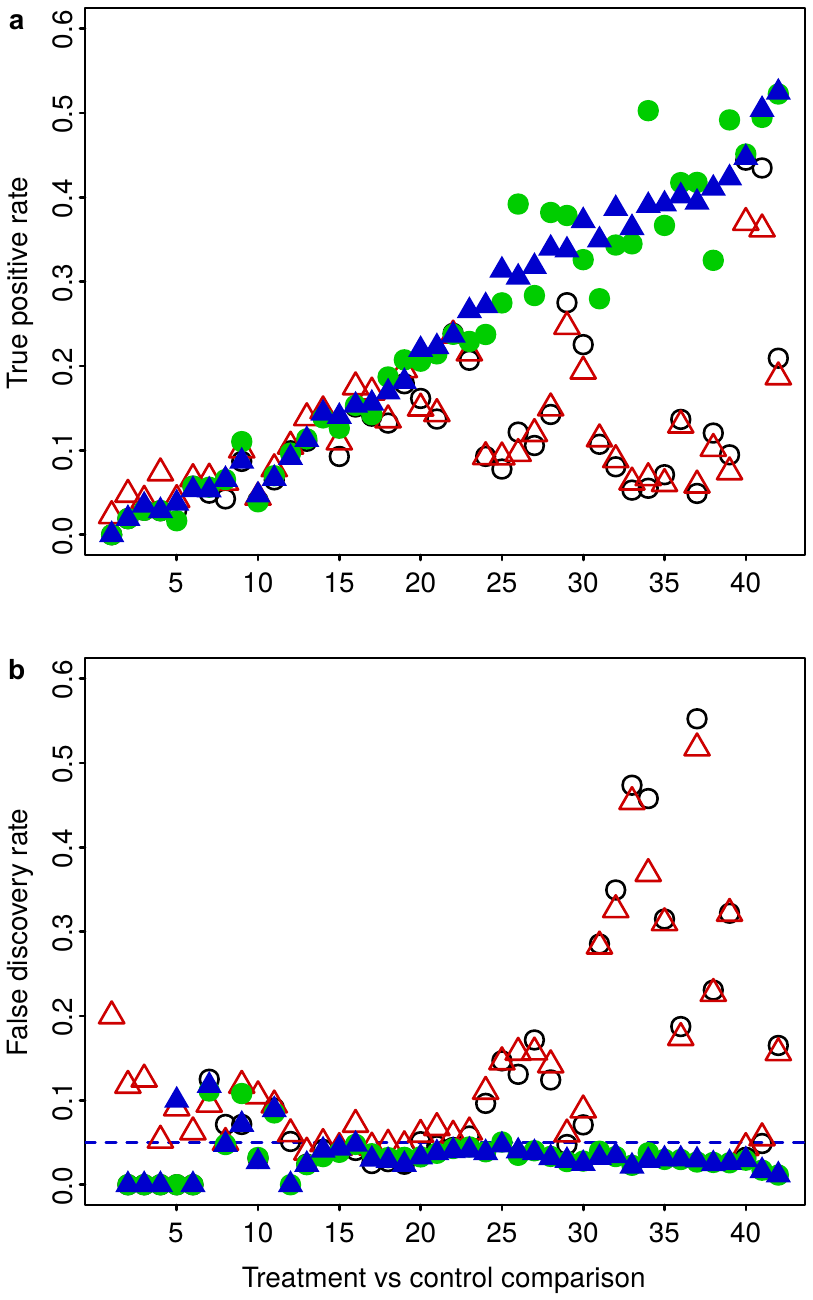}
\end{figure}

\begin{center}
Figure 6
\end{center}

\clearpage

\begin{figure}[ht]
\caption{
In the synthetic dataset with differential gene expression, and 
with more than 10\% of treatment positives (Fig.~\ref{fig:03}b), 
Median and Quantile normalization resulted in less statistical power and 
uncontrolled false discovery rate. 
The panels display the true positive rate (a) and false discovery rate (b), 
for each treatment compared to the corresponding control, 
obtained after applying the four normalization methods 
(same symbols as in Fig.~\ref{fig:03}; 
empty black circles, Median normalization; 
empty red up triangles, Quantile normalization; 
filled green circles, MedianCD normalization; 
filled blue up triangles, SVCD normalization). 
Both panels have the same order of treatments as in Fig.~\ref{fig:03}b, 
i.e.\ in increasing number of differentially expressed genes identified with 
SVCD normalization. 
Differential gene expression was analyzed with R/Bioconductor package limma. 
The dashed horizontal line in (b) indicates the desired bound on the false 
discovery rate at 0.05. 
}
\label{fig:06}
\end{figure}

\clearpage

\begin{figure}[ht!]
\centering
\includegraphics[width=80mm]{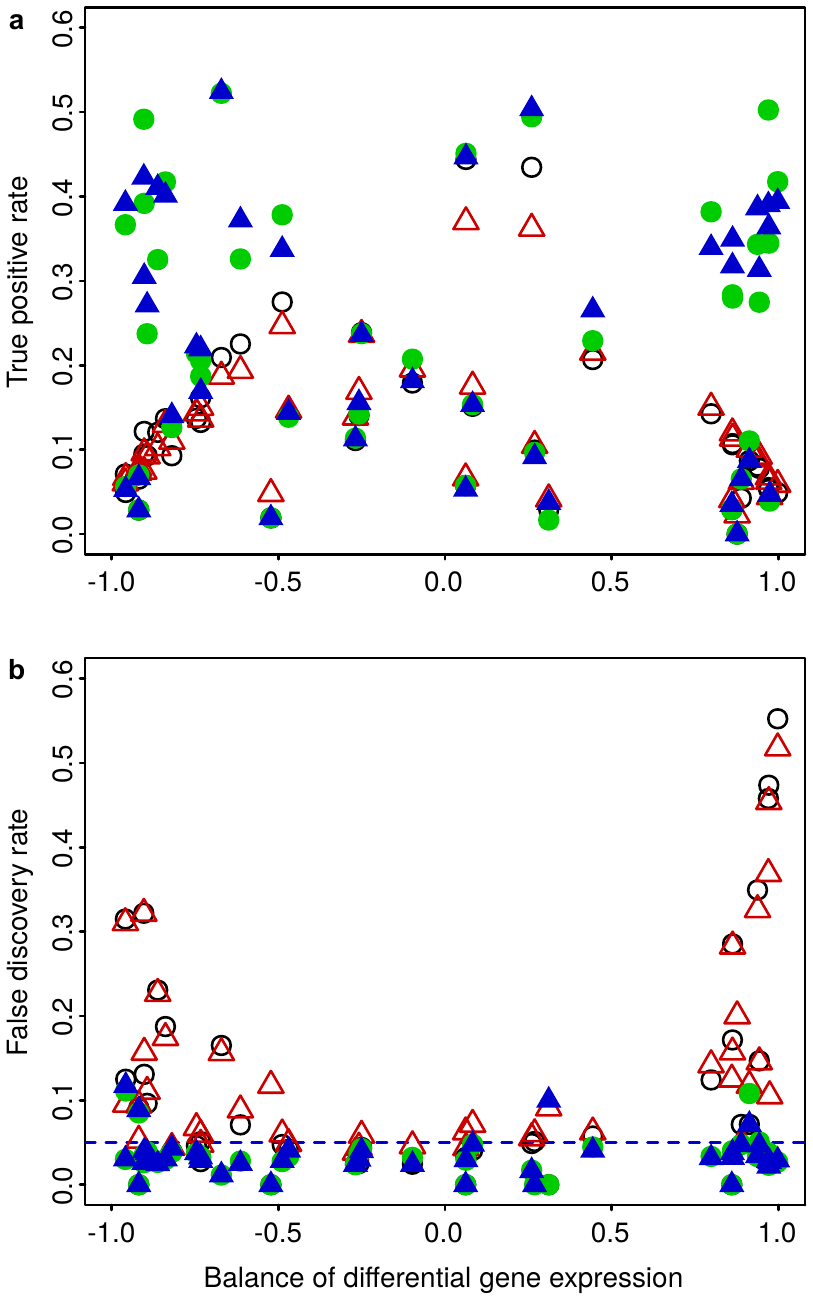}
\end{figure}

\begin{center}
Figure 7
\end{center}

\clearpage

\begin{figure}[ht]
\caption{
In the synthetic dataset with differential gene expression (Fig.~\ref{fig:06}), 
the unbalance between over- and under-expressed genes was a key factor in the 
lowered true positive rate and uncontrolled false discovery rate obtained after 
Median and Quantile normalization. 
Panels show the true positive rate (a) and false discovery rate (b) 
as a function of the balance of differential gene expression, $\overline{B}$ 
($\overline{B}=0$, same number of over- and under-expressed genes; 
$\overline{B}=+1$, all DEGs over-expressed; 
$\overline{B}=-0.5$, 75\% DEGs under-expressed). 
Each point in both panels represents the results for one treatment compared to 
the corresponding control, 
obtained after applying the four normalization methods 
(same symbols as in Figs.~\ref{fig:03}, \ref{fig:06}; 
empty black circles, Median normalization; 
empty red up triangles, Quantile normalization; 
filled green circles, MedianCD normalization; 
filled blue up triangles, SVCD normalization). 
Differential gene expression was analyzed with R/Bioconductor package limma. 
The dashed horizontal line in (b) indicates the desired bound on the false 
discovery rate at 0.05. 
}
\label{fig:07}
\end{figure}

\clearpage

\begin{figure}[ht!]
\centering
\includegraphics[width=80mm]{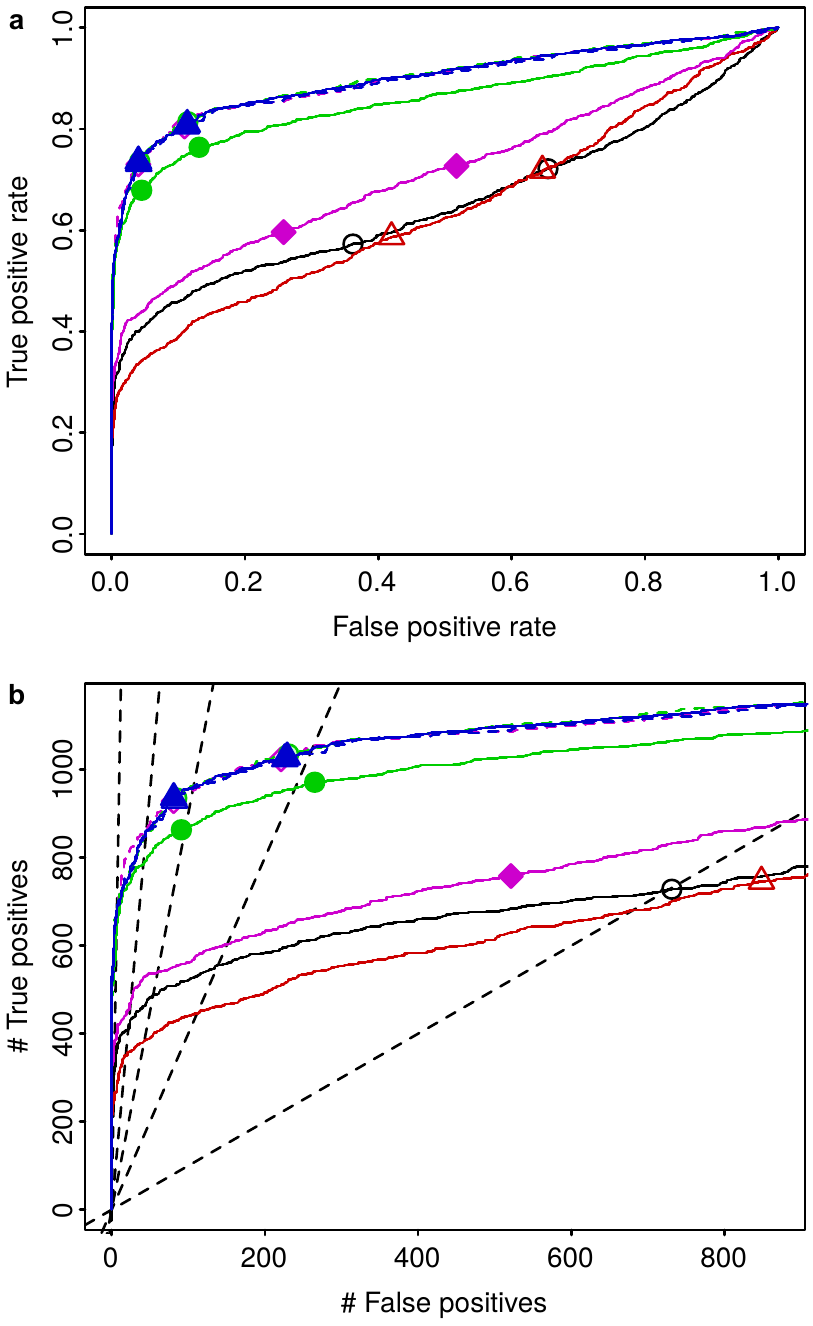}
\end{figure}

\begin{center}
Figure 8
\end{center}

\clearpage

\begin{figure}[ht]
\caption{
In the Golden Spike dataset, 
the best detection of differential gene expression was achieved after using 
MedianCD and, specially, SVCD normalization. 
Panels display ROC curves, 
with the true positive rate versus the false positive rate (a), 
or the number of true positives versus the number of false positives (b). 
Each curve shows the results obtained after applying the four normalization 
methods plus Cyclic Loess normalization 
(same colors and symbols as in Figs.~\ref{fig:03}, \ref{fig:06}, \ref{fig:07}; 
black curve with empty black circles, Median normalization; 
red curve with empty red up triangles, Quantile normalization; 
green curve with filled green circles, MedianCD normalization; 
blue curve with filled blue up triangles, SVCD normalization; 
magenta curve with filled magenta diamonds, Cyclic Loess normalization). 
Dashed curves with lightly filled symbols, 
overlapping the response of SVCD normalization, 
show results when the list of known negatives was provided to MedianCD, SVCD, 
and Cyclic Loess normalization. 
The two points per normalization method show results when controlling the 
false discovery rate (FDR) to be below 0.01 (left point) or 0.05 (right point). 
Dashed lines in (b) show references for actual FDR equal to 0.01, 0.05, 0.1, 
0.2, or 0.5 (from left to right). 
Compared to MedianCD and SVCD normalization, 
the other normalization methods resulted in notably more severe degradation of 
the FDR. 
}
\label{fig:08}
\end{figure}

\clearpage

\begin{figure}[ht!]
\centering
\includegraphics[width=80mm]{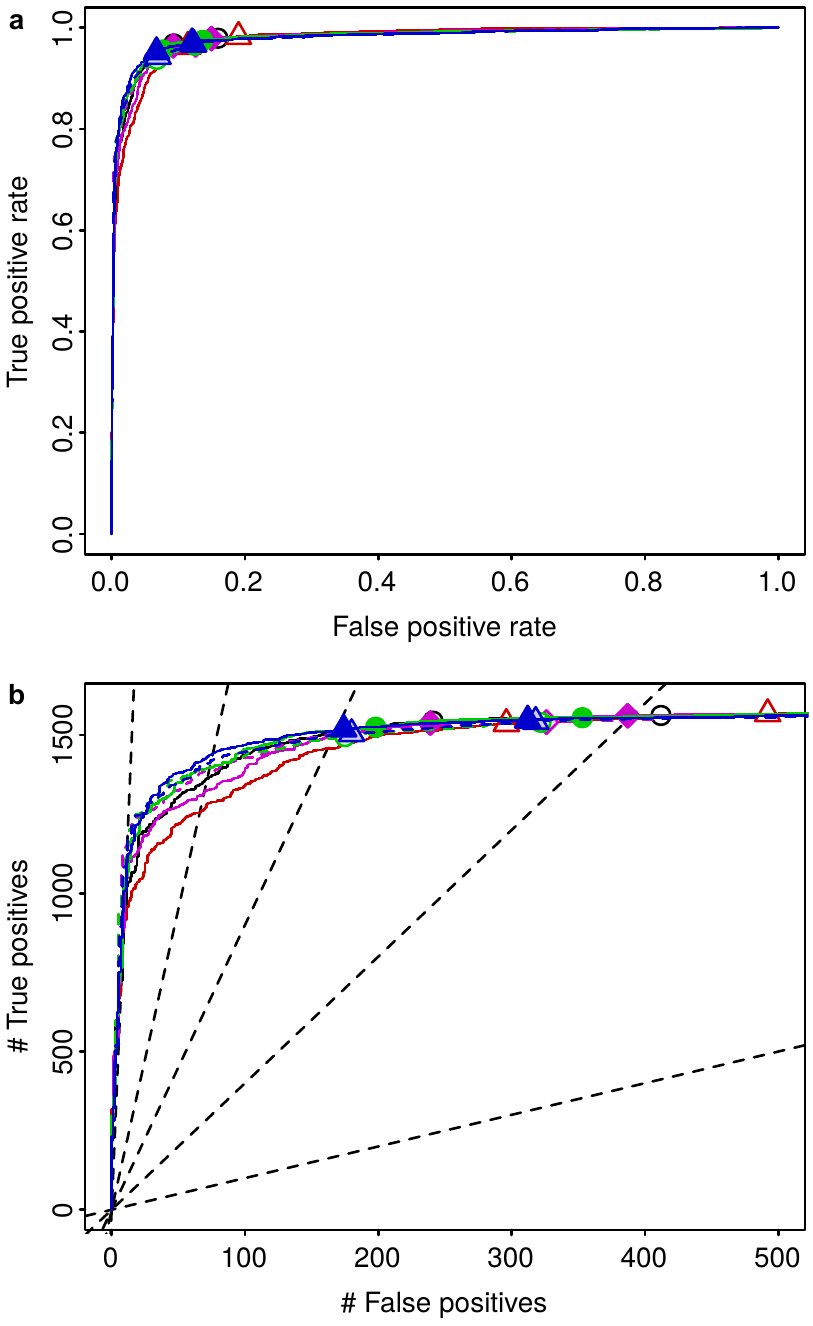}
\end{figure}

\begin{center}
Figure 9
\end{center}

\clearpage

\begin{figure}[ht]
\caption{
In the Platinum Spike dataset, 
all normalization methods resulted in similar detection of differential 
gene expression, 
with MedianCD and SVCD normalization being only marginally better. 
Panels display ROC curves, 
with the true positive rate versus the false positive rate (a), 
or the number of true positives versus the number of false positives (b). 
Each curve shows the results obtained after applying the four normalization 
methods plus Cyclic Loess normalization 
(same colors and symbols as in Figs.~\ref{fig:03}, \ref{fig:06}--\ref{fig:08}; 
black curve with empty black circles, Median normalization; 
red curve with empty red up triangles, Quantile normalization; 
green curve with filled green circles, MedianCD normalization; 
blue curve with filled blue up triangles, SVCD normalization; 
magenta curve with filled magenta diamonds, Cyclic Loess normalization). 
Dashed curves with lightly filled symbols show results when the list of known 
negatives was provided to MedianCD, SVCD, and Cyclic Loess normalization. 
As in Fig.~\ref{fig:08}, 
the two points per normalization method show results when controlling the 
false discovery rate (FDR) to be below 0.01 (left point) or 0.05 (right point). 
Dashed lines in (b) show references for actual FDR equal to 0.01, 0.05, 0.1, 
0.2, or 0.5 (from left to right). 
Compared to the Golden Spike dataset (Fig.~\ref{fig:08}), 
the difference between normalization methods in the resulting degradation of 
the FDR was smaller for this dataset. 
}
\label{fig:09}
\end{figure}

\clearpage

\section*{Supplementary Tables}

\renewcommand{\tablename}{Supplementary Table}
\renewcommand{\thetable}{S\arabic{table}}

\begin{table}[ht!]
\caption{
Experimental conditions of the toxicity experiment on \emph{E.\ crypticus}, 
listed in the same order as they appear in each panel of Fig.~1, 
from left to right.
}
\begin{center}
\includegraphics[height=194mm]{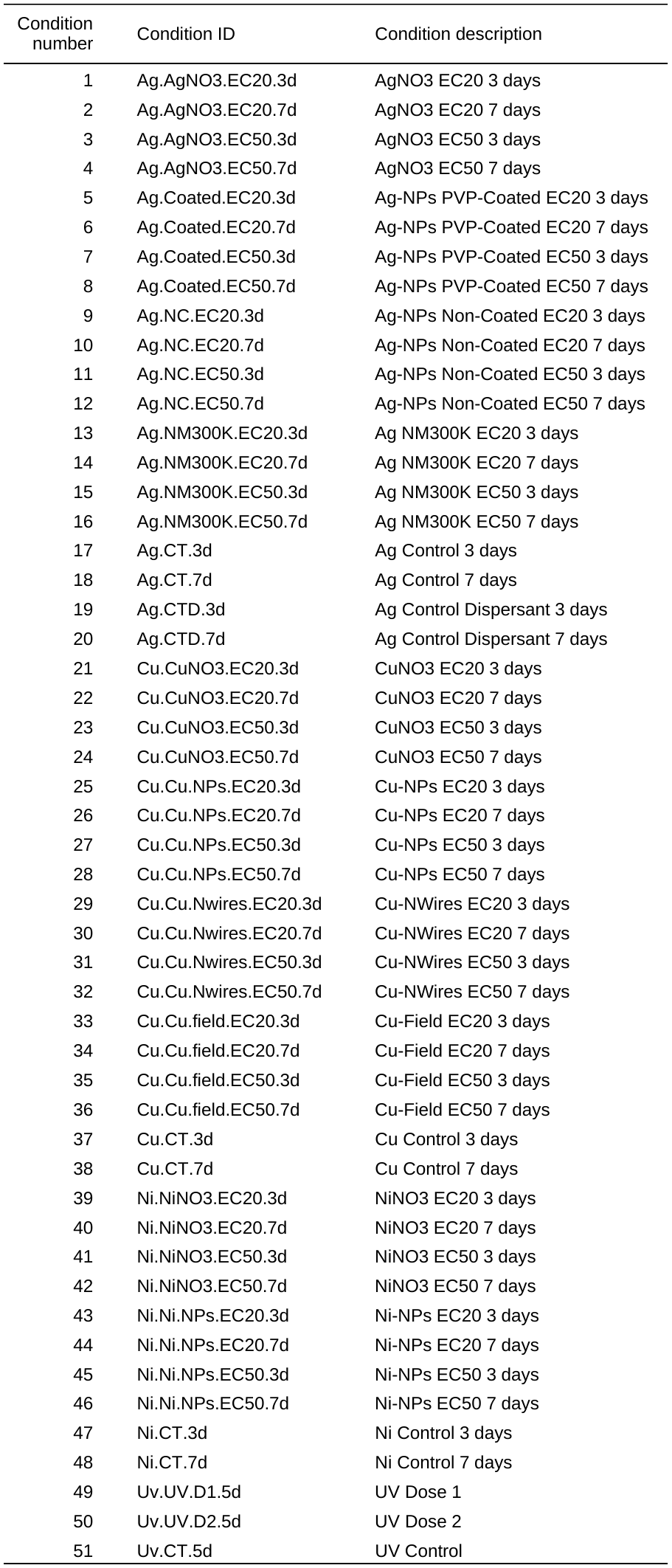}
\end{center}
\end{table}

\clearpage

\begin{table}[ht!]
\caption{
Treatment vs control comparisons, listed in increasing number of 
differentially expressed genes (DEGs), obtained for the real 
\emph{E. crypticus} dataset with SVCD normalization and limma analysis. 
This is the same order as in Figs.~3a, 4, 5, from left to right. 
}
\begin{center}
\includegraphics[height=161mm]{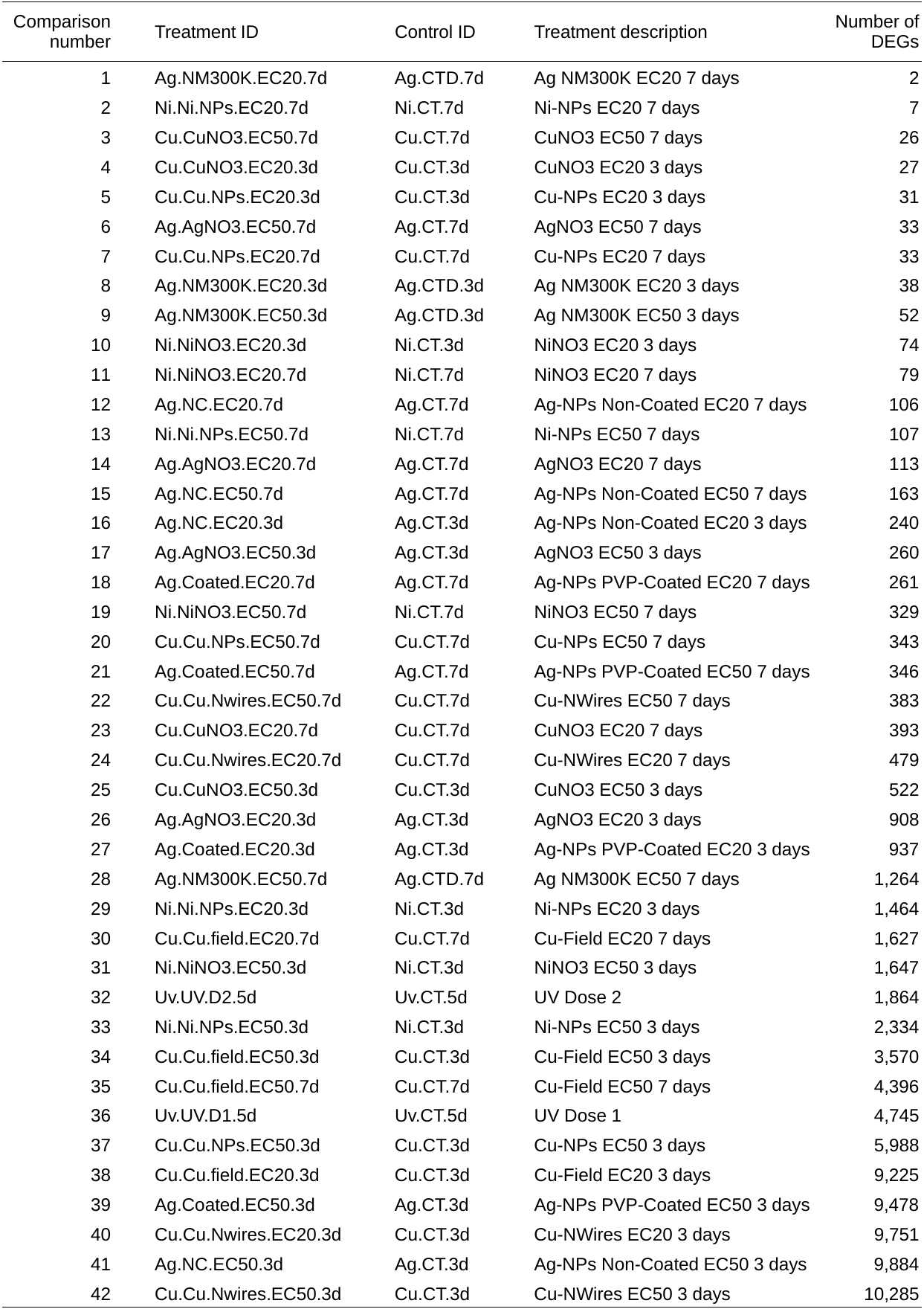}
\end{center}
\end{table}

\clearpage

\section*{Supplementary Figures}

\setcounter{figure}{0}
\renewcommand{\figurename}{Supplementary Figure}
\renewcommand{\thefigure}{S\arabic{figure}}

\begin{figure}[ht!]
\centering
\includegraphics[width=80mm]{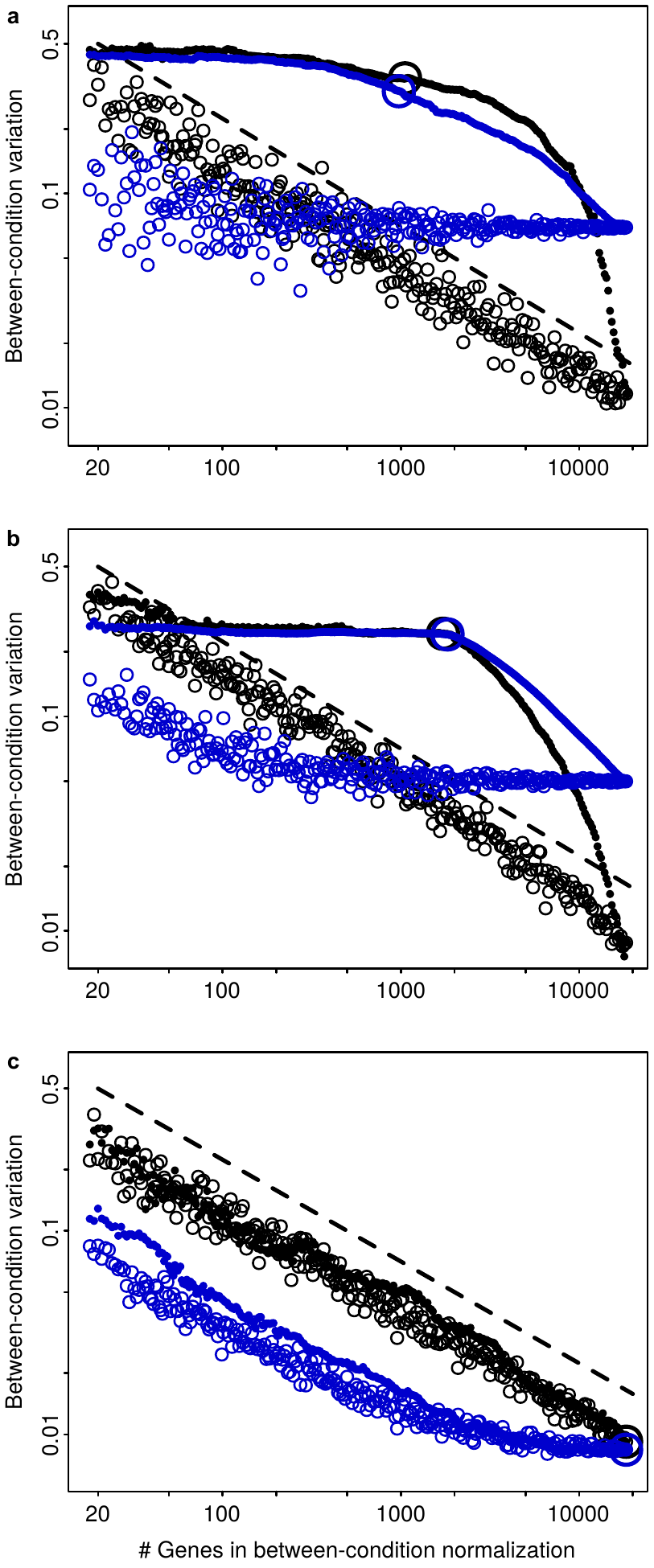}
\end{figure}

\begin{center}
Supplementary Figure S1
\end{center}

\clearpage

\begin{figure}[ht]
\caption{
Representing between-condition variation as the standard deviation of 
the within-condition median averages 
(averages of sample medians, for all samples of the condition) 
produced similar results to those obtained with within-condition mean 
averages (Fig.~2). 
Panels show detected variation as a function of the number of genes used in the 
between-condition normalization, 
for the real dataset (a), 
synthetic dataset with differential gene expression (b), 
and synthetic dataset without differential gene expression (c). 
Labeling is the same as in Fig.~2. 
Each point in each panel indicates the variation obtained with one complete 
normalization 
(black circles, MedianCD normalization; 
blue circles, SVCD normalization). 
Gene were selected in two ways: 
randomly (empty circles) or in decreasing order of $p$-values from a test for 
detecting no-variation genes (filled circles). 
Big circles show the working points corresponding to the results depicted in 
Fig.~1j--o. 
Black dashed lines show references for $n^{-1/2}$ decays, 
with the same values in all panels. 
}
\end{figure}

\clearpage

\begin{figure}[ht!]
\centering
\includegraphics[width=80mm]{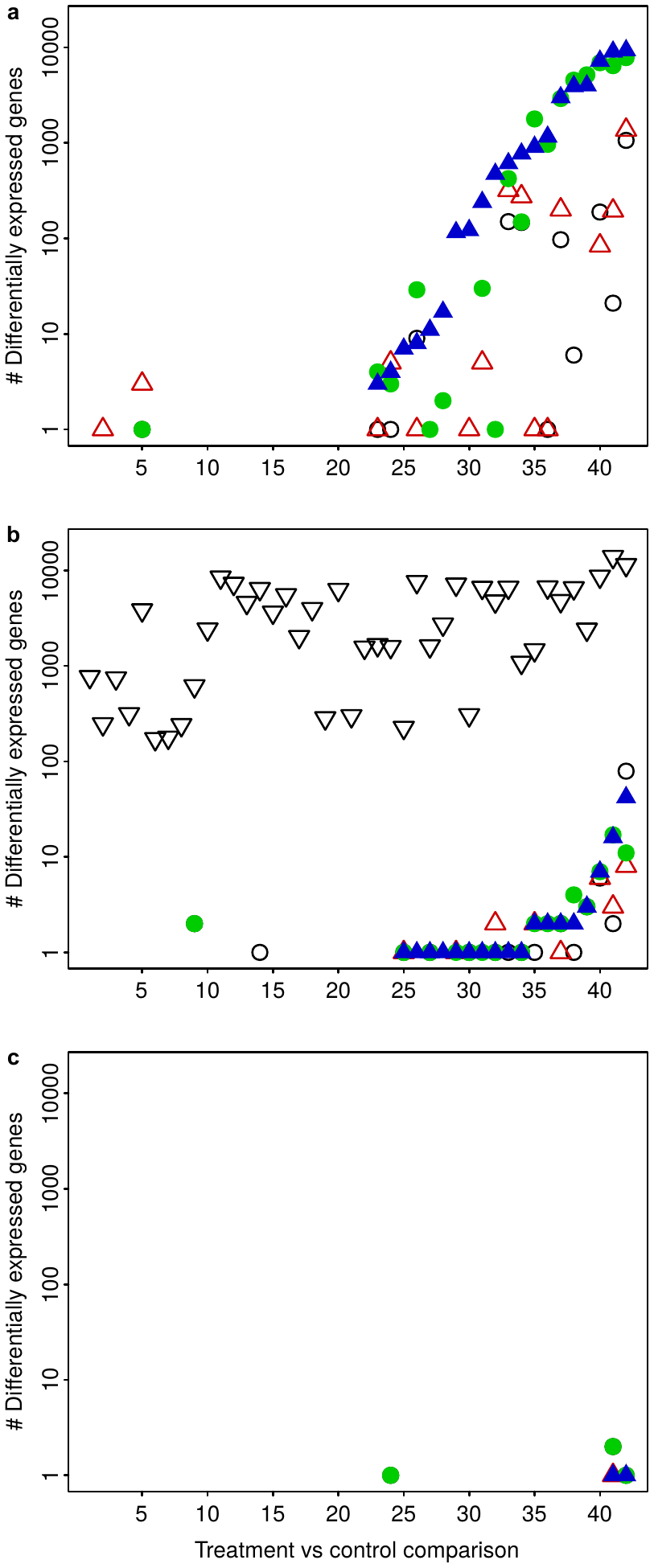}
\end{figure}

\begin{center}
Supplementary Figure S2
\end{center}

\clearpage

\begin{figure}[ht]
\caption{
With $t$-tests instead of limma analysis (Fig.~3a), 
MedianCD and SVCD normalization also allowed to detect larger numbers of 
differentially expressed genes (DEGs), 
compared to Median and Quantile normalization. 
Panels show the number of DEGs obtained for the real dataset (a), 
synthetic dataset with differential gene expression (b), 
and synthetic dataset without differential gene expression (c). 
Symbols are the same as in Fig.~3 
(empty black circles, Median normalization; 
empty red up triangles, Quantile normalization; 
filled green circles, MedianCD normalization; 
filled blue up triangles, SVCD normalization;
empty black down triangles, number of treatment positives (b)). 
In each panel, treatments are ordered according to the number of DEGs 
identified with SVCD normalization, 
increasing from left to right. 
}
\end{figure}

\clearpage

\begin{figure}[ht!]
\centering
\includegraphics[width=80mm]{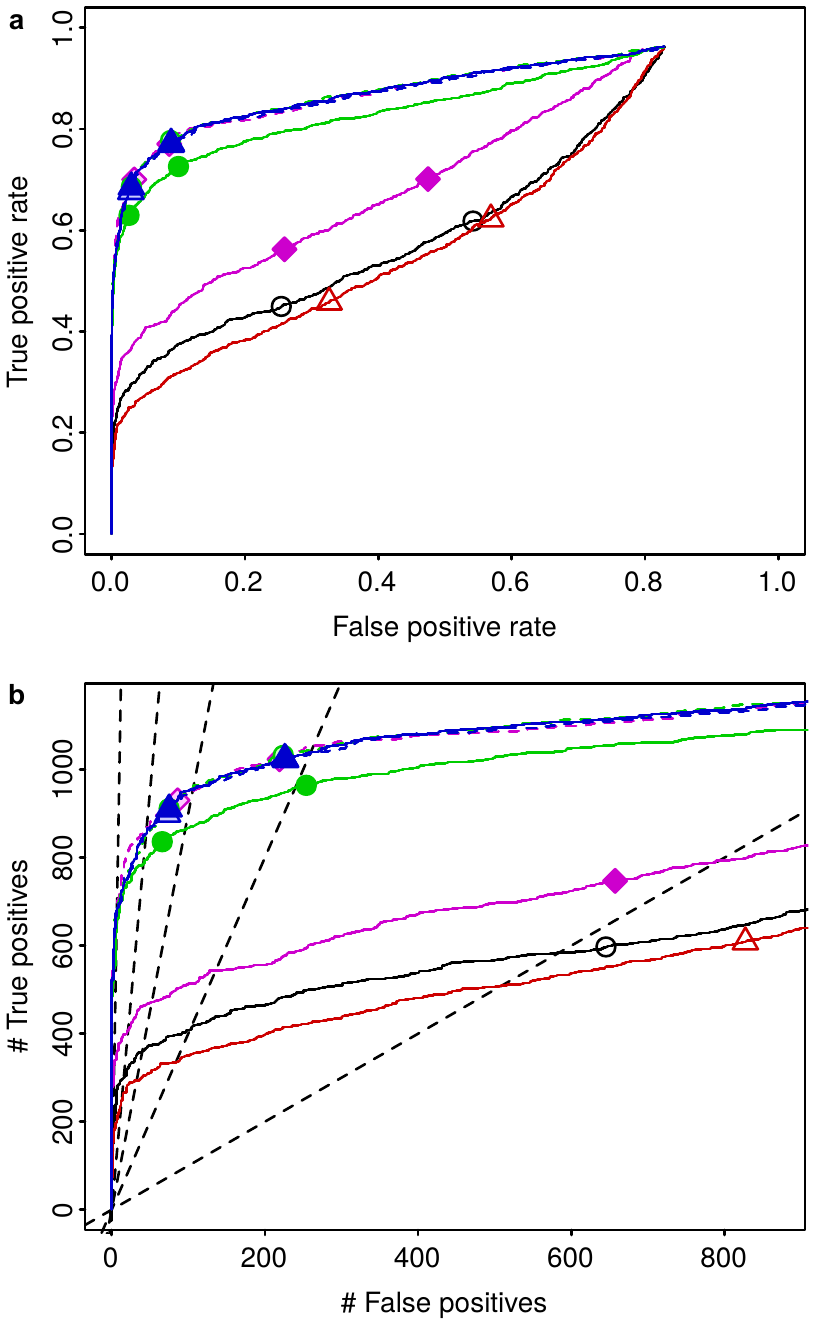}
\end{figure}

\begin{center}
Supplementary Figure S3
\end{center}

\clearpage

\begin{figure}[ht]
\caption{
In the Golden Spike dataset, 
and without restricting probe sets to those with signal in all samples 
(Fig.~8), 
MedianCD and SVCD normalization also allowed the best detection of differential 
gene expression. 
Both panels display ROC curves, 
with the true positive rate versus the false positive rate (a), 
or the number of true positives versus the number of false positives (b). 
Each curve shows the results obtained after applying the four normalization 
methods plus Cyclic Loess normalization 
(same colors and symbols as in Fig.~8; 
black curve with empty black circles, Median normalization; 
red curve with empty red up triangles, Quantile normalization; 
green curve with filled green circles, MedianCD normalization; 
blue curve with filled blue up triangles, SVCD normalization; 
magenta curve with filled magenta diamonds, Cyclic Loess normalization). 
Dashed curves with lightly filled symbols, 
overlapping the response of SVCD normalization, 
show results when the list of known negatives was provided to MedianCD, SVCD, 
and Cyclic Loess normalization. 
The two points per normalization method show results when controlling the 
false discovery rate (FDR) to be below 0.01 (left point) or 0.05 (right point). 
Dashed lines in (b) show references for actual FDR equal to 0.01, 0.05, 0.1, 
0.2, or 0.5 (from left to right). 
As in Fig.~8, compared to MedianCD and SVCD normalization, 
the other normalization methods resulted in notably more severe degradation of 
the FDR. 
}
\end{figure}

\clearpage

\begin{figure}[ht!]
\centering
\includegraphics[width=80mm]{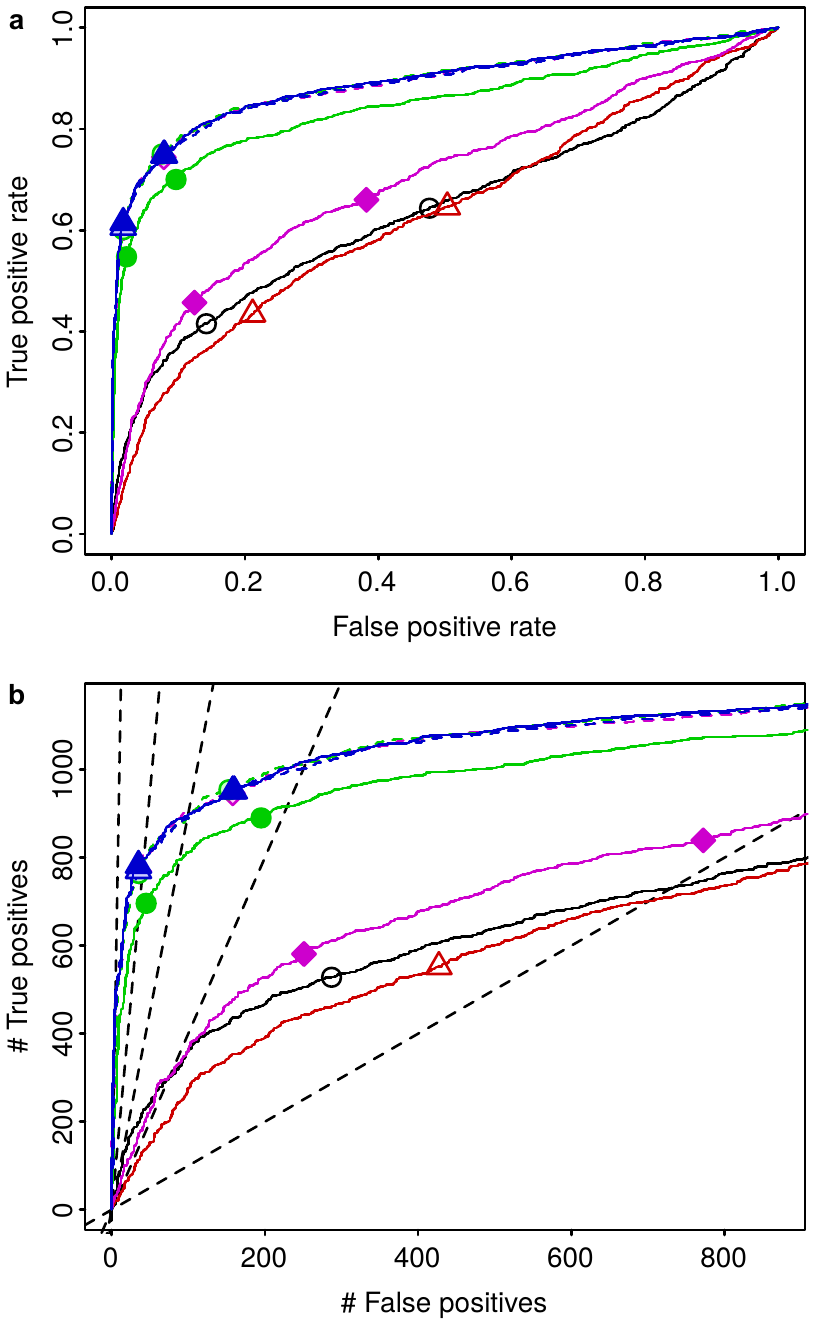}
\end{figure}

\begin{center}
Supplementary Figure S4
\end{center}

\clearpage

\begin{figure}[ht]
\caption{
In the Golden Spike dataset, 
and with $t$-tests instead of limma analysis (Fig.~8), 
MedianCD and SVCD normalization also allowed the best detection of differential 
gene expression. 
Both panels display ROC curves, 
with the true positive rate versus the false positive rate (a), 
or the number of true positives versus the number of false positives (b). 
Each curve shows the results obtained after applying the four normalization 
methods plus Cyclic Loess normalization 
(same colors and symbols as in Figs.~8, S3; 
black curve with empty black circles, Median normalization; 
red curve with empty red up triangles, Quantile normalization; 
green curve with filled green circles, MedianCD normalization; 
blue curve with filled blue up triangles, SVCD normalization; 
magenta curve with filled magenta diamonds, Cyclic Loess normalization). 
Dashed curves with lightly filled symbols, 
overlapping the response of SVCD normalization, 
show results when the list of known negatives was provided to MedianCD, SVCD, 
and Cyclic Loess normalization. 
The two points per normalization method show results when controlling the 
false discovery rate (FDR) to be below 0.01 (left point) or 0.05 (right point). 
Dashed lines in (b) show references for actual FDR equal to 0.01, 0.05, 0.1, 
0.2, or 0.5 (from left to right). 
Compared to results obtained with limma analysis (Figs.~8, S3), 
the degradation of FDR was slightly less severe with $t$-tests. 
MedianCD and SVCD normalization resulted again in the least degradation of the 
FDR. 
}
\end{figure}

\clearpage

\begin{figure}[ht!]
\centering
\includegraphics[width=80mm]{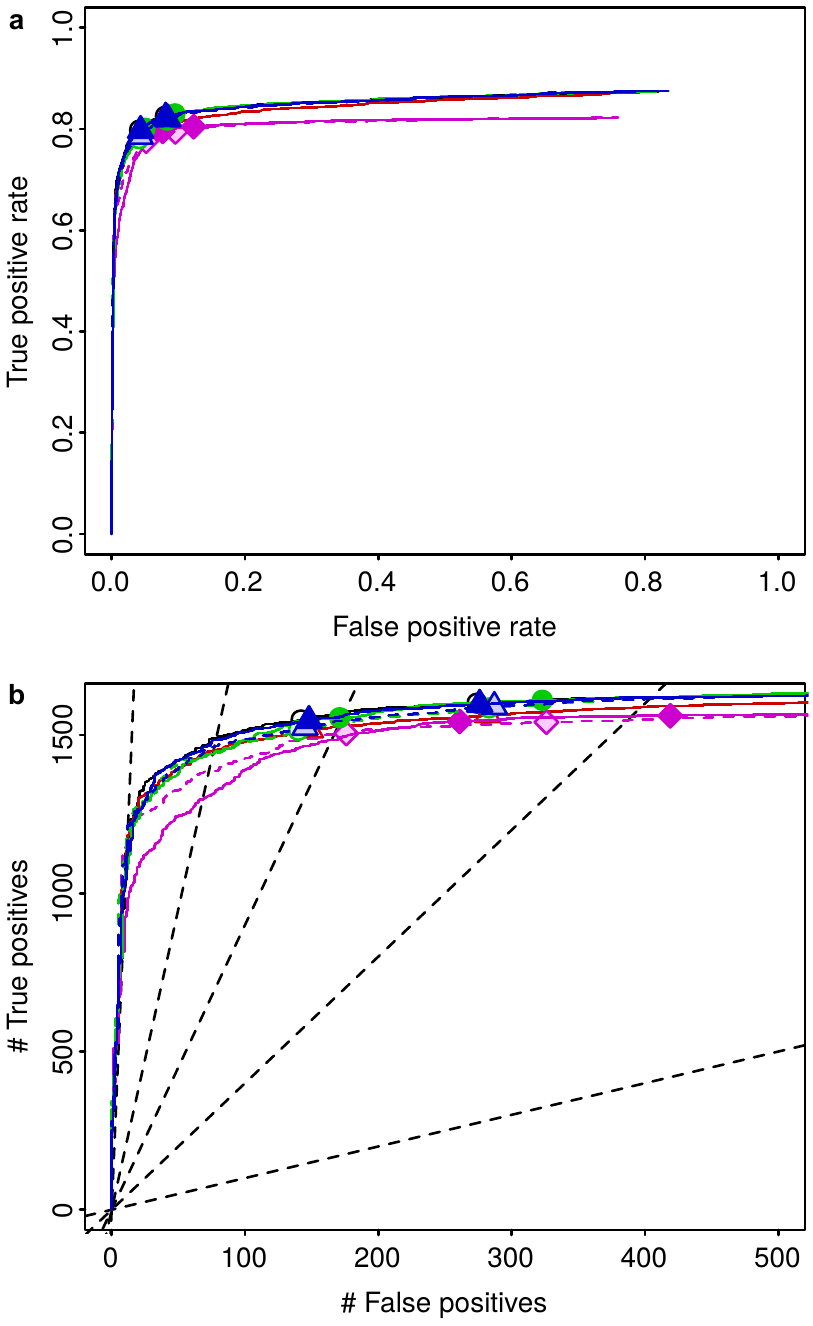}
\end{figure}

\begin{center}
Supplementary Figure S5
\end{center}

\clearpage

\begin{figure}[ht]
\caption{
In the Platinum Spike dataset, 
and without restricting probe sets to those with signal in all samples 
(Fig.~9), 
all normalization methods resulted in similar detection of differential gene 
expression, 
with the exception of Cyclic Loess normalization (magenta curve/symbols), 
whose number of detected positives was slightly smaller. 
Both panels display ROC curves, 
with the true positive rate versus the false positive rate (a), 
or the number of true positives versus the number of false positives (b). 
Each curve shows the results obtained after applying the four normalization 
methods plus Cyclic Loess normalization 
(same colors and symbols as in Fig.~9; 
black curve with empty black circles, Median normalization; 
red curve with empty red up triangles, Quantile normalization; 
green curve with filled green circles, MedianCD normalization; 
blue curve with filled blue up triangles, SVCD normalization; 
magenta curve with filled magenta diamonds, Cyclic Loess normalization). 
Dashed curves with lightly filled symbols show results when the list of known 
negatives was provided to MedianCD, SVCD, and Cyclic Loess normalization. 
The two points per normalization method show results when controlling the 
false discovery rate (FDR) to be below 0.01 (left point) or 0.05 (right point). 
Dashed lines in (b) show references for actual FDR equal to 0.01, 0.05, 0.1, 
0.2, or 0.5 (from left to right). 
As in Fig.~9, 
the difference between normalization methods in the resulting degradation of 
the FDR was smaller for this dataset than for the Golden Spike dataset 
(Figs.~8, S3, S4). 
}
\end{figure}

\clearpage

\begin{figure}[ht!]
\centering
\includegraphics[width=80mm]{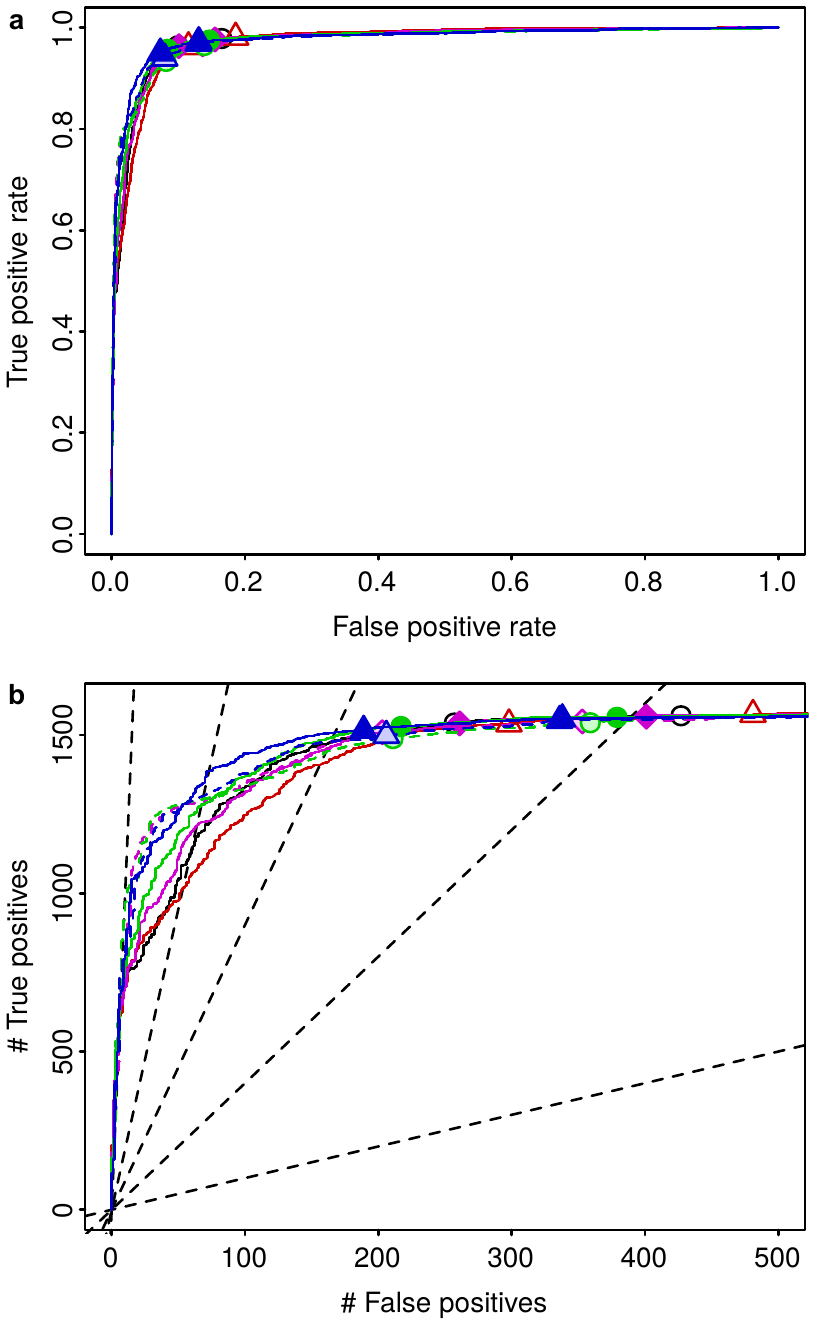}
\end{figure}

\begin{center}
Supplementary Figure S6
\end{center}

\clearpage

\begin{figure}[ht]
\caption{
In the Platinum Spike dataset, 
and with $t$-tests instead of limma analysis (Fig.~9), 
all normalization methods resulted in similar detection of differential gene 
expression, 
with MedianCD and SVCD normalization being marginally better. 
Both panels display ROC curves, 
with the true positive rate versus the false positive rate (a), 
or the number of true positives versus the number of false positives (b). 
Each curve shows the results obtained after applying the four normalization 
methods plus Cyclic Loess normalization 
(same colors and symbols as in Figs.~9, S5; 
black curve with empty black circles, Median normalization; 
red curve with empty red up triangles, Quantile normalization; 
green curve with filled green circles, MedianCD normalization; 
blue curve with filled blue up triangles, SVCD normalization; 
magenta curve with filled magenta diamonds, Cyclic Loess normalization). 
Dashed curves with lightly filled symbols show results when the list of known 
negatives was provided to MedianCD, SVCD, and Cyclic Loess normalization. 
The two points per normalization method show results when controlling the 
false discovery rate (FDR) to be below 0.01 (left point) or 0.05 (right point). 
Dashed lines in (b) show references for actual FDR equal to 0.01, 0.05, 0.1, 
0.2, or 0.5 (from left to right). 
Compared to results obtained with limma analysis (Figs.~9, S5), 
and in contrast to the Golden Spike dataset (Figs.~8, S3, S4), 
the degradation of the FDR was slightly more severe with $t$-tests in this 
dataset. 
}
\end{figure}

\clearpage

\section*{Legends of Supplementary Movies}

\textbf{Supplementary Movie~S1.} 
Example of one within-condition Standard-Vector normalization, 
for the real (\emph{E.\ crypticus}) dataset. 
The movie shows the 14 steps of the convergence of Standard-Vector 
normalization performed for the condition Ag.NM300K.EC20.3d 
(exposure to Ag NM300K nanoparticles, 
with an $\mathrm{EC}_{50}$ dose for three days). 
Left panels show a subset of 10,000 randomly-chosen sample standard vectors, 
with one gray line per gene, 
in the plane of residual vectors, 
i.e.\ the plane perpendicular to the vector of coordinates $(1,1,1)$. 
The lines labeled s1--s3 indicate the projection of the axes onto this 
plane, the number 1--3 being the sample number. 
The red line is the estimated vector of normalization factors at each step, 
with length $\|\textrm{offset}\|$, 
which results from the bias of the standard vectors towards that 
direction. 
Right panels show the polar distribution of vector angles 
(black solid curve), 
compared to the distribution of vector angles after all six possible 
permutations of the sample labels (blue dashed curve). 
The Watson $U^2$ statistic provides a measure of the difference between 
both distributions. 
In the initial step, there is a large bias towards the first and second 
sample, compared to the third one. 
The bias is reduced in each step by subtracting the normalization factor 
estimate, 
which makes the distribution of standard vectors more permutationally 
symmetric and with a correspondingly smaller $U^2$. 
After convergence in 14 steps, there is no detectable bias left. 

\textbf{Supplementary Movie~S2.} 
Example of one within-condition Standard-Vector normalization, 
for the synthetic dataset without differential gene expression. 
The movie displays the Standard-Vector normalization performed for the 
condition Ag.NM300K.EC20.3d, 
on the synthetic dataset generated with the standard normal 
$\mathcal{N}(0,1)$ as base distribution. 
Format and labels are the same as in Supplementary Movie~S1. 
Note the uniform distribution of angles after normalizing, 
which corresponds to a parametric family of probability distributions with 
spherical symmetry. 
The corresponding movie for the synthetic dataset with differential gene 
expression is virtually identical, 
given that standard vectors are independent of sample averages. 

\textbf{Supplementary Movie~S3.} 
Example of one within-condition Standard-Vector normalization, 
for synthetic log-normal data. 
The movie shows the standard-vector normalization performed for the 
condition Ag.NM300K.EC20.3d, 
on a synthetic dataset generated in the same way as that of Supplementary 
Movie~S2, 
except for using as base distribution the log-normal 
$\log\mathcal{N}(0,0.5^2)$, 
which has large positive skewness ($\approx 1.75$). 
Format and labels are the same as in Supplementary Movies~S1, S2. 
Note that the distribution of standard vector angles after normalizing is not 
uniform, 
but it has permutation symmetry. 

\textbf{Supplementary Movie~S4.} 
Identification of no-variation genes (non-differentially expressed genes) for 
the real (\emph{E.\ crypticus}) dataset. 
The movie shows the 27 steps of the corresponding between-condition 
normalization, with SVCD normalization. 
Both panels show the empirical distribution function of $p$-values obtained 
from ANOVA tests on expression levels, 
per gene and grouped by experimental condition. 
The left panel shows the complete interval $[0,1]$, 
while the right panel depicts the interval close to 1 where the first 
goodness-of-fit (GoF) test was not rejected. 
The black portion of the distribution corresponds to $p$-values at 
which the GoF test was rejected, 
the big black circle indicates the first $p$-value at which the GoF test 
was not rejected, 
and the red portion shows the range of $p$-values whose genes, 
as a result, were identified as no-variation genes. 
The dashed blue line and the dotted blue line indicate, respectively, 
the theoretical distribution function of the uniform distribution and 
the threshold of the one-sided Kolmogorov-Smirnov test 
($\alpha=0.001$, $n$~equal to the number of $p$-values for the first GoF test 
that was not rejected). 
Convergence criteria was met from steps 18 to 27. 
These last ten steps ensured stability of the detected set of no-variation 
genes, 
by cumulative intersection of the successive sets identified, 
each one with $\#H_0$ no-variation genes, as shown. 
The resulting final set had 974 no-variation genes. 

\textbf{Supplementary Movie~S5.} 
Identification of no-variation genes for the synthetic dataset with 
differential gene expression. 
The movie shows the 15 steps of the corresponding between-condition 
normalization, with SVCD normalization. 
Format and labels are the same as in Supplementary Movie~S4. 
Note the similarity with the behavior observed for the real dataset 
(Supplementary Movie~S4). 

\textbf{Supplementary Movie~S6.} 
Identification of no-variation genes for the synthetic dataset without 
differential gene expression. 
The movie shows the 14 steps of the corresponding between-condition 
normalization, with SVCD normalization. 
Format and labels are the same as in Supplementary Movies~S4, S5. 
Note that the distribution of $p$-values at convergence (steps 5--14) is 
uniform in the whole interval $[0,1]$, 
up to the level detected by the goodness-of-fit test. 
This corresponds to a dataset with no differentially expressed genes. 

\newpage

\section*{Supplementary Mathematical Methods}

\renewcommand{\thesubsection}{S\arabic{subsection}}

\tableofcontents

\subsection{Vectorial representation of sample data}

Let $x_1, \dots, x_n$ be the samples of $n$ independent and 
identically distributed random variables $X_1,\dots,X_n$. 
Let us represent the samples $x_1, \dots, x_n$ with the 
$\mathbb{R}^n$ column vector 
$\mathbf{x} = (x_1, \dots , x_n)'$, 
and let us denote the sample mean by 
$\bar{x} = \sum_{i=1}^n x_i / n$. 

Let us define the $\mathbb{R}^n \to \mathbb{R}^n$ vectorial operators 
mean ($\overline{\,\mathbf{\cdot}\,}$) and residual 
($\widetilde{\mathbf{\,\cdot\,}}$), respectively, as
\begin{align}
\label{eq:vec-mean}
\overline{\mathbf{x}} & \;=\; (\, \bar{x}, \,\dots\,, \bar{x} \,)' 
    \;=\; \bar{x} \mathbf{1} , \\
\label{eq:vec-resid}
\widetilde{\mathbf{x}} 
    & \;=\; \mathbf{x} - \overline{\mathbf{x}} 
    \;=\; \mathbf{x} - \bar{x} \mathbf{1} , 
\end{align}
$\mathbf{1}$ being the all-ones column vector of dimension $n$. 

Thus, any sample vector $\mathbf{x} \in \mathbb{R}^n$ can be decomposed as
\begin{equation}
\mathbf{x} \;=\; \overline{\mathbf{x}} + \widetilde{\mathbf{x}} .
\end{equation}
The mean vector $\overline{\mathbf{x}}$ contains the sample mean, 
while the residual vector $\widetilde{\mathbf{x}}$ carries the sample 
variation around the mean. 

The vectorial operators mean \eqref{eq:vec-mean} and residual 
\eqref{eq:vec-resid} are linear. \\
\emph{Proposition}. 
For any two sample vectors $\mathbf{x}, \mathbf{y} \in \mathbb{R}^n$ and 
any two numbers $\alpha, \beta \in \mathbb{R}$,
\begin{align}
\label{eq:linear-mean}
\overline{\alpha\mathbf{x} + \beta\mathbf{y}} 
    & \;=\; \alpha\overline{\mathbf{x}} + \beta\overline{\mathbf{y}} , \\
\label{eq:linear-resid}
\widetilde{\alpha\mathbf{x} + \beta\mathbf{y}} 
    & \;=\; \alpha\widetilde{\mathbf{x}} + \beta\widetilde{\mathbf{y}} .
\end{align}

\emph{Proof}.
Let us denote $\mathbf{x} = (x_1, \dots, x_n)'$ and
$\mathbf{y} = (y_1, \dots, y_n)'$.
\begin{align*}
\overline{\alpha\mathbf{x} + \beta\mathbf{y}} 
    & \;=\; \frac{\sum_{i=1}^n 
    \left( \alpha x_i + \beta y_i \right) }{n} \mathbf{1} 
    \;=\; \alpha \frac{\sum_{i=1}^n x_i}{n} \mathbf{1} + 
    \beta \frac{\sum_{i=1}^n y_i}{n} \mathbf{1} 
    \;=\; \alpha\overline{\mathbf{x}} + \beta\overline{\mathbf{y}} , \\
\widetilde{\alpha\mathbf{x} + \beta\mathbf{y}} 
    & \;=\; \alpha\mathbf{x} + \beta\mathbf{y} - 
    \overline{\alpha\mathbf{x} + \beta\mathbf{y}} 
    \;=\; \alpha\mathbf{x} + \beta\mathbf{y} - 
    ( \alpha\overline{\mathbf{x}} + \beta\overline{\mathbf{y}} ) , \\
    & \;=\; \alpha(\mathbf{x} - \overline{\mathbf{x}}) + 
    \beta(\mathbf{y} - \overline{\mathbf{y}}) 
    \;=\; \alpha\widetilde{\mathbf{x}} + \beta\widetilde{\mathbf{y}} .
    \quad \square
\end{align*}

An essential property of the mean and residual vectors is that they 
belong to subspaces that are orthogonal complements \citep{eaton:2007}. 
Hence, for any sample vector $\mathbf{x} \in \mathbb{R}^n$, 
the mean vector $\overline{\mathbf{x}}$ belongs to the subspace 
of dimension 1 spanned by the unit vector 
$\widehat{\mathbf{1}} = \mathbf{1}/\sqrt{n}$, 
while the residual vector $\widetilde{\mathbf{x}}$ belongs to the 
$(n-1)$-dimensional hyperplane orthogonal to $\widehat{\mathbf{1}}$. 

The lengths of the mean vector and residual vector are equal, 
up to a scaling factor, to the sample mean and sample standard 
deviation, respectively. 
For a set of samples $x_1, \dots, x_n$,  where $n \ge 2$, 
let us denote the sample mean as before by 
$\bar{x} = \sum_{i=1}^n x_i / n$, 
and the sample variance as 
$s_x^2 = \sum_{i=1}^n ( x_i - \bar{x} )^2 / (n-1)$. 
Then, the lengths of the mean and residual vectors obtained from 
the sample vector $\mathbf{x} = (x_1, \dots, x_n)'$ are
\begin{align}
\label{eq:mean-vec-len}
\| \overline{\mathbf{x}} \| & \;=\; \sqrt{ n \, \bar{x}^2 } 
    \;=\; \sqrt{n} \, |\bar{x}| , \\
\label{eq:resid-vec-len}
\| \widetilde{\mathbf{x}} \| 
    & \;=\; \sqrt{ \sum_{i=1}^n (x_i - \bar{x})^2 } 
    \;=\; \sqrt{n-1} \, s_x .
\end{align}

Finally, let us define the standard vector of the sample vector 
$\mathbf{x} = (x_1, \dots, x_n)'$ ($n \ge 2$), as 
\begin{equation} 
\mathrm{stdvec}( \mathbf{x} ) 
    \;=\; \sqrt{n-1} \frac{ \widetilde{\mathbf{x}} }
    { \| \widetilde{\mathbf{x}} \| } ,
\end{equation}
whenever $\widetilde{\mathbf{x}} \ne \mathbf{0}$, 
or otherwise as $\mathrm{stdvec}( \mathbf{x} ) = \mathbf{0}$. 
$\mathbf{0}$ is the all-zeros column vector of dimension $n$. 

For a given number of samples $n$, 
all the non-zero standard vectors belong to the ($n-2$)-sphere of 
radius $\sqrt{n-1}$, embedded in the $(n-1)$-dimensional hyperplane 
perpendicular to $\widehat{\mathbf{1}}$. 
Besides, all the components of a standard vector 
are equal to the corresponding standardized samples, 
\begin{equation}
\sqrt{n-1} \frac{ \widetilde{x}_i }{ \|\widetilde{\mathbf{x}}\| }
    \;=\; \frac{ x_i - \bar{x} }{ s_x } .
\end{equation} 

For the degenerate case of having only two samples ($n=2$), 
the only possible values of a non-zero standard vector are 
$\pm(\, 1/\sqrt{2},\, -1/\sqrt{2} \:)'$.

\subsection{Linear decomposition of the normalization problem}

Let us consider a gene expression dataset, 
with $g$ genes and $c$ experimental conditions. 
Each condition $k$ has $s_k$ samples. 
The total number of samples is $s = \sum_{k=1}^c s_k$. 

Let us denote the \emph{observed} expression level of gene $j$ in the 
sample $i$ of condition $k$ by $y_{ij}^{(k)}$. 
We assume that the observed level $y_{ij}^{(k)}$ is equal, 
in the usual $\log_2$-scale, 
to the addition of the normalization factor $a_i^{(k)}$ to the 
\emph{true} gene expression level $x_{ij}^{(k)}$, 
\begin{equation}
\label{eq:norm-def}
y_{ij}^{(k)} \;=\; x_{ij}^{(k)} + a_i^{(k)} .
\end{equation}

Solving the \emph{normalization problem} amounts to finding the 
normalization factors $a_i^{(k)}$ from the observed values 
$y_{ij}^{(k)}$. 
The normalization factors can be understood as sample-wide changes in the 
concentration of mRNA molecules by multiplicative factors equal to 
$2^{a_i^{(k)}}$. 
These changes are caused by technical reasons in the assay and 
are independent of the biological variation in the true levels 
$x_{ij}^{(k)}$.

Let us represent the true and observed expression levels, 
$x_{ij}^{(k)}$ and $y_{ij}^{(k)}$, of gene $j$ in the samples 
$i = 1 \dots s_k$ of condition $k$, by the $s_k$-dimensional vectors 
\begin{align}
\label{eq:true-explev}
\mathbf{x}_j^{(k)} 
    & \;=\; (\, x_{1j}^{(k)}, \dots, x_{s_kj}^{(k)} \,)' , \\
\label{eq:obs-explev}
\mathbf{y}_j^{(k)} 
    & \;=\; (\, y_{1j}^{(k)}, \dots, y_{s_kj}^{(k)} \,)' .
\end{align}
Let us also represent the unknown normalization factors of condition $k$ 
by the $s_k$-dimensional vector
\begin{equation}
\label{eq:norm-factor}
\mathbf{a}^{(k)} \;=\; (\, a_1^{(k)}, \dots, a_{s_k}^{(k)} \,)' .
\end{equation}

From \eqref{eq:norm-def}--\eqref{eq:norm-factor}, 
the normalization problem can be written in vectorial form as
\begin{equation}
\label{eq:norm-def-vec}
\mathbf{y}_j^{(k)} \;=\; \mathbf{x}_j^{(k)} + \mathbf{a}^{(k)} .
\end{equation}
Applying the vectorial operators mean \eqref{eq:vec-mean} and residual 
\eqref{eq:vec-resid}, we obtain
\begin{align}
\label{eq:norm-vec-mean}
\overline{\mathbf{y}}_j^{(k)} 
    & \;=\; \overline{\mathbf{x}}_j^{(k)} + \overline{\mathbf{a}}^{(k)} , \\
\label{eq:norm-vec-resid}
\widetilde{\mathbf{y}}_j^{(k)} 
    & \;=\; \widetilde{\mathbf{x}}_j^{(k)} + \widetilde{\mathbf{a}}^{(k)} .
\end{align}

The residual-vector equations \eqref{eq:norm-vec-resid} correspond to 
the $c$ within-condition normalizations. 
Each within-condition normalization uses the equations 
\eqref{eq:norm-vec-resid} 
particular to a condition $k$, 
for the subset of genes $\mathcal{G}_k \subseteq \{1,\dots,g\}$ that 
have expression level available and of enough quality in that 
experimental condition. 

Let us denote the condition means for each gene as 
\begin{align}
\overline{x}_j^{(k)} 
    & \;=\; \frac{ \sum_{i=1}^{s_k} x_{ij}^{(k)} }{s_k} , \\
\overline{y}_j^{(k)} 
    & \;=\; \frac{ \sum_{i=1}^{s_k} y_{ij}^{(k)} }{s_k} , \\
\overline{a}^{(k)} 
    & \;=\; \frac{ \sum_{i=1}^{s_k} a_i^{(k)} }{s_k} ,
\end{align}
so that 
\begin{align}
\label{eq:cond-mean-x}
\overline{\mathbf{x}}_j^{(k)} 
    & \;=\; \overline{x}_j^{(k)} \mathbf{1}_{s_k} , \\
\label{eq:cond-mean-y}
\overline{\mathbf{y}}_j^{(k)} 
    & \;=\; \overline{y}_j^{(k)} \mathbf{1}_{s_k} , \\
\label{eq:cond-mean-a}
\overline{\mathbf{a}}^{(k)} 
    & \;=\; \overline{a}^{(k)} \mathbf{1}_{s_k} , 
\end{align}
$\mathbf{1}_{s_k}$ being the all-ones column vector of dimension $s_k$. 

Then, the mean-vector equations \eqref{eq:norm-vec-mean} can be 
written as
\begin{equation}
\overline{y}_j^{(k)} \mathbf{1}_{s_k} 
    \;=\; \overline{x}_j^{(k)} \mathbf{1}_{s_k} + 
    \overline{a}^{(k)} \mathbf{1}_{s_k} ,
\end{equation}
so they reduce to the scalar equations
\begin{equation}
\label{eq:cond-mean-scalar}
\overline{y}_j^{(k)} \;=\; \overline{x}_j^{(k)} + \overline{a}^{(k)} .
\end{equation}

Let us define the vectors of conditions means as
\begin{align}
\label{eq:cond-mean-vec-x}
\mathbf{x}_j^* 
    & \;=\; (\, \overline{x}_j^{(1)}, \dots, \overline{x}_j^{(c)} \,)' , \\
\label{eq:cond-mean-vec-y}
\mathbf{y}_j^* 
    & \;=\; (\, \overline{y}_j^{(1)}, \dots, \overline{y}_j^{(c)} \,)' , \\
\label{eq:cond-mean-vec-a}
\mathbf{a}^* 
    & \;=\; (\, \overline{a}^{(1)}, \dots, \overline{a}^{(c)} \,)' ,
\end{align}
and let us express the condition-mean equations in vectorial form as
\begin{equation}
\label{eq:cond-mean-vec}
\mathbf{y}_j^* \;=\; \mathbf{x}_j^* + \mathbf{a}^* .
\end{equation}
Applying again the mean and variance operators, we obtain
\begin{align}
\label{eq:norm-vec-cond-mean}
\overline{\mathbf{y}}_j^* 
    & \;=\; \overline{\mathbf{x}}_j^* + \overline{\mathbf{a}}^* , \\
\label{eq:norm-vec-cond-resid}
\widetilde{\mathbf{y}}_j^* 
    & \;=\; \widetilde{\mathbf{x}}_j^* + \widetilde{\mathbf{a}}^* .
\end{align}

The residual-vector equations on condition means 
\eqref{eq:norm-vec-cond-resid} correspond to the single 
between-condition normalization, 
in a similar way as \eqref{eq:norm-vec-resid} do for 
the each of the within-condition normalizations. 
There is one equation \eqref{eq:norm-vec-cond-resid} per gene. 
The only equations used in the between-condition normalization are those 
of the subset of genes $\mathcal{G}^* \subseteq \{1,\dots,g\}$ 
that show no evidence of variation across experimental conditions, 
according to a statistical test. 

Given that $\overline{\mathbf{a}}^* = \overline{a}^* \mathbf{1}_c$, 
\eqref{eq:norm-vec-cond-mean} has the only unknown $\overline{a}^*$. 
The meaning of $\overline{a}^*$ is a conversion factor between the 
scale the true and observed expression levels. 
This factor depends on the technology used to measure the expression 
levels and finding it is out of the scope of the normalization problem. 
Therefore, without loss of generality, we assume $\overline{a}^* = 0$, 
so
\begin{align}
\label{eq:cond-norm-mean}
\overline{\mathbf{a}}^* & \;=\; \mathbf{0}_c , \\
\label{eq:cond-norm-resid}
{\mathbf{a}}^* & \;=\; \widetilde{\mathbf{a}}^* .
\end{align}

The solution of the between-condition normalization, 
$\widetilde{\mathbf{a}}^*$, allows to find the mean vectors of 
the normalization factors $\overline{\mathbf{a}}^{(k)}$, 
via \eqref{eq:cond-mean-a}, \eqref{eq:cond-mean-vec-a} and 
\eqref{eq:cond-norm-resid}. 
The within-condition normalizations yield the residual vectors 
$\widetilde{\mathbf{a}}^{(k)}$. 
The complete solution to the normalization problem is finally obtained, 
for each condition $k$, with
\begin{equation}
\label{eq:norm_solution}
\mathbf{a}^{(k)} \;=\; \overline{\mathbf{a}}^{(k)} + 
    \widetilde{\mathbf{a}}^{(k)} .
\end{equation} 

Thus, the original normalization problem \eqref{eq:norm-def-vec} has 
been divided in $c+1$ normalization sub-problems on residual vectors, 
stated by \eqref{eq:norm-vec-resid} and 
\eqref{eq:norm-vec-cond-resid}. 
In fact, this linear decomposition is possible for any partition of 
the set of $s$ samples. 
The choice of the partition as the one defined by the experimental 
conditions is motivated by the need to control the biological variation 
among the genes used in each normalization. 
All the $c+1$ normalizations face the same kind of 
\emph{normalization of residuals problem},
which we define in general as follows. 

\textbf{Normalization of Residuals Problem}. 
Let $y_{ij}$ be the $i$-th observed value of feature $j$, 
in a dataset with $n \ge 2$ observations for each of the $m$ features. 
The observed values $y_{ij}$ are equal to the true values $x_{ij}$ 
plus the normalization factors $a_i$, 
which are constant across features. 
In vectorial form, there are $m$ equations
\begin{equation}
\label{eq:unit-norm-def}
\mathbf{y}_j \;=\; \mathbf{x}_j + \mathbf{a} ,
\end{equation}
where the vectors belong to $\mathbb{R}^n$. 
As a consequence
\begin{equation}
\label{eq:unit-norm-def-var}
\widetilde{\mathbf{y}}_j \;=\; \widetilde{\mathbf{x}}_j + 
    \widetilde{\mathbf{a}} .
\end{equation}

Solving the \emph{normalization of residuals problem} amounts to 
finding the residual vector of normalization factors 
$\widetilde{\mathbf{a}}$ 
from the observed residual vectors $\widetilde{\mathbf{y}}_j$. 
In the within-condition normalizations, 
the features are gene expression levels, 
with one observation per sample of the corresponding experimental 
condition. 
In the between-condition normalization, 
the features are means of gene expression levels, 
with one observation per condition. 

There is, however, an additional requirement imposed by the methods 
with which we propose to solve the between-condition normalization. 
We would like to consider the condition means $\overline{x}_j^{(k)}$ in 
\eqref{eq:cond-mean-scalar} as sample data across conditions. 
This only holds when all the conditions have the same 
number of samples. 
Otherwise, we balance the condition means so that they result from the 
same number of samples in all conditions, 
according to the procedure described in the following.

Let $s^*$ be the minimum number of samples across conditions, 
$s^* = \min \{s_1,\dots,s_c\}$. 
Let $\mathcal{S}_j^{(k)}$ be independent random samples (without 
replacement) of size $s^*$ from the set of indexes $\{1,\dots,s_k\}$, 
with one sample per gene $j$ and condition $k$. 
Then, the balanced condition means are defined as
\begin{align}
\overline{x}_j^{(k)*} 
    & \;=\; \frac{\displaystyle \sum_{i \in \mathcal{S}_j^{(k)}} 
    x_{ij}^{(k)} }{s^*} , \\[1ex] 
\overline{y}_j^{(k)*} 
    & \;=\; \frac{\displaystyle \sum_{i \in \mathcal{S}_j^{(k)}} 
    y_{ij}^{(k)} }{s^*} , \\[1ex] 
\overline{a}_j^{(k)*} 
    & \;=\; \frac{\displaystyle \sum_{i \in \mathcal{S}_j^{(k)}} 
    a_i^{(k)} }{s^*} .
\end{align} 

From \eqref{eq:norm-def}, the balanced condition means verify a 
relationship similar to \eqref{eq:cond-mean-scalar},
\begin{equation}
\label{eq:bal-cond-mean-scalar}
\overline{y}_j^{(k)*} \;=\; \overline{x}_j^{(k)*} + \overline{a}_j^{(k)*} .
\end{equation}
Moreover, the average of $\overline{a}_j^{(k)*}$ across the sampling subsets 
$\mathcal{S}_j^{(k)}$ is equal to the unknown $\overline{a}^{(k)}$. 
This implies that \eqref{eq:bal-cond-mean-scalar} are, on average, 
equivalent to \eqref{eq:cond-mean-scalar}.
Hence, we use the following vectors of balanced conditions means 
\begin{align}
\label{eq:bal-cond-mean-vec-x}
\mathbf{x}_j^* 
    & \;=\; (\, \overline{x}_j^{(1)*}, \dots, \overline{x}_j^{(c)*} \,) , \\
\label{eq:bal-cond-mean-vec-y}
\mathbf{y}_j^* 
    & \;=\; (\, \overline{y}_j^{(1)*}, \dots, \overline{y}_j^{(c)*} \,) ,
\end{align}
instead of \eqref{eq:cond-mean-vec-x}, \eqref{eq:cond-mean-vec-y}, 
in order to build the condition-mean equations \eqref{eq:cond-mean-vec}. 
This balancing of the condition means is only required when the 
experimental conditions have different number of samples.

\subsection{Permutation invariance of multivariate data}

Let $x_{ij}$ and $y_{ij}$ be, respectively, the true and observed values 
of a dataset with $n$ observations of $m$ features, 
as defined in the \emph{normalization of residuals problem} above. 

We have assumed that the $n$ true values $x_{1j},\dots,x_{nj}$ of 
feature $j$ are samples of independent and identically 
distributed random variables $X_{1j},\dots,X_{nj}$. 
These random variables can be represented with the random vector 
$\mathbf{X}_j = ( X_{1j}, \dots, X_{nj} )'$, 
carried by the probability space 
$ ( \Omega, \mathcal{F}, \mathrm{P} ) $ 
and with induced space $ ( \mathbb{R}^n, \mathbb{B}^n, \mathbb{P} )$. 
Let us define the random vectors $\overline{\mathbf{X}}_j$ and 
$\widetilde{\mathbf{X}}_j$ with the vectorial operators mean 
\eqref{eq:vec-mean} and residual \eqref{eq:vec-resid}, respectively, 
\begin{align}
\overline{X}_j & \;=\; \sum_{i=1}^n \frac{ X_{ij} }{ n } , \\
\label{eq:vec-mean-vec}
\overline{\mathbf{X}}_j & 
    \;=\; (\, \overline{X}_j, \,\dots\,, \overline{X}_j \,)' 
    \;=\; \overline{X}_j \mathbf{1} , \\
\label{eq:vec-resid-vec}
\widetilde{\mathbf{X}}_j 
    & \;=\; \mathbf{X}_j - \overline{\mathbf{X}}_j 
    \;=\; \mathbf{X}_j - \overline{X}_j \mathbf{1} .
\end{align}

$\mathbf{X}_j = \overline{\mathbf{X}}_j + \widetilde{\mathbf{X}}_j$ 
holds for any random vector $\mathbf{X}_j$, 
as well as the other properties presented above. 
Let us assume that $ \mathrm{E} (\, \| \mathbf{X}_j \| \,) < \infty $ 
and that $ \mathrm{P} (\, \| \widetilde{\mathbf{X}}_j \| = 0 \,) = 0 $, 
which imply that 
$ \widetilde{\mathbf{X}}_j / \| \widetilde{\mathbf{X}}_j \| $ 
has length 1 almost surely. 

The standard random vector 
$\sqrt{n-1} \, \widetilde{\mathbf{X}}_j / \| \widetilde{\mathbf{X}}_j \|$ 
is a pivotal quantity, 
where the location (mean) and scale (standard deviation) of feature 
$j$ have been removed. 
The probability distribution of 
$\widetilde{\mathbf{X}}_j / \| \widetilde{\mathbf{X}}_j \|$ 
across the remaining degrees of freedom over the unit $(n-2)$-sphere is 
governed by the parametric family of the random 
variables $X_{1j}, \dots, X_{nj}$. 
Moreover, the independence and identity of distribution across the $n$ 
observations implies that the distribution of $ \mathbf{X}_j $ is 
\emph{exchangeable}, 
i.e.\ invariant with respect to permutations of the observation labels. 
As a result, 
$\widetilde{\mathbf{X}}_j / \| \widetilde{\mathbf{X}}_j \|$ 
is also permutation invariant, 
which geometrically corresponds to symmetries with respect to the $n!$ 
permutations of the axes in the $n$-dimensional space of random vectors, 
projected onto the $(n-1)$-dimensional hyperplane of residual vectors. 

Residual vectors and standard vectors have been widely studied, 
especially in relation to elliptically symmetric distributions and 
linear models \citep{fang:1990,gupta:2013}, 
and to the invariances of probability 
distributions \citep{kallenberg:2005}. 
Here, we consider these vectors from the viewpoint of the problem 
of normalizing multivariate data, 
and its relationship with permutation invariance. 

It is well know that, for a multivariate distribution with independent 
and identically distributed components, 
the expected value of the standard vector is 
zero \citep{eaton:2007}, 
given that it is so for each component. 
We prove this here for completeness, and to show that it is also a 
necessary consequence of the permutation invariance of the distribution. 

\emph{Proposition}. 
The expected value of any true 
(i.e.\ without normalization issues) standard vector is zero. 
If the $n \ge 2$ samples of feature $j$ are independent and 
identically distributed, then
\begin{equation}
\label{eq:x-expval}
\mathrm{E} \left( \sqrt{n-1}\, \frac{ \widetilde{\mathbf{X}}_j }
    { \| \widetilde{\mathbf{X}}_j \| } \right) 
    \;=\; \mathbf{0} .
\end{equation}

\emph{Proof}. Let $\mathcal{P}_n$ be the set of all the 
permutation matrices in $\mathbb{R}^{n \times n}$. 
Then, for any $P \in \mathcal{P}_n$, 
$ \widetilde{\mathbf{X}}_j / 
    \| \widetilde{\mathbf{X}}_j \| $ is 
equal in distribution to 
$ P \, \widetilde{\mathbf{X}}_j / 
    \| \widetilde{\mathbf{X}}_j \|$. 
This implies that 
\begin{equation*}
\mathrm{E} \left( \frac{ \widetilde{\mathbf{X}}_j }
    { \| \widetilde{\mathbf{X}}_j \| } \right) 
\;=\; \mathrm{E} \left( P \frac{ \widetilde{\mathbf{X}}_j }
    { \| \widetilde{\mathbf{X}}_j \| } \right) 
\;=\; P \, \mathrm{E}  \left( \frac{ \widetilde{\mathbf{X}}_j }
    { \| \widetilde{\mathbf{X}}_j \| } \right) .
\end{equation*}

The only vectors that are invariant with respect to all possible 
permutations are those that have all components identical. 
Therefore, $\mathrm{E} (\, \widetilde{\mathbf{X}}_j 
    / \| \widetilde{\mathbf{X}}_j \| \,) = \alpha \widehat{\mathbf{1}}$, 
with $\alpha \in \mathbb{R}$. 
However, $ \widetilde{\mathbf{X}}'_j \widehat{\mathbf{1}} = 0$, 
so that $ \alpha = \mathrm{E} (\, \widetilde{\mathbf{X}}_j 
    / \| \widetilde{\mathbf{X}}_j \| \,)' \, \widehat{\mathbf{1}} = 0 $. 
Hence $\mathrm{E} (\, \widetilde{\mathbf{X}}_j 
    / \| \widetilde{\mathbf{X}}_j \| \,) = \mathbf{0} $. 
$\quad \square$

For each true random vector $\mathbf{X}_j$, 
there is an observed random vector 
$\mathbf{Y}_j = \mathbf{X}_j + \mathbf{A}$, 
where $\mathbf{A}$ is the random vector of normalization factors. 
The random vectors $\mathbf{X}_j$ and $\mathbf{A}$ 
are independent, 
representing biological and technical variation, respectively. 
Therefore, and without loss of generality, 
we assume in what follows a fixed vector of normalization 
factors $\mathbf{a}$, 
i.e.\ we condition on the event $\{ \mathbf{A} = \mathbf{a} \,\}$. 
We also assume that 
$\mathrm{P}(\, \| \widetilde{\mathbf{Y}}_j \| = 0 \,) = 0$, 
which implies that $\widetilde{\mathbf{Y}}_j / 
    \| \widetilde{\mathbf{Y}}_j \|$ has length 1 almost surely. 

In contrast to the true standard vector 
$ \sqrt{n-1} \, \widetilde{\mathbf{X}}_j / 
    \| \widetilde{\mathbf{X}}_j \|$, 
the observed standard vector \\
$ \sqrt{n-1} \, \widetilde{\mathbf{Y}}_j / 
    \| \widetilde{\mathbf{Y}}_j \|$ 
is biased toward the direction of $\widetilde{\mathbf{a}}$, 
with the result that the expected value is not zero. 

\emph{Proposition}. 
If the $n \ge 2$ samples of feature $j$ are independent and 
identically distributed, 
whenever $\widetilde{\mathbf{a}} \ne \mathbf{0}$, 
\begin{equation}
\label{eq:y-expval}
\mathrm{E} \left( \sqrt{n-1}\, \frac{ \widetilde{\mathbf{Y}}_j }
    { \| \widetilde{\mathbf{Y}}_j \| } \right) 
    \;\ne\; \mathbf{0} .
\end{equation}
When $n=2$, there is the additional requirement that 
$\mathrm{P}(\, \|\widetilde{\mathbf{X}}_i\| < 
    \|\widetilde{\mathbf{a}}\| \,) > 0$. 
This threshold of detection only occurs for the degenerate case of 
$n=2$. 

\emph{Proof}. Let us consider the projection of 
$\widetilde{\mathbf{Y}}_j / \| \widetilde{\mathbf{Y}}_j \|$ 
on $\widetilde{\mathbf{a}}$, 
compared to the projection of 
$\widetilde{\mathbf{X}}_j / \| \widetilde{\mathbf{X}}_j \|$. 

When the vectors $\widetilde{\mathbf{X}}_j$ and 
$\widetilde{\mathbf{a}}$ are collinear, 
\begin{equation*}
\frac{ \widetilde{\mathbf{X}}'_j \, \widetilde{\mathbf{a}} } 
    { \| \widetilde{\mathbf{X}}_j \| \, 
    \| \widetilde{\mathbf{a}} \| } 
\;=\; \pm 1 , \quad \text{and} \quad 
\frac{ \widetilde{\mathbf{Y}}'_j \, \widetilde{\mathbf{a}} } 
    { \| \widetilde{\mathbf{Y}}_j \| \, 
    \| \widetilde{\mathbf{a}} \| } \;=\; \pm 1 , 
\end{equation*}
with 
\begin{equation*}
\frac{ \widetilde{\mathbf{Y}}'_j \, \widetilde{\mathbf{a}} } 
    { \| \widetilde{\mathbf{Y}}_j \| \, 
    \| \widetilde{\mathbf{a}} \| }
\;\ge\; \frac{ \widetilde{\mathbf{X}}'_j \, \widetilde{\mathbf{a}} } 
    { \| \widetilde{\mathbf{X}}_j \| \, 
    \| \widetilde{\mathbf{a}} \| } .
\end{equation*}

This is the only case when $n=2$. 
The additional requirement ensures that, for $n=2$, 
\begin{equation*}
\mathrm{P} \left( 
\frac{ \widetilde{\mathbf{Y}}'_j \, \widetilde{\mathbf{a}} } 
    { \| \widetilde{\mathbf{Y}}_j \| \, 
    \| \widetilde{\mathbf{a}} \| }
\;>\; \frac{ \widetilde{\mathbf{X}}'_j \, \widetilde{\mathbf{a}} } 
    { \| \widetilde{\mathbf{X}}_j \| \, 
    \| \widetilde{\mathbf{a}} \| } \right) 
\;>\; 0, 
\end{equation*}
which implies 
\begin{equation*}
\mathrm{E} \left( 
    \frac{ \widetilde{\mathbf{Y}}'_j \, \widetilde{\mathbf{a}} } 
    { \| \widetilde{\mathbf{Y}}_j \| \, 
    \| \widetilde{\mathbf{a}} \| } 
\right) \;>\; \mathrm{E} \left( 
    \frac{ \widetilde{\mathbf{X}}'_j \, \widetilde{\mathbf{a}} } 
    { \| \widetilde{\mathbf{X}}_j \| \, 
    \| \widetilde{\mathbf{a}} \| } 
\right) .
\end{equation*}

Otherwise, when $n > 2$ and the vectors $\widetilde{\mathbf{X}}_j$ and 
$\widetilde{\mathbf{a}}$ are not collinear, 
they lie on a plane. 
The vector $\widetilde{\mathbf{Y}}_j 
    = \widetilde{\mathbf{X}}_j + \widetilde{\mathbf{a}}$ 
is the diagonal of the parallelogram defined by 
$\widetilde{\mathbf{X}}_j$ and $\widetilde{\mathbf{a}}$. 
Hence the angle between $\widetilde{\mathbf{Y}}_j$ and 
$\widetilde{\mathbf{a}}$ is strictly less than 
the angle between $\widetilde{\mathbf{X}}_j$ and 
$\widetilde{\mathbf{a}}$, 
so the cosine of the angle is strictly greater. 
Thus, 
\begin{equation*}
\frac{ \widetilde{\mathbf{Y}}'_j \, \widetilde{\mathbf{a}} } 
    { \| \widetilde{\mathbf{Y}}_j \| \, 
    \| \widetilde{\mathbf{a}} \| }
\;>\; \frac{ \widetilde{\mathbf{X}}'_j \, \widetilde{\mathbf{a}} } 
    { \| \widetilde{\mathbf{X}}_j \| \, 
    \| \widetilde{\mathbf{a}} \| } . 
\end{equation*}

Due to the permutation symmetries in the distribution of 
$\widetilde{\mathbf{X}}_j / \| \widetilde{\mathbf{X}}_j \|$, 
when $n>2$ the vector $\widetilde{\mathbf{X}}_j$ has non-zero 
probability of being not collinear with $\widetilde{\mathbf{a}}$, 
i.e.\ $\mathrm{P} (\, | \widetilde{\mathbf{X}}'_j \, 
    \widetilde{\mathbf{a}} | < 1 \,) > 0$. 

Therefore, 
\begin{equation*}
\mathrm{P} \left( 
\frac{ \widetilde{\mathbf{Y}}'_j \, \widetilde{\mathbf{a}} } 
    { \| \widetilde{\mathbf{Y}}_j \| \, 
    \| \widetilde{\mathbf{a}} \| }
\;>\; \frac{ \widetilde{\mathbf{X}}'_j \, \widetilde{\mathbf{a}} } 
    { \| \widetilde{\mathbf{X}}_j \| \, 
    \| \widetilde{\mathbf{a}} \| } \right) 
\;>\; 0, 
\end{equation*}
which again implies 
\begin{equation*}
\mathrm{E} \left( 
    \frac{ \widetilde{\mathbf{Y}}'_j \, \widetilde{\mathbf{a}} } 
    { \| \widetilde{\mathbf{Y}}_j \| \, 
    \| \widetilde{\mathbf{a}} \| } 
\right) \;>\; \mathrm{E} \left( 
    \frac{ \widetilde{\mathbf{X}}'_j \, \widetilde{\mathbf{a}} } 
    { \| \widetilde{\mathbf{X}}_j \| \, 
    \| \widetilde{\mathbf{a}} \| } 
\right) .
\end{equation*}

Finally, 
\begin{equation*}
\left\|\, \mathrm{E} \left( 
    \frac{ \widetilde{\mathbf{Y}}_j }{ \| \widetilde{\mathbf{Y}}_j \| } 
    \right) \right\|
\;\ge\; \mathrm{E} \left( 
    \frac{ \widetilde{\mathbf{Y}}_j }
    { \| \widetilde{\mathbf{Y}}_j \| } \right)' 
    \frac{ \widetilde{\mathbf{a}} }{ \| \widetilde{\mathbf{a}} \| } 
\;>\; \mathrm{E} \left( 
    \frac{ \widetilde{\mathbf{X}}_j }
    { \| \widetilde{\mathbf{X}}_j \| } \right)' 
    \frac{ \widetilde{\mathbf{a}} }{ \| \widetilde{\mathbf{a}} \| } 
\;=\; 0 .
\quad \square
\end{equation*}

As a consequence, the \emph{normalization of residuals problem} may be 
restated as the problem of finding the normalization factors 
$\widetilde{\mathbf{a}}$ from the observed vectors 
$\widetilde{\mathbf{y}}_j$, such that the standard vectors 
$ \sqrt{n-1} \, ( \widetilde{\mathbf{y}}_j - \widetilde{\mathbf{a}}) 
/ \|\widetilde{\mathbf{y}}_j - \widetilde{\mathbf{a}}\| $ are invariant 
against permutations of the observation labels.
Or equivalently, such that the standard vectors 
$ \sqrt{n-1} \, ( \widetilde{\mathbf{y}}_j - \widetilde{\mathbf{a}}) 
/ \|\widetilde{\mathbf{y}}_j - \widetilde{\mathbf{a}}\| $ 
have zero mean. 
The following property provides an approach to the solution. 

\emph{Proposition}. Whenever $\widetilde{\mathbf{a}} \ne \mathbf{0}$, 
the component of the expected value of 
$\widetilde{\mathbf{Y}}_j / \| \widetilde{\mathbf{Y}}_j \|$ 
parallel to $\widetilde{\mathbf{a}}$ verifies
\begin{equation}
\label{eq:y-aproj}
0 \;<\; \mathrm{E} \left( 
    \frac{ \widetilde{\mathbf{Y}}_j }
    { \| \widetilde{\mathbf{Y}}_j \| } \right)' 
    \frac{ \widetilde{\mathbf{a}} }{ \| \widetilde{\mathbf{a}} \| } 
\;<\; \mathrm{E} \left( 
    \frac{ 1 }{ \| \widetilde{\mathbf{Y}}_j \| } \right) 
    \| \widetilde{\mathbf{a}} \| . 
\end{equation}
As in \eqref{eq:y-expval}, when $n=2$ we also assume that 
$\mathrm{P}(\, \|\widetilde{\mathbf{X}}_j\| 
    < \|\widetilde{\mathbf{a}}\| \,) > 0$. 

\emph{Proof}. The first inequality holds from the previous proof. 
Concerning the second inequality, let us consider
\begin{equation*} 
\frac{ \widetilde{\mathbf{Y}}_j' \, \widetilde{\mathbf{a}} }
    { \| \widetilde{\mathbf{Y}}_j \| \, \| \widetilde{\mathbf{a}} \| } 
\;=\; \frac{ ( \widetilde{\mathbf{X}}_j + \widetilde{\mathbf{a}} )' 
    \, \widetilde{\mathbf{a}} } 
    { \| \widetilde{\mathbf{X}}_j + \widetilde{\mathbf{a}} \| \, 
    \| \widetilde{\mathbf{a}} \| } 
\;=\; \frac{ \| \widetilde{\mathbf{X}}_j \| }
    { \| \widetilde{\mathbf{X}}_j + \widetilde{\mathbf{a}} \| } \,
    \frac{ \widetilde{\mathbf{X}}_j' \, \widetilde{\mathbf{a}} }
    { \| \widetilde{\mathbf{X}}_j \| \, \| \widetilde{\mathbf{a}} \| } 
    \;+\; \frac{ \| \widetilde{\mathbf{a}} \| }
    { \| \widetilde{\mathbf{Y}}_j \| } .
\end{equation*}

We need to prove that the first term on the RHS has negative expected 
value. 
Let us decompose this term into the positive and negative parts, 
\begin{equation*} 
\frac{ \| \widetilde{\mathbf{X}}_j \| }
    { \| \widetilde{\mathbf{X}}_j + \widetilde{\mathbf{a}} \| } \,
    \frac{ \widetilde{\mathbf{X}}_j' \, \widetilde{\mathbf{a}} }
    { \| \widetilde{\mathbf{X}}_j \| \, \| \widetilde{\mathbf{a}} \| } 
\;=\; \left( \frac{ \| \widetilde{\mathbf{X}}_j \| }
    { \| \widetilde{\mathbf{X}}_j + \widetilde{\mathbf{a}} \| } \,
    \frac{ \widetilde{\mathbf{X}}_j' \, \widetilde{\mathbf{a}} }
    { \| \widetilde{\mathbf{X}}_j \| \, \| \widetilde{\mathbf{a}} \| } 
\right)^+ \,-\; \left( \frac{ \| \widetilde{\mathbf{X}}_j \| }
    { \| \widetilde{\mathbf{X}}_j + \widetilde{\mathbf{a}} \| } \,
    \frac{ \widetilde{\mathbf{X}}_j' \, \widetilde{\mathbf{a}} }
    { \| \widetilde{\mathbf{X}}_j \| \, \| \widetilde{\mathbf{a}} \| } 
    \right)^- ,
\end{equation*}
where $X^+ = \max(X,0)$ and $X^- = -\min(X,0)$. 

Because $\| \widetilde{\mathbf{X}}_j + \widetilde{\mathbf{a}} \|^2 = 
    \| \widetilde{\mathbf{X}}_j \|^2 + \|\widetilde{\mathbf{a}} \|^2 + 
    2 \widetilde{\mathbf{X}}'_j \widetilde{\mathbf{a}}$, 
\begin{align*}
\left( \frac{ \| \widetilde{\mathbf{X}}_j \| }
    { \| \widetilde{\mathbf{X}}_j + \widetilde{\mathbf{a}} \| } \,
    \frac{ \widetilde{\mathbf{X}}_j' \, \widetilde{\mathbf{a}} }
    { \| \widetilde{\mathbf{X}}_j \| \, \| \widetilde{\mathbf{a}} \| } 
\right)^+ & \;\le\; \left( 
    \frac{ \| \widetilde{\mathbf{X}}_j \| }
    { \sqrt{ \| \widetilde{\mathbf{X}}_j \|^2 + 
    \| \widetilde{\mathbf{a}} \|^2 } } \,
    \frac{ \widetilde{\mathbf{X}}_j' \, \widetilde{\mathbf{a}} }
    { \| \widetilde{\mathbf{X}}_j \| \, \| \widetilde{\mathbf{a}} \| } 
    \right)^+ , \\ 
\left( \frac{ \| \widetilde{\mathbf{X}}_j \| }
    { \| \widetilde{\mathbf{X}}_j + \widetilde{\mathbf{a}} \| } \,
    \frac{ \widetilde{\mathbf{X}}_j' \, \widetilde{\mathbf{a}} }
    { \| \widetilde{\mathbf{X}}_j \| \, \| \widetilde{\mathbf{a}} \| } 
\right)^- & \;\ge\; \left( \frac{ \| \widetilde{\mathbf{X}}_j \| }
    { \sqrt{ \| \widetilde{\mathbf{X}}_j \|^2 + 
    \| \widetilde{\mathbf{a}} \|^2 } }
\frac{ \widetilde{\mathbf{X}}_j' \, \widetilde{\mathbf{a}} }
    { \| \widetilde{\mathbf{X}}_j \| \, \| \widetilde{\mathbf{a}} \| }
    \right)^- .
\end{align*}

These inequalities are identities when 
$\widetilde{\mathbf{X}}_j' \, \widetilde{\mathbf{a}}$ is of 
opposite sign to $(\, \cdot \,)^\pm $, 
or when $\widetilde{\mathbf{X}}_j' \, \widetilde{\mathbf{a}} = 0$. 
Because of the permutation symmetries of 
$ \widetilde{\mathbf{X}}_j / \| \widetilde{\mathbf{X}}_j \| $, 
it follows that 
$\mathrm{P}(\, \widetilde{\mathbf{X}}'_j  \, 
    \widetilde{\mathbf{a}} \ne 0 \,) > 0$, 
which implies 
\begin{align*}
\mathrm{P} \left(\, \left( \frac{ \| \widetilde{\mathbf{X}}_j \| }
    { \| \widetilde{\mathbf{X}}_j + \widetilde{\mathbf{a}} \| } \,
    \frac{ \widetilde{\mathbf{X}}_j' \, \widetilde{\mathbf{a}} }
    { \| \widetilde{\mathbf{X}}_j \| \, \| \widetilde{\mathbf{a}} \| } 
\right)^+ \,<\, \left( \frac{ \| \widetilde{\mathbf{X}}_j \| }
    { \sqrt{ \| \widetilde{\mathbf{X}}_j \|^2 + 
    \| \widetilde{\mathbf{a}} \|^2 } } \,
    \frac{ \widetilde{\mathbf{X}}_j' \, \widetilde{\mathbf{a}} }
    { \| \widetilde{\mathbf{X}}_j \| \, \| \widetilde{\mathbf{a}} \| } 
    \right)^+ \;\right) & \;>\; 0 , \\
\mathrm{P} \left(\, \left( \frac{ \| \widetilde{\mathbf{X}}_j \| }
    { \| \widetilde{\mathbf{X}}_j + \widetilde{\mathbf{a}} \| } \,
    \frac{ \widetilde{\mathbf{X}}_j' \, \widetilde{\mathbf{a}} }
    { \| \widetilde{\mathbf{X}}_j \| \, \| \widetilde{\mathbf{a}} \| } 
\right)^- \,>\, \left( \frac{ \| \widetilde{\mathbf{X}}_j \| }
    { \sqrt{ \| \widetilde{\mathbf{X}}_j \|^2 + 
    \| \widetilde{\mathbf{a}} \|^2 } } \,
    \frac{ \widetilde{\mathbf{X}}_j' \, \widetilde{\mathbf{a}} }
    { \| \widetilde{\mathbf{X}}_j \| \, \| \widetilde{\mathbf{a}} \| }
    \right)^- \;\right) & \;>\; 0 ,
\end{align*}
and hence 
\begin{align*}
\mathrm{E} \left(\, \left( \frac{ \| \widetilde{\mathbf{X}}_j \| }
    { \| \widetilde{\mathbf{X}}_j + \widetilde{\mathbf{a}} \| } \,
    \frac{ \widetilde{\mathbf{X}}_j' \, \widetilde{\mathbf{a}} }
    { \| \widetilde{\mathbf{X}}_j \| \, \| \widetilde{\mathbf{a}} \| } 
    \right)^+ \;\right) & \,<\, 
\mathrm{E} \left(\, \left( \frac{ \| \widetilde{\mathbf{X}}_j \| }
    { \sqrt{ \| \widetilde{\mathbf{X}}_j \|^2 + 
    \| \widetilde{\mathbf{a}} \|^2 } } \,
    \frac{ \widetilde{\mathbf{X}}_j' \, \widetilde{\mathbf{a}} }
    { \| \widetilde{\mathbf{X}}_j \| \, \| \widetilde{\mathbf{a}} \| } 
    \right)^+ \;\right) , \\
\mathrm{E} \left(\, \left( \frac{ \| \widetilde{\mathbf{X}}_j \| }
    { \| \widetilde{\mathbf{X}}_j + \widetilde{\mathbf{a}} \| } \,
    \frac{ \widetilde{\mathbf{X}}_j' \, \widetilde{\mathbf{a}} }
    { \| \widetilde{\mathbf{X}}_j \| \, \| \widetilde{\mathbf{a}} \| } 
\right)^- \;\right) & \,>\, 
\mathrm{E} \left(\, \left( \frac{ \| \widetilde{\mathbf{X}}_j \| }
    { \sqrt{ \| \widetilde{\mathbf{X}}_j \|^2 + 
    \| \widetilde{\mathbf{a}} \|^2 } } \,
    \frac{ \widetilde{\mathbf{X}}_j' \, \widetilde{\mathbf{a}} }
    { \| \widetilde{\mathbf{X}}_j \| \, \| \widetilde{\mathbf{a}} \| } 
    \right)^- \;\right) .
\end{align*}

For any permutation matrix $P \in \mathcal{P}_n$, 
\begin{align*}
\frac{ \| \widetilde{\mathbf{X}}_j \| }
    { \sqrt{ \| \widetilde{\mathbf{X}}_j \|^2 + 
    \| \widetilde{\mathbf{a}} \|^2 } }
& \;=\; \frac{ \| P \, \widetilde{\mathbf{X}}_j \| }
    { \sqrt{ \| P \, \widetilde{\mathbf{X}}_j \|^2 + 
    \| \widetilde{\mathbf{a}} \|^2 } } \quad \text{surely,} \\ 
\frac{ \widetilde{\mathbf{X}}_j } 
    { \| \widetilde{\mathbf{X}}_j \| } 
& \;=\; P \, \frac{ \widetilde{\mathbf{X}}_j } 
    { \| \widetilde{\mathbf{X}}_j \| } \quad \text{in distribution,}
\end{align*}
so that
\begin{equation*}
\frac{ \| \widetilde{\mathbf{X}}_j \| }
    { \sqrt{ \| \widetilde{\mathbf{X}}_j \|^2 + 
    \| \widetilde{\mathbf{a}} \|^2 } } \, 
    \frac{ \widetilde{\mathbf{X}}_j } 
    { \| \widetilde{\mathbf{X}}_j \| }
\;=\; P \, \frac{ \| \widetilde{\mathbf{X}}_j \| }
    { \sqrt{ \| \widetilde{\mathbf{X}}_j \|^2 + 
    \| \widetilde{\mathbf{a}} \|^2 } }  \, 
    \frac{ \widetilde{\mathbf{X}}_j } 
    { \| \widetilde{\mathbf{X}}_j \| }\quad \text{in distribution,} 
\end{equation*}
which together with 
\begin{equation*}
\left(\, \frac{ \| \widetilde{\mathbf{X}}_j \| }
    { \sqrt{ \| \widetilde{\mathbf{X}}_j \|^2 + 
    \| \widetilde{\mathbf{a}} \|^2 } } \, 
    \frac{ \widetilde{\mathbf{X}}_j } 
    { \| \widetilde{\mathbf{X}}_j \| } \,\right)' \, \widehat{\mathbf{1}}
\;=\; 0 \quad \text{surely,} 
\end{equation*}
implies, as in \eqref{eq:x-expval}, that 
\begin{equation*}
\mathrm{E} \left(\, \frac{ \| \widetilde{\mathbf{X}}_j \| } 
    { \sqrt{ \| \widetilde{\mathbf{X}}_j \|^2 + 
    \| \widetilde{\mathbf{a}} \|^2 } } \, 
    \frac{ \widetilde{\mathbf{X}}_j } 
    { \| \widetilde{\mathbf{X}}_j \| } \,\right) 
\;=\; \mathbf{0} . 
\end{equation*}

Therefore, 
\begin{equation*}
\mathrm{E} \left(\, \left( \frac{ \| \widetilde{\mathbf{X}}_j \| }
    { \sqrt{ \| \widetilde{\mathbf{X}}_j \|^2 + 
    \| \widetilde{\mathbf{a}} \|^2 } }
    \frac{ \widetilde{\mathbf{X}}_j' \, \widetilde{\mathbf{a}} }
    { \| \widetilde{\mathbf{X}}_j \| \, \| \widetilde{\mathbf{a}} \| }
    \right)^+ \;\right) 
\;=\; \mathrm{E} \left(\, \left( \frac{ \| \widetilde{\mathbf{X}}_j \| }
    { \sqrt{ \| \widetilde{\mathbf{X}}_j \|^2 + 
    \| \widetilde{\mathbf{a}} \|^2 } }
    \frac{ \widetilde{\mathbf{X}}_j' \, \widetilde{\mathbf{a}} }
    { \| \widetilde{\mathbf{X}}_j \| \, \| \widetilde{\mathbf{a}} \| } 
    \right)^- \;\right) .
\end{equation*}

Back to the initial expected values, it follows that 
\begin{equation*}
\mathrm{E} \left(\, \left( \frac{ \| \widetilde{\mathbf{X}}_j \| }
    { \| \widetilde{\mathbf{X}}_j + \widetilde{\mathbf{a}} \| }
    \frac{ \widetilde{\mathbf{X}}_j' \, \widetilde{\mathbf{a}} }
    { \| \widetilde{\mathbf{X}}_j \| \, \| \widetilde{\mathbf{a}} \| } 
    \right)^+ \;\right) 
\;<\; \mathrm{E} \left(\, \left( \frac{ \| \widetilde{\mathbf{X}}_j \| }
    { \| \widetilde{\mathbf{X}}_j + \widetilde{\mathbf{a}} \| }
    \frac{ \widetilde{\mathbf{X}}_j' \, \widetilde{\mathbf{a}} }
    { \| \widetilde{\mathbf{X}}_j \| \, \| \widetilde{\mathbf{a}} \| } 
\right)^- \;\right) , 
\end{equation*}
which implies 
\begin{equation*}
\mathrm{E} \left( 
    \frac{ \| \widetilde{\mathbf{X}}_j \| }
    { \| \widetilde{\mathbf{X}}_j + \widetilde{\mathbf{a}} \| } 
    \frac{ \widetilde{\mathbf{X}}_j' \, \widetilde{\mathbf{a}} }
    { \| \widetilde{\mathbf{X}}_j \| \, \| \widetilde{\mathbf{a}} \| }
    \right) \;<\; 0 .
\quad \square
\end{equation*}

The Gaussian multivariate distribution, among others, has spherical 
symmetry besides permutation symmetry. 
For parametric families with spherical symmetry, 
the true standard vector 
$ \sqrt{n-1} \, \widetilde{\mathbf{X}}_j / 
    \| \widetilde{\mathbf{X}}_j \|$ 
has uniform distribution over the $(n-2)$-sphere. 
As a result, the components of 
$ \widetilde{\mathbf{Y}}_j / \| \widetilde{\mathbf{Y}}_j \|$ 
perpendicular to $\widetilde{\mathbf{a}}$ 
are antisymmetric with respect to the direction of 
$\widetilde{\mathbf{a}}$, so that they cancel out in expectation. 
That is, for parametric families with spherical symmetry, 
and as long as $\widetilde{\mathbf{a}} \ne \mathbf{0}$, 
\begin{equation}
\label{eq:y-aproj-spher-sym}
\mathrm{E} \left( 
    \frac{ \widetilde{\mathbf{Y}}_j }{ \| \widetilde{\mathbf{Y}}_j \| } 
    \right) \;=\; \lambda \widetilde{\mathbf{a}} , 
    \quad \text{with} \quad
    0 < \lambda < \mathrm{E} \left( 
    \frac{ 1 }{ \| \widetilde{\mathbf{Y}}_j \| } \right) . 
\end{equation}

\subsection{Standard-vector normalization}

The properties \eqref{eq:y-aproj}, \eqref{eq:y-aproj-spher-sym} suggest 
the use of 
\begin{equation}
\widehat{\mathbf{b}} \;=\; \frac{\displaystyle \sum_{j=1}^m 
    \frac{ \widetilde{\mathbf{y}}_j }
    { \| \widetilde{\mathbf{y}}_j \| } }
    {\displaystyle  \sum_{j=1}^m \frac{ 1 }
    { \| \widetilde{\mathbf{y}}_j \| } } 
\end{equation} 
to approximate the unknown residual vector of normalization factors 
$\widetilde{\mathbf{a}}$. 
The following iterative method implements this approach to solve the 
\emph{normalization of residuals problem}.

Let us define the following recursive sequence, 
where each step $t$ comprises $m$ vectors 
$\widehat{\mathbf{y}}_j^{(t)}$ ($j \in \{1,\dots,m\}$) and one vector 
$\widehat{\mathbf{b}}^{(t)}$,
\begin{align}
\label{eq:y0-seq-def}
\widehat{\mathbf{y}}_j^{(0)} & \;=\; \widetilde{\mathbf{y}}_j , \\
\label{eq:yt-seq-def}
\widehat{\mathbf{y}}_j^{(t)} & \;=\; \widehat{\mathbf{y}}_j^{(t-1)} - 
    \, \widehat{\mathbf{b}}^{(t-1)} ,
    \quad \text{for } t \ge 1 , \\
\label{eq:b-seq-def}
\widehat{\mathbf{b}}^{(t)} & \;=\; \frac{\displaystyle 
     \sum_{i=1}^m  \frac{ \widehat{\mathbf{y}}_j^{(t)} }
    { \| \widehat{\mathbf{y}}_j^{(t)} \| } }
    {\displaystyle \sum_{i=1}^m \frac{1}
    { \| \widehat{\mathbf{y}}_j^{(t)} \| } } ,
    \quad \text{for } t \ge 0 .
\end{align}

We assume that $\widehat{\mathbf{y}}_j^{(t)} \ne \mathbf{0}_n$, 
for all $j \in \{1,\dots,m\}$ and all $t \ge 0$. 
Nonetheless, an implementation of this algorithm benefits from 
trimming out a small fraction (e.g.\ 1\%) of the features with lesser 
$\| \widehat{\mathbf{y}}_j^{(t)} \|$ in \eqref{eq:b-seq-def}, 
in order to avoid numerical singularities. 

Let us write $\widehat{\mathbf{y}}_j^{(t)}$ as a function of the unknowns 
$\widetilde{\mathbf{x}}_j$ and $\widetilde{\mathbf{a}}$. 
For any $t \ge 1$,
\begin{align}
\widehat{\mathbf{y}}_j^{(t)} & \;=\; \widehat{\mathbf{y}}_j^{(t-1)} - 
    \, \widehat{\mathbf{b}}^{(t-1)} , \\
& \;=\; \widehat{\mathbf{y}}_j^{(t-2)}
    - \,\widehat{\mathbf{b}}^{(t-2)} - \,\widehat{\mathbf{b}}^{(t-1)} , \\
& \;\;\vdots \\
& \;=\; \widehat{\mathbf{y}}_j^{(0)} 
    - \,\sum_{r=0}^{t-1} \widehat{\mathbf{b}}^{(r)} , \\
& \;=\; \widetilde{\mathbf{y}}_j 
    - \,\sum_{r=0}^{t-1} \widehat{\mathbf{b}}^{(r)} , \\
\label{eq:x-tele}
& \;=\; \widetilde{\mathbf{x}}_j + \widetilde{\mathbf{a}} 
    - \,\sum_{r=0}^{t-1} \widehat{\mathbf{b}}^{(r)} .
\end{align}
Note that \eqref{eq:x-tele} is also valid for $t = 0$. 

Let us also define the vectors $\widehat{\mathbf{a}}^{(t)}$, 
for $t \ge 0$, 
which describe the vector of normalization factors still to be 
removed at step $t$, 
\begin{equation}
\label{eq:a-def}
\widehat{\mathbf{a}}^{(t)} \;=\; \widetilde{\mathbf{a}} 
    - \,\sum_{r=0}^{t-1} \widehat{\mathbf{b}}^{(r)} , 
\end{equation}
so that, by \eqref{eq:x-tele}, for $t \ge 0$, 
\begin{equation}
\label{eq:y-x-a}
\widehat{\mathbf{y}}_j^{(t)} \;=\; 
    \widetilde{\mathbf{x}}_j + \widehat{\mathbf{a}}^{(t)} .
\end{equation}

Therefore, the recursive sequence 
\eqref{eq:y0-seq-def}--\eqref{eq:b-seq-def} faces a new, 
weaker \emph{normalization of residuals problem} at each step $t$, 
with true residual vectors $\widetilde{\mathbf{x}}_j$, 
observed residual vectors $\widehat{\mathbf{y}}_j^{(t)}$ and 
unknown normalization factors $\widehat{\mathbf{a}}^{(t)}$. 
The step $t$ results in the estimation of normalization factors 
$\widehat{\mathbf{b}}^{(t)}$, 
which are removed from $\widehat{\mathbf{y}}_j^{(t)}$, 
generating the next step. 
At the beginning, 
$\widehat{\mathbf{y}}_j^{(0)} = \widetilde{\mathbf{y}}_j$ 
and $\widehat{\mathbf{a}}^{(0)} = \widetilde{\mathbf{a}}$. 

At convergence, 
$\lim_{t \to \infty} \widehat{\mathbf{b}}^{(t)} = \mathbf{0}$. 
Equations~\eqref{eq:x-expval}, \eqref{eq:y-expval}, \eqref{eq:b-seq-def} 
imply that, in such a case, 
$\lim_{t \to \infty} \widehat{\mathbf{y}}_j^{(t)} = 
    \widetilde{\mathbf{x}}_j$ and 
$\sum_{t=0}^\infty \widehat{\mathbf{b}}^{(t)} = 
    \widetilde{\mathbf{a}}$. 
Convergence is optimal when the parametric family of the $m$ 
features has spherical symmetry, 
Gaussian being the most prominent case. 
Otherwise, the more uniform the distribution of standard vectors 
$ \sqrt{n-1} \, \widetilde{\mathbf{x}}_j / 
    \| \widetilde{\mathbf{x}}_j \| $ is on the $(n-2)$-sphere, 
the faster the sequence \eqref{eq:y0-seq-def}--\eqref{eq:b-seq-def} 
converges. 
See examples of convergence in Supplementary Movies~S1--S3.

\subsection{Identification of no-variation genes}

Let us consider a gene expression dataset, 
with $g$ genes and $c$ experimental conditions. 
Each condition $k$ has $s_k$ samples. 
The total number of samples is $s = \sum_{k=1}^c s_k$. 
Let us assume that $c \ge 2$ and that $s_k \ge 2$, for all conditions 
$k \in \{1,\dots,c\}$. 
Let us also assume that, among the $g$ genes, 
there is a fraction $\pi_0$ of non-differentially expressed genes (non-DEGs), 
with $0 \le \pi_0 \le 1$, 
while the remaining fraction $1-\pi_0$ comprises the differentially expressed 
genes (DEGs) \citep{storey:2003a}. 

Let us consider the usual ANOVA test comparing average expression levels
across conditions, gene-by-gene. 
Under the null hypothesis of a non-DEG, 
the corresponding $F$-statistic follows the $F$-distribution with 
$c-1$ and $s-c$ degrees of freedom. 
The test of this hypothesis yields a $p$-value $p_j$ for each gene 
$j \in \{1,\dots,g\}$. 
The obtained $p$-values $p_j$ follow a probability distribution that can 
be considered as the mixture of two probability distributions, 
$F_0$ and $F_1$, 
for the non-DEGs and the DEGs, 
respectively \citep{storey:2003b}. 
The fraction $\pi_0$ of non-DEGs follows the uniform distribution on 
the interval $[0,1]$, 
\begin{equation}
F_0(p) \;=\; p ,
\end{equation}
while the fraction $1-\pi_0$ of DEGs follows a distribution that 
verifies, for any $p \in (0,1)$, 
\begin{equation}
F_1(p) \;>\; p ,
\end{equation}
and the mixture distribution is 
\begin{equation}
F(p) \;=\; \pi_0 F_0(p) + (1-\pi_0) F_1(p) .
\end{equation}

Let us further assume that there exists a $p^*$, with $0 < p^* < 1$, 
such that $F_1(p) = 1$ for every $p \ge p^*$. 
In other words, all DEGs have $p$-value $p_j$ from the ANOVA test 
such that $p_j \le p^*$, 
while only some genes among the non-DEGs have p-value with 
$p_j > p^*$. 
This implies that the mixture distribution of $p$-values is uniform on 
the interval $[p^*,1]$, 
\begin{align}
F(p) & \;=\; \pi_0 p + 1 - \pi_0 , \quad \text{for } p^* \le p \le 1 , \\
f(p) & \;=\; \pi_0 , \quad \text{for } p^* < p < 1 .
\end{align}

On the other hand, for any set of $n$ samples 
$x_{(1)} \le x_{(2)} \le \dots \le x_{(n)}$ 
obtained from $n$ independent and identically distributed uniform random 
variables on the interval $[a,b]$, 
all the distances between consecutive ordered samples (including 
boundaries), 
$x_{(1)}-a, x_{(2)}-x_{(1)}, \dots, x_{(n)}-x_{(n-1)}, b-x_{(n)}$, 
obey the same distribution \citep{feller:1971}. 
Then, it can be realized that, 
for any $j$ such that $2 \le j \le n-1$, 
the two subsets of samples $x_{(1)}, \dots, x_{(j-1)}$ and 
$x_{(j+1)}, \dots, x_{(n)}$ follow uniform distributions on the 
intervals $[a,x_{(j)}]$ and $[x_{(j)},b]$, respectively. 

Based on these facts, 
to identify no-variation genes we propose finding the minimum 
$p_{(j)}$, from the ordered sequence of $p$-values 
$p_{(1)} \le p_{(2)} \le \dots \le p_{(g)}$, 
such that a goodness-of-fit test for the uniform 
distribution on the interval $[p_{(j)},1]$, 
performed on $p_{(j+1)}, \dots, p_{(g)}$, is not rejected. 
As a result, the genes corresponding to the $p$-values 
$p_{(j)}, p_{(j+1)}, \dots, p_{(g)}$ are considered as no-variation genes. 

Given the concavity of $F(p)$, 
the goodness-of-fit test used is the one-sided Kolmogorov-Smirnov test 
on positive deviations of the empirical distribution function. 

See Supplementary Movies~S4--S6 for examples of this approach to identifying 
no-variation genes.

\end{document}